\documentclass[fleqn,usenatbib,useAMS]{mnras}

\usepackage{graphicx}	% Including figure files\usepackage[T1]{fontenc}
\usepackage{amsmath}	% Advanced maths commands
\usepackage{amssymb}	% Extra maths symbols
\usepackage{multicol}        % Multi-column entries in tables
\usepackage{bm}		% Bold maths symbols, including upright Greek
\usepackage{pdflscape}	% Landscape pages
\usepackage{caption}
\usepackage[table]{xcolor}

\usepackage[T1]{fontenc}
\usepackage{ae,aecompl}

\usepackage{newtxtext,newtxmath}
\title[Time delays between radio wavelengths and X-ray of Sgr A*]
{Improved analysis of time delays between radio wavelengths and X-ray of Sgr A* flares in the shock oscillation model}
\author[ Aktar, Okuda \& Singh]{Ramiz Aktar$^{1}$ \thanks{Contact e-mail: \href{ramizaktar@gmail.com}{ramizaktar@gmail.com}}, Toru Okuda$^2$, Chandra B. Singh$^{3}$%
\\
$^{1}${Institute of Astronomy and Astrophysics, Academia Sinica, Taipei 10617, Taiwan, R.O.C.}\\
$^{2}${Hakodate Campus, Hokkaido University of Education, Hachiman-Cho 1-2, Hakodate
040-8567, Japan}\\
$^{3}${Calle 134A No. 148A 21 San Pedro de Tibabuyes, Suba - Bogotá, Colombia }\\
}
\pubyear{2022}

\begin{document}
\label{firstpage}
\pagerange{\pageref{firstpage}--\pageref{lastpage}}
\maketitle

%%%%%%%%%%%%%%%%%%%%%%%%%%%%%%%%%%%%%%%%%%%%%%%%%%%%%%%%%%%%%%%%%%%%
\begin{abstract}

We present an improved analysis of the time delays between radio and X-ray flares from the supermassive black hole Sagittarius A* (Sgr A*), extending our previous shock oscillation model by incorporating Compton scattering. Our simulations predict delay times of 1–4 h between X-ray and radio flares, 20–40 min and 0.5–1 h between 334 GHz and 4.5 $\mu$m and 2–8 keV flares, and 7–20 min and 0.5–1 h between 22–43 GHz and 350 GHz radio flares, respectively, in good agreement with recent observations. The delay time between the radio emissions (synchrotron) and X-ray emissions (bremsstrahlung) is interpreted as the transit time for sound waves traveling from the central core near the event horizon to the location of the oscillating shock. Additionally, the delay time observed between different radio wavelengths corresponds to the transit time of sound waves moving between the frequency-dependent effective radii \( R_{\rm eff}(\nu) \), where the optical thickness \( \tau_{\nu}(R) \) at frequency \( \nu \) equals one on the surface of the accretion disc. The X-ray emission and the radio emission at lower frequencies peak later than the radio emission at higher frequencies. This delay occurs because the acoustic perturbations created by synchrotron emission peak near the event horizon and then expand outward. Conversely, an inverse delay happens when the acoustic waves generated by the upstream shock wave travel toward the event horizon. The predicted delay times from this interpretation qualitatively match the numerical results obtained in simulations.

\end{abstract}

%%%%%%%%%%%%%%%%%%%%%%%%%%%%%%%%%%%%%%%%%%%%%%%%%%

%%%%%%%%%%%%%%%%% BODY OF PAPER %%%%%%%%%%%%%%%%%%

% The MNRAS class isn't designed to include a table of contents, but for this document one is useful.
% I therefore have to do some kludging to make it work without masses of blank space.
\begingroup
\let\clearpage\relax
%\tableofcontents
\endgroup
\newpage

%%%%%%%%%%%%%%%%%%%%%%%%%%%%%%%%%%%%%%%%%%%%%%%%%%%%%%%%%%%%%%%%%%%%%
\section{Introduction} \label{sec:intro}

Sagittarius A$^*$ (Sgr A$^*$), the supermassive black hole located at the center of our galaxy, has displayed some unusual characteristics since its discovery, including an exceptionally low luminosity and complex spectral properties \citep{Genzel-etal-03, Genzel-etal-10}. The observed luminosity is five orders of magnitude lower than the predictions made by the standard thin disk model \citep{Shakura-Sunyaev73} (hereafter referred to as the SS73 model). Additionally, the spectrum of Sgr A$^*$ significantly deviates from the multi-temperature blackbody spectra expected from the SS73 model. Over the past three decades, various theoretical models have been proposed to explain this low luminosity and the unique spectral characteristics. The advection-dominated accretion flow (ADAF) model, which involves high angular momentum, has been particularly successful in accounting for the observations \citep{Narayan-McClintock-08, Yuan-Narayan14}. Alongside high angular momentum flows such as ADAF, models incorporating low angular momentum flows and the formation of standing shocks have been developed to explain the quasi-periodic oscillations (QPOs) observed around black holes. The study of standing shocks in this astrophysical context was initially pioneered by \citep{Fukue-87}, followed by \citep{Chakrabarti89}. Several subsequent analytical studies have also investigated this phenomenon using axisymmetric \citep{Chakrabarti-90a, Chakrabarti-1990, Chakrabarti-Titarchuk95, Lu-etal99, Das-etal01, Das-Chattopadhyay-Nandi-01, Chakrabarti-Das-01, Fukumura-Tsuruta04, Chakrabarti-Das04, Aktar-etal15, Sarkar-Das16, Aktar-etal17, Aktar-etal18, Dihingia-etal18, Aktar-etal19, Mitra-etal-22,  Sarkar-etal-23, Sarkar-25} and non-axisymmetric \citep{Spruit-87, Chakrabarti-90b, Aktar-etal21, Aktar-etal-23} assumption. On the other hand, various independent numerical simulations have examined shock-induced accretion flows around black holes \citep{Chakrabarti-Molteni-93, Molteni-etal-94, Chakrabarti-Molteni-95, Molteni-etal-96a, Molteni-etal-96b, Ryu-etal-97, Das-etal14, Okuda-etal04, Okuda-etal07, Okuda-Das15, Kim-etal-19, Okuda-etal19, Okuda-etal22, Garain-Kim-23, Debnath-etal-24, Dihingia-etal-25, Mao-etal-25}.

Since the pioneering work on magnetized discs with shear instability \citep{Balbus-Hawley91}, multidimensional magnetohydrodynamic (MHD) simulations have shown that the magnetic field and the resulting magnetorotational instability (MRI) are crucial not only for the structure of accretion flows but also for the time variations in luminosity and spectra around black holes \citep{Machida-etal00, Machida-etal01, Stone-Pringle01, Igumenshchev-etal-03, Narayan-etal-03, Narayan-etal12, Yuan-etal-12, Yuan-etal-15}. Several magnetohydrodynamics (MHD) simulation studies have aimed to investigate the rapid flares of Sgr A* \citep{Proga-Begelman-03, Chan-etal-09, Dexter-etal-09, Yuan-etal-09, Dodds-Eden-etal-10, Ball-etal-16, Ressler-etal-17, Roberts-etal-17, Li-etal-17, Okuda-etal19, Singh-etal-21}. Recent observations by \citet{Event-Horizon-etal-2022a, Event-Horizon-etal-2022b} have detected near-infrared flares close to the event horizon of Sgr A* and confirm the crucial role of the magnetic field. These studies also provide a physical interpretation of the Event Horizon Telescope (EHT) data, spanning radio to X-ray wavelengths, by comparing it with predictions derived from time-dependent general relativistic magnetohydrodynamics (GRMHD) simulations. A recent observational campaign involving simultaneous radio, near-infrared (NIR), and X-ray data collected from CHANDRA, HST NICMOS, XMM-Newton, and VLA revealed flares occurring on a timescale of approximately several hours between the radio and X-ray emissions. Additionally, there was a time lag of a few minutes between the radio bands, infrared, and X-ray signals from Sgr A* \citep{Yusef-Zadeh-etal-06a, Yusef-Zadeh-etal-09, Degenaar-etal-13, Neilsen-etal-13, Neilsen-etal-15, Ponti-etal-15, Ponti-etal-17, Capellupo-etal-17, Witzel-etal-21, Michail-etal-21a, Michail-etal-21b, Boyce-etal-22}.

Motivated by these observations, we aimed to explain the delay times between radio and X-ray emissions, as well as between different radio wavelengths, by examining the shock oscillation model for Sgr A* using 2D radiation, relativistic (2D-Rad-RMHD) simulations. Our findings indicated that the magnetized flows cause significant fluctuations in luminosity, with timescales of approximately 5 to 10 days \citep{Okuda-etal19}. Additionally, we observed a time delay of 1 to 4 hours between radio and X-ray flares and a delay of several tens of minutes between different radio wavelengths of Sgr A* \citep{Okuda-etal22, Okuda-etal23}. In this paper, we further investigate the time delay relationships of Sgr A* flares, taking into account Compton scattering, and provide a more comprehensive explanation based on improved physical interpretations, revising our previous interpretations of the time delay phenomena.

%%%%%%%%%%%%%%%%%%%%%%%%%%%%%%%%%%%%%%%%%%%%%%%%%%%%%%%%%%%%%%%%%%%%%
\section{Numerical method}

Our simulation method for analyzing time delays involves the following steps. First, we conduct 2D relativistic radiation magnetohydrodynamic (2D-Rad-RMHD) simulations of the accretion flow around Sgr A$^*$ using a single-temperature model. We apply appropriate initial and boundary conditions, utilizing the Rad-RMHD module of the publicly available code PLUTO\footnote{\url{http://plutocode.ph.unito.it}} \citep{Migone-etal07, Fuksman-Mignone-19}. From these simulations, we extract the time variations of the primitive variables, which include density ($\rho$), velocity ($\textbf{v}$), magnetic field ($\textbf{B}$), gas pressure ($p_{\rm g}$), radiation energy density ($E_{r}$), and radiation flux ($\textbf{F}$). In the post-processing phase, we implement a two-temperature model to derive the electron and ion temperatures of the gas. Additionally, we calculate the monochromatic radiation flux on the surface of the accretion disc. Importantly, this model incorporates the effects of Compton scattering on luminosity for the first time, which influences the analysis of time delays.

%Our numerical method for analyzing time delays is as follows. First, we perform 2D relativistic radiation magnetohydrodynamic (2D-Rad-RMHD) simulations of the accretion flow around Sgr A$^*$ using a single temperature model. We apply appropriate initial and boundary conditions, utilizing the Rad-RMHD module of the publicly available code PLUTO\footnote{\url{http://plutocode.ph.unito.it}} \citep{Migone-etal07, Fuksman-Mignone-19}. From these simulations, we obtain the time variations of the primitive variables: density ($\rho$), velocity ($\textbf{v}$), magnetic field ($\textbf{B}$), gas pressure ($p_{\rm g}$), radiation energy density ($E_{r}$), and radiation flux ($\textbf{F}$). In the post-processing phase, we apply a two-temperature model to derive the electron and ion temperatures of the gas. Additionally, we calculate the monochromatic radiation flux on the surface of the accretion disc. Notably, this model incorporates the effects of Compton scattering on luminosity for the first time, which influences the time delay analysis.

%%%%%%%%%%%%%%%%%%%%%%%%%%%%%%%%%%%%%%%%%%%%%%%%%%%%%%%%%%%%%%%%%%%%% 
\subsection{Initial and boundary conditions}

In this study, we utilize the initial conditions derived from our previous numerical model of 2D relativistic radiation magnetohydrodynamics (MHD) simulations \citep{Okuda-etal23}. We generally divide the outer radial boundary at \( R = R_{\text{out}} \) into two parts. The first part is the disc boundary, through which matter enters from the outer radius. At this disc boundary (defined by \( -h_{\text{out}} \leq z \leq h_{\text{out}} \) at \( R = R_{\text{out}} \), where \( h \) is the disc height), we impose a continuous inflow of matter with constant variables. The second part of the outer boundary region is located above the accretion disc, where we impose free-floating conditions that allow for the outflow of matter, while prohibiting any inflow. At the outer vertical boundary defined by \( z = \pm z_{\text{out}} \), we also apply free-floating conditions. Along the rotating axis, all variables are set to be symmetric with respect to the axis. The inner boundary, at \( R = R_{\text{in}} (= 2.2 R_{\text{g}}) \), is treated as an absorbing boundary since it lies below the last stable circular orbital radius, which is \( 3 R_{\text{g}} \). Here, the velocity is set to the free-fall velocity, while other variables are interpolated from the values just outside the inner boundary \citep{Okuda-etal22, Okuda-etal23, Aktar-etal24a, Aktar-etal24b}. The Schwarzschild radius \( R_{\text{g}} \) is given as \( R_{\text{g}} = \frac{2GM}{c^2} \).

%Here we use the initial conditions based on our previous numerical model of 2D relativistic radiation MHD simulations \citep{Okuda-etal23}. In general, we divide the outer radial boundary at $R=R_{\rm out}$ into two parts. One is the disk boundary through which matter is entering  from the outer radius. At the disc boundary ( $-h_{\rm out} \leq z \leq h_{\rm out}$ at $R$ =$R_{\rm out}$) where  $h$ is the disc height, we impose continuous inflow of matter with constant variables. In the other outer boundary region above the accretion disk. we impose free floating conditions and allow for outflow of matter, whereas any inflow is prohibited. At the outer vertical boundary $z=\pm z_{\rm out}$, we also impose the free floating conditions. On the rotating axis, all variables are set to be symmetric relative to the axis. The inner boundary at $R = R_{\rm in} (=2.2R_{\rm g}$) is treated as the absorbing boundary since it is below the last stable circular orbital radius 3$R_{\rm g}$, that is, the velocity is given by the free fall velocity and other variables are given by interpolation of the variables just outside the inner boundary \citep{Okuda-etal22, Okuda-etal23, Aktar-etal24a, Aktar-etal24b}. $R_{\rm g}$ is the Schwarzschild radius given by $R_{\rm g} = 2GM/c^2$.

In the present model, the specific angular momentum \(\lambda\) and the plasma beta \(\beta_0 = \frac{8\pi p_{\text{out}}}{B_{\text{out}}^2}\) at the radial outer boundary are particularly significant. Here, \(p_{\text{out}}\) represents the gas pressure, and \(B_{\text{out}}\) denotes the strength of the magnetic field at the outer boundary. These parameters play a crucial role in determining the characteristics of the oscillating shock. If \(\lambda\) is large, the resulting standing shock extends beyond the outer boundary. Conversely, if \(\beta_0\) is excessively large—indicating a weak magnetic field—the shock stabilizes, making it difficult to achieve an adequate oscillating shock. Referring to our previous studies on models with \(\lambda = 1.3\) and \(\lambda = 1.35\) \citep{Okuda-etal19, Okuda-etal23, Singh-etal-21}, we select \(\lambda = 1.28\) and \(\beta_{\text{out}} = 50\). The outer radial boundary conditions at \(R_{\text{out}}\) within the disc, defined by \(-h_{\text{out}} \le z \le h_{\text{out}}\), are summarized in Table \ref{Table_1}. The computational domain extends from \(0 \leq R \leq 200\) and \(-200 \leq z \leq 200\), with a resolution of \(410 \times 820\) cells in cylindrical coordinates \((R, \phi, z)\), where \(R\) and \(z\) are measured in units of the Schwarzschild radius (\(R_{\text{g}}\)). In the PLUTO code, the inner boundary conditions are treated as if the coordinates were spherical, even though cylindrical coordinates are used \citep{Aktar-etal-25, Aktar-etal-26}.

%%%%%%%%%%%%%%%%%%%%%%%%%%%%%%%%%%%%%%%%%%%%%%%%%%%%%%%%%%%%%%%%%%%%%
%                         Table 1
%%%%%%%%%%%%%%%%%%%%%%%%%%%%%%%%%%%%%%%%%%%%%%%%%%%%%%%%%%%%%%%%%%%%%
\begin{table*}
\centering
\caption{Specific angular momentum $\lambda_{\rm out}$, plasma beta $\beta_{\rm out}$, radial speed $v_{\rm out}$, the sound speed $a_{\rm out}$, the input density $\rho_{\rm out}$, the temperature $T_{\rm out}$, the injection height of accretion flow $h_{\rm out}$, the radiation energy density $(E_{\rm r})_{\rm out}$, the input accretion rate ${\dot M}_{\rm input}$ at the outer radial boundary $R_{\rm out}$= 200, where $\rho_{0} = 10^{-20}$ g  cm$^{-3}$ is the unit density used in this simulation.}
\label{Table_1}
\vspace{3mm}
\begin{tabular}{@{}ccccccccc} \hline \hline
&&&&&& \\
 $\lambda_{\rm out}$ & $\beta_{\rm out}$ & $v_{\rm out}$ & $a_{\rm out}$ & $\rho_{\rm out}$ & $T_{\rm out}$ & $h_{\rm out}$ & $(E_{\rm r})_{\rm out}$ & ${\dot M}_{\rm input}$  \\
  ($R_{g}c$)& & ($c$) & ($c$) & (${\rho}_0$) & ($ \rm K$) & ($R_{\rm out}$) & (${\rho}_0 c^2$) & ($M_{\odot} \rm{yr}^{-1}$)  \\
 \hline \\
1.28        &      50 & $-4.98 \times 10^{-2}$ & $2.73 \times 10^{-2}$ & 58.7    & $2.538 \times 10^9$ & 0.4315 & $9.1 \times 10^{-8}$  & $4.0 \times 10^{-6}$ \\ 

\hline
\end{tabular}
\end{table*}
%%%%%%%%%%%%%%%%%%%%%%%%%%%%%%%%%%%%%%%%%%%%%%%%%%%%%%%%%%%%%%%%%%%%%

%%%%%%%%%%%%%%%%%%%%%%%%%%%%%%%%%%%%%%%%%%%%%%%%%%%%%%%%%%%%%%%%%%%%%
\subsection{Two-temperature model}
\label{two-temp}

To evaluate the synchrotron, bremsstrahlung, and monochromatic luminosities, we need to know the electron and ion temperatures of the gas. We utilize a two-temperature model based on the assumption of radiation energy equilibrium. In this model, the energy transfer rate from ions to electrons due to Coulomb collisions, represented as \( q^{\rm ie} \), is equal to the sum of the synchrotron cooling rate \( q_{\rm syn} \) and the bremsstrahlung cooling rate \( q_{\rm br} \) affecting the electrons. It is given by
\begin{equation} \label{Coulmb_syn_brem_eqn}
 q^{\rm ie} =  q_{\rm syn} +  q_{\rm br}.
\end{equation}
where $q^{\rm ie}$ is given as follows \citep{Stepney-Guilbert-83}.
 \begin{eqnarray}
 &q^{\rm ie}&=5.61\times 10^{-32} {{n_{\rm e}n_{\rm i}(T_{\rm i}-T_{\rm e})}\over {K_2(1/\theta_{\rm e})
  K_2(1/\theta_{\rm i})}} \nonumber \\
  && \times \left[{{2(\theta_{\rm e}+\theta_{\rm i})^2+1}\over (\theta_{\rm e}+\theta_{\rm i})}
  K_1({{\theta_{\rm e}+\theta_{\rm i}}\over {\theta_{\rm e}\theta_{\rm i}}}) 
  +2K_0({{\theta_{\rm e}+\theta_{\rm i}}\over {\theta_{\rm e}\theta_{\rm i}}})\right] \nonumber \\
   &&  \hspace{3.8cm} {\rm erg \ cm^{-3} \; s^{-1}},
 \end{eqnarray}
where $n_{\rm e}$ and $n_{\rm i}$ are the number density of electrons and ions, $K_0$, $K_1$ and $K_2$ are modified Bessel functions, and the dimensionless electron and ion temperature are defined by

\begin{equation}
  \theta_{\rm e}= {kT_{\rm e}\over {m_{\rm e}c^2}}, \;\;\; \theta_{\rm i}={kT_{\rm i}\over {m_{\rm p}c^2}},
  \end{equation}
where $k$, $m_{\rm e}$, $m_{\rm p}$ and $c$ are Boltzmann constant, electron mass, proton mass and the light velocity.

The synchrotron cooling rate  $q_{\rm syn}$ is given by \citep{Narayan-Yi95, Esin-etal96}
\begin{eqnarray}
  q_{\rm syn}  &=& 
    {{2{\rm \pi} kT_{\rm e}\nu_{\rm c}^3 }\over {3Hc^2}}  
     +6.76\times 10^{-28} {n_{\rm i}\over{K_2(1/\theta_{\rm e})a_1^{1/6}}} \nonumber \\ 
  & \times &  [{1\over a_4^{11/2}}\Gamma({11\over 2},a_4\nu_{\rm c}^{1/3})
     +{a_2\over a_4^{19/4}} \Gamma({19\over 4},a_4\nu_{\rm c}^{1/3}) \nonumber \\        
  &+& {a_3\over a_4^4}( a_4^3\nu_{\rm c}+3a_4^2\nu_{\rm c}^{2/3}    
   +6a_4\nu_{\rm c}^{1/3}+6) {\rm e}^{-a_4\nu_{\rm c}^{1/3}}] \nonumber \\   
  && \hspace{3.8cm} {\rm erg \ cm^{-3} \ s^{-1}},
   \end{eqnarray}
   where $H$ is a scale height of the disc,
    \begin{eqnarray}
   a_1={2\over {3\nu_0\theta_{\rm e}^2}},\; a_2={0.4\over a_1^{1/4}},\; a_3={0.5316\over a_1^{1/2}},
   \; a_4={1.8899a_1^{1/3}}, \nonumber \\                                    
   \Gamma(a,x)= \int_{x}^{\infty}t^{a-1}{\rm e}^{-t}dt, 
   \nu_0={eB\over {2{\rm \pi} m_{\rm e}c}} {\; \rm and} \;\nu_{\rm c}={3\over 2}\nu_0\theta_{\rm e}^2x_{\rm M}.
     \end{eqnarray}
Here, $e$ and $B$ are the electron charge and the strength of the magnetic field and $x_{\rm M}$  is determined from the next equation as
\begin{eqnarray}
      {\rm exp}(1.8899x_{\rm M}^{1/3} ) &=&  2.49\times 10^{-10}{4{\rm \pi} n_{\rm e}R\over
       B} {1\over {\theta_{\rm e}^3 K_2(1/\theta_e) }} \nonumber \\
       &\times&    \left({1\over  x_{\rm M}^{7/6} }
        + {0.40 \over x_{\rm M}^{17/12}}+{0.5316\over {x_{\rm M}^{5/3}}}\right). \nonumber \\        
\end{eqnarray}

The bremsstrahlung cooling rate $q_{\rm br}$ is given as follows \citep{Stepney-Guilbert-83}.
\begin{eqnarray}
  q_{\rm br} = q_{\rm ei} + q_{\rm ee},
\end{eqnarray}
  
\begin{eqnarray}
  q_{\rm ei}=1.48\times 10^{-22}n_{\rm e}^2 F_{\rm ei}(\theta_{\rm e}) \hspace{1cm}{\rm erg \ cm^{-3}\; s^{-1}},
\end{eqnarray}
  
\begin{eqnarray}
   \;\;\;F_{\rm ei}(\theta_{\rm e}) =\left\{\begin{array}{ll}
    1.02\theta_{\rm e}^{1/2}(1+1.78\theta_{\rm e}^{1.34}) & \mbox{ for $\theta_{\rm e} <1$},  \\
    1.43\theta_{\rm e}[{\rm ln}(1.12\theta_{\rm e}+0.48)+1.5]
                                                          & \mbox{ for  $\theta_{\rm e}>1$},
\end{array}
                                      \right. 
\end{eqnarray}
  
\begin{eqnarray}
  q_{\rm ee} = \left\{\begin{array}{llll}
     2.56\times 10^{-22}n_{\rm e}^2\theta_{\rm e}^{1.5}
       (1+1.10\theta_{\rm e}+\theta_{\rm e}^2-1.25\theta_{\rm e}^{2.5})  & \\
     \hspace{3cm}  {\rm erg \ cm^{-3} \; s^{-1} }  \mbox{ for $\theta_{\rm e} <1$}, 
  & \\
    3.40\times 10^{-22}n_{\rm e}^2\theta_{\rm e} [{\rm ln}(1.123\theta_{\rm e})+1.28 ]
  & \\                                                                                       
     \hspace{3cm}   {\rm erg \ cm^{-3} \; s^{-1}}  \mbox{ for  $\theta_{\rm e} >1$}. & \\
                               \end{array}
                      \right.
\end{eqnarray}

Our two-temperature model is based on the assumption that the gas is a fully ionized hydrogen gas, where the radiative energy gain of electrons from ions is equal to the radiative energy loss from synchrotron and bremsstrahlung emissions by the electrons. We consider the two-temperature model using the following relation \citep{Aktar-etal24b, Aktar-etal-26}:
\begin{equation} 
   T_{\rm e} + T_{\rm i} = 2T.
\end{equation}
This is derived from our assumption that the gas pressure \( p_{\rm g} \) is given by \( 2nkT = n_{\rm e}kT_{\rm e} + n_{\rm i}kT_{\rm i} \), where \( n \), the number density, is equal to both \( n_{\rm e} \) and \( n_{\rm i} \) in the one-temperature model. We numerically solve the equations \( q^{\rm ie} = q_{\rm syn} + q_{\rm br} \) and \( T_{\rm e} + T_{\rm i} = 2T \) to obtain the ion temperature \( T_{\rm i} \) and the electron temperature \( T_{\rm e} \). The assumption of radiation energy equilibrium used here is reasonable because the hydrodynamic and magnetohydrodynamic time scales are significantly longer than the radiation time scale in most regions, even when advective or magnetic stress energy is present. 

We note that the relationship described by equation (\ref{Coulmb_syn_brem_eqn}) in our simplified treatment is not a fundamental connection between the relevant physical timescales. The electron-ion energy exchange timescale \(t_{ie}\) is governed by Coulomb coupling, while \(t_{\rm syn}\) and \(t_{\rm br}\) represent distinct radiative cooling processes. As a result, we do not expect a strict relationship between these timescales in a fully ionized plasma, and it is possible for this assumption to be violated. Therefore, we consider this relationship to be a simplifying assumption within our current model rather than a general physical constraint.

%%%%%%%%%%%%%%%%%%%%%%%%%%%%%%%%%%%%%%%%%%%%%%%%%%%%%%%%%%%%%%%%%%%%%
%                        Figure 1
%%%%%%%%%%%%%%%%%%%%%%%%%%%%%%%%%%%%%%%%%%%%%%%%%%%%%%%%%%%%%%%%%%%%%
\begin{figure} 
    \begin{center}
        \includegraphics[width=0.50\textwidth]{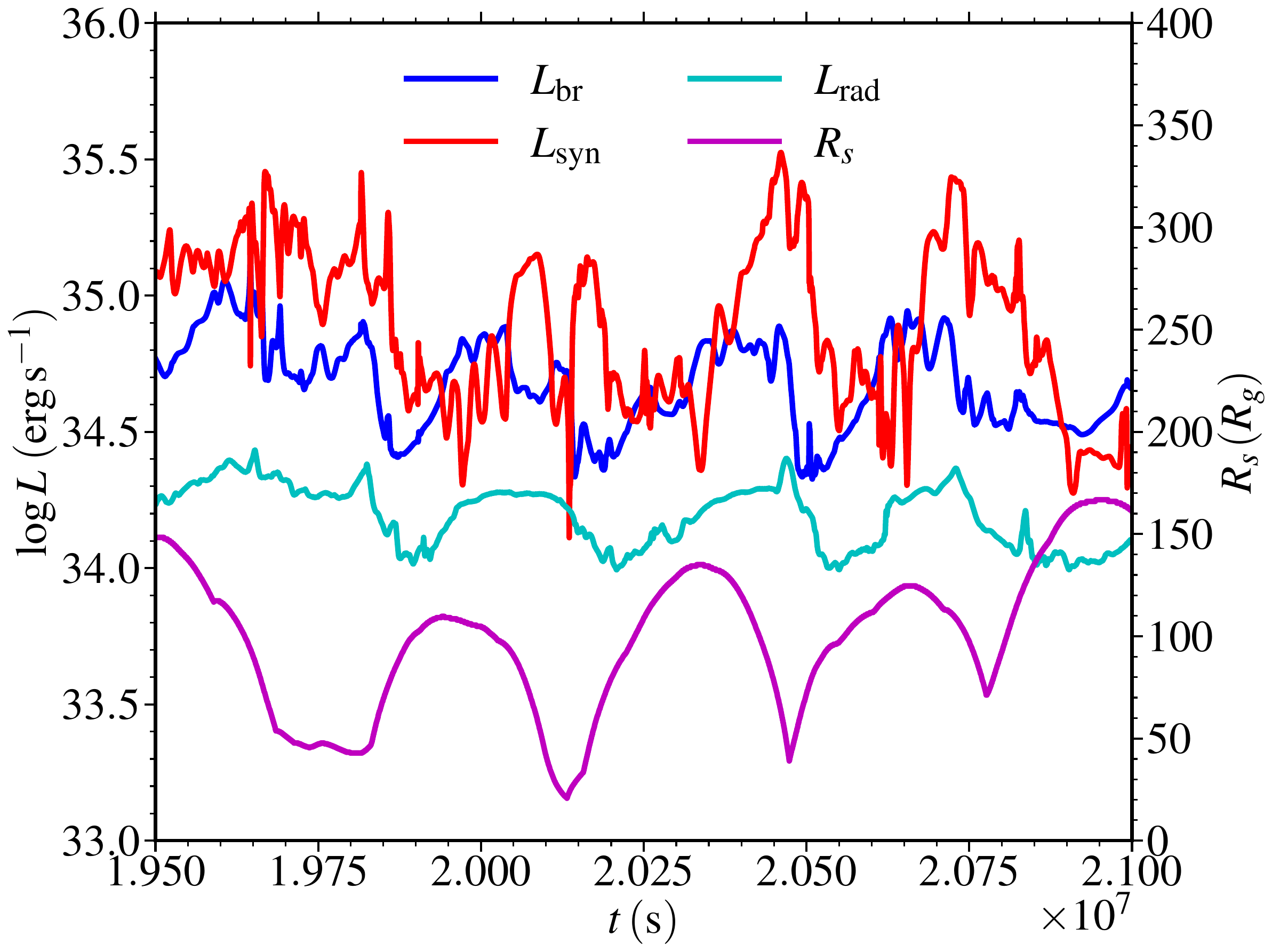}
\caption {The time variations of bremsstrahlung luminosity \(L_{\rm br}\) (blue line), synchrotron luminosity \(L_{\rm syn}\) (red line), radiative luminosity \(L_{\rm rad}\) (cyan line), and shock location (magenta line) are shown. During four successive shock oscillations, the luminosities change dramatically over a timescale of several hours. See the text for details.}
\label{Figure_1}
\end{center} 
\end{figure}
%%%%%%%%%%%%%%%%%%%%%%%%%%%%%%%%%%%%%%%%%%%%%%%%%%%%%%%%%%%%%%%%%%%%%

%%%%%%%%%%%%%%%%%%%%%%%%%%%%%%%%%%%%%%%%%%%%%%%%%%%%%%%%%%%%%%%%%%%%%
%                        Figure 2
%%%%%%%%%%%%%%%%%%%%%%%%%%%%%%%%%%%%%%%%%%%%%%%%%%%%%%%%%%%%%%%%%%%%%
\begin{figure}
    \begin{center}
        \includegraphics[width=0.50\textwidth]{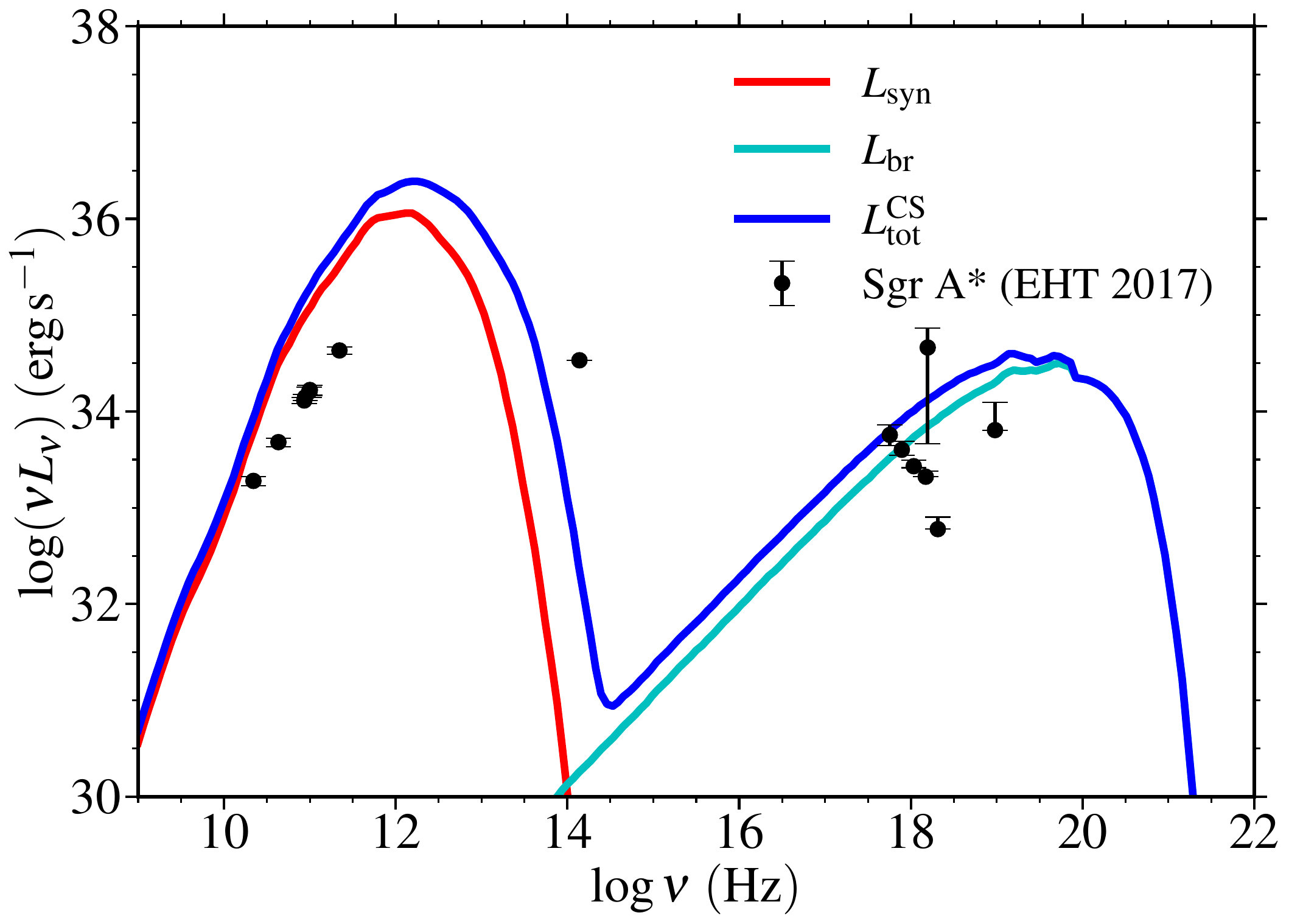}
\caption {The spectral energy density of various luminosities at time \( t = 2.0 \times 10^7 \) s is presented. The synchrotron radiation is illustrated with red lines, while the bremsstrahlung radiation, without scattering, is shown with cyan lines. The total spectrum affected by Compton scattering is represented by blue lines. Black solid circles with error bars indicate the observational data from the EHT 2017 observations \citep{Event-Horizon-etal-2022b}. See the text for details.}
\label{Figure_2}
\end{center} 
\end{figure}
%%%%%%%%%%%%%%%%%%%%%%%%%%%%%%%%%%%%%%%%%%%%%%%%%%%%%%%%%%%%%%%%%%%%%

%%%%%%%%%%%%%%%%%%%%%%%%%%%%%%%%%%%%%%%%%%%%%%%%%%%%%%%%%%%%%%%%%%%%%
%                        Figure 3
%%%%%%%%%%%%%%%%%%%%%%%%%%%%%%%%%%%%%%%%%%%%%%%%%%%%%%%%%%%%%%%%%%%%%
\begin{figure*}
    \begin{tabular}{cc}
     \begin{minipage}{0.5\linewidth}
        \centering
        \includegraphics[width=1.0\textwidth]{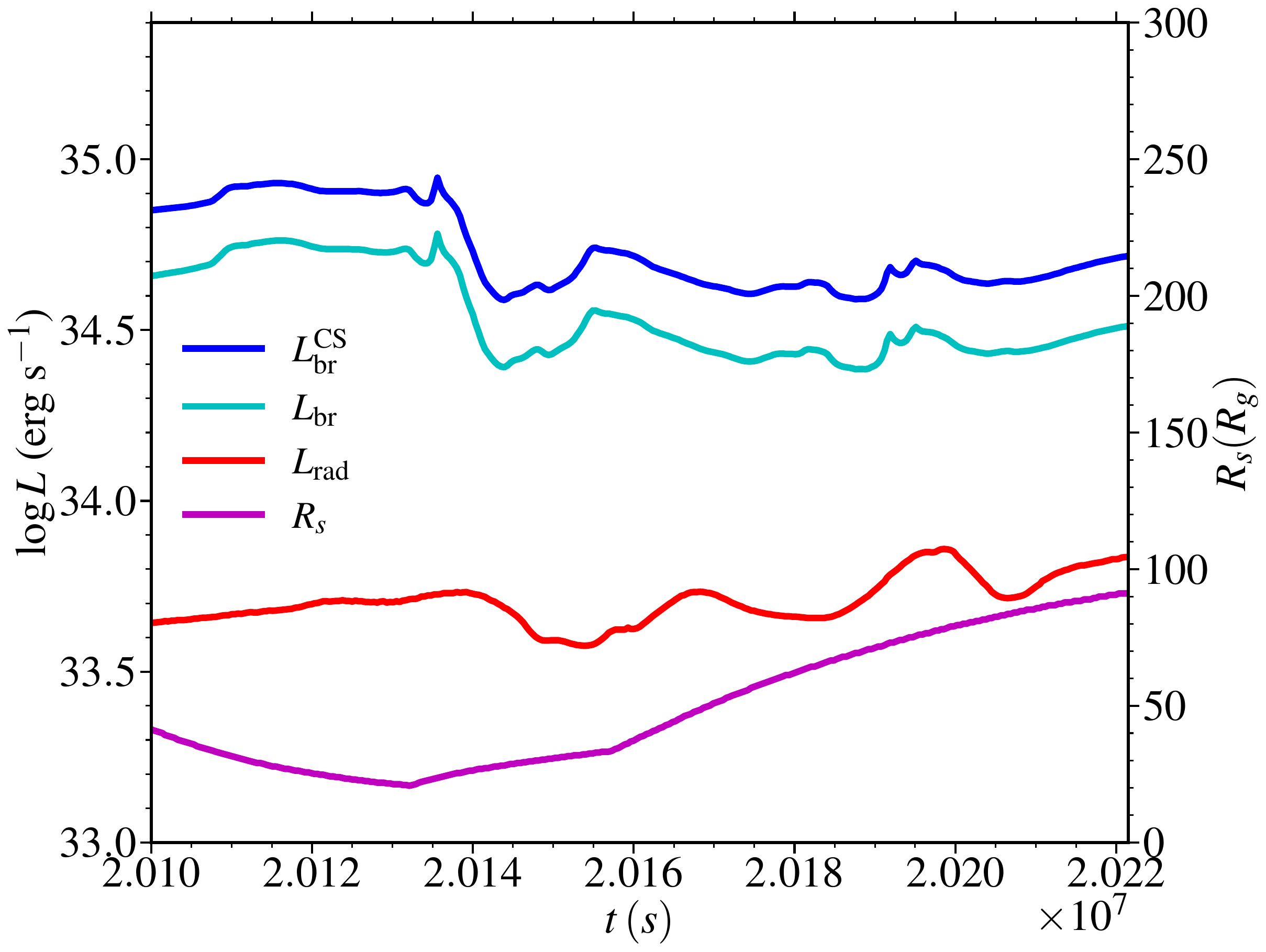}
        \label{fig3a}
      \end{minipage}

      \begin{minipage}{0.5\linewidth}
        \centering
        \includegraphics[width=1.0\textwidth]{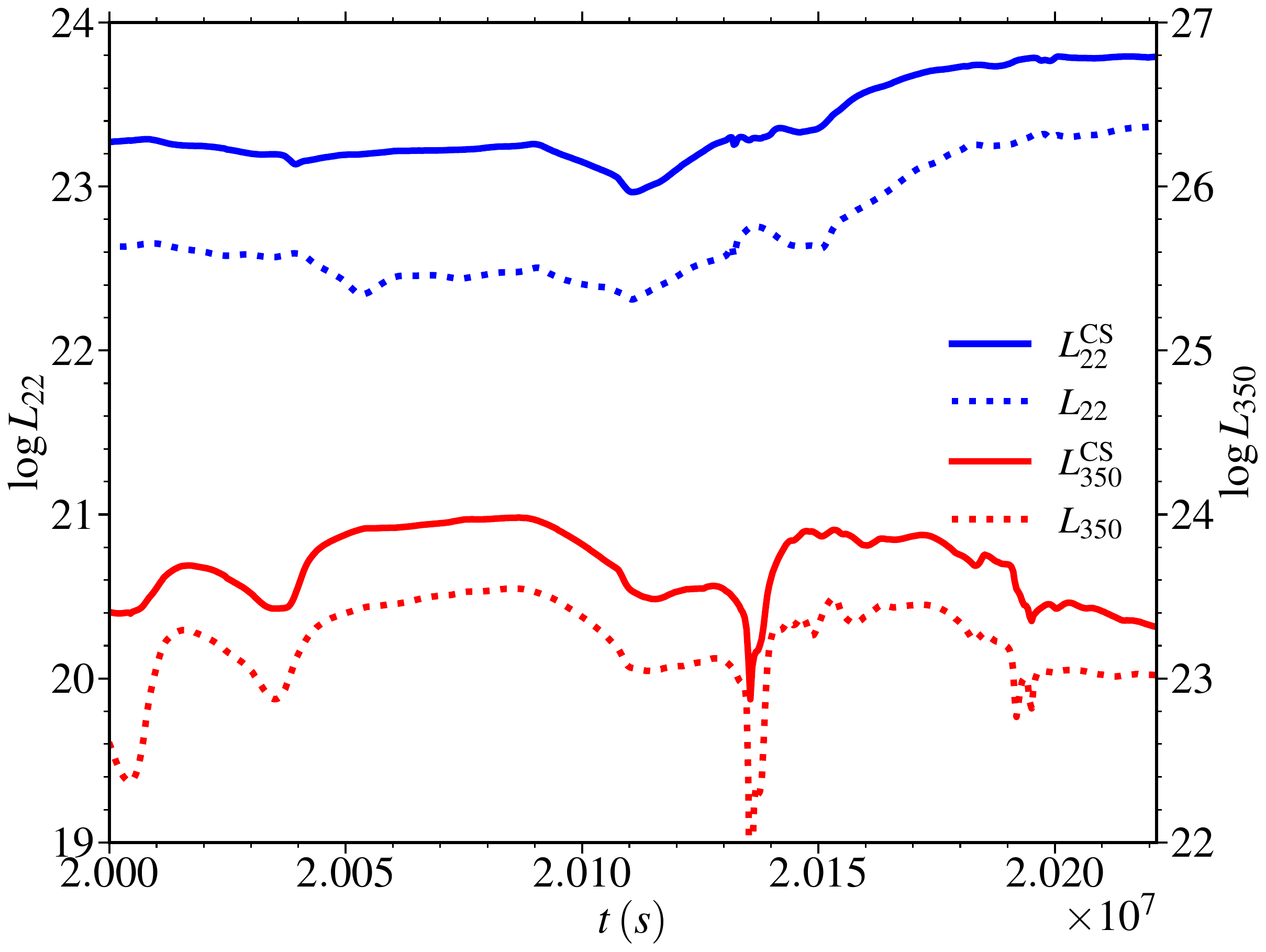}
        \label{fig3b}
      \end{minipage} \\
    \end{tabular}
\caption {Left: The time variations of the radiative luminosity \(L_{\rm rad}\) (red line), the Comptonized bremsstrahlung luminosity \(L_{\rm br}^{\rm CS}\) (blue line), the non-Comptonized bremsstrahlung luminosity \(L_{\rm br}\) (cyan line), and the shock location \(R_{\rm s}\) (purple line) are presented for the time interval \(t = 2.010-2.022 \times 10^7\) s. Right: The time variations of monochromatic luminosities \(L_{\rm 22}\) (blue line) at 22 GHz and \(L_{\rm 350}\) (red line) at 350 GHz are shown for the time interval \(t = 2.000-2.021 \times 10^7\) s. The solid and dotted lines represent luminosities with and without Compton scattering.
See the text for details.}
\label{Figure_3}
\end{figure*}
%%%%%%%%%%%%%%%%%%%%%%%%%%%%%%%%%%%%%%%%%%%%%%%%%%%%%%%%%%%%%%%%%%%%%

%%%%%%%%%%%%%%%%%%%%%%%%%%%%%%%%%%%%%%%%%%%%%%%%%%%%%%%%%%%%%%%%%%%%%
\subsection{Calculation of luminosities}

After determining the electron and ion temperatures, we calculate the monochromatic radiation flux (\( F_{\nu} \)) at the surface of the disc, as well as the luminosity (\( L_{\nu} \)), bremsstrahlung luminosity (\( L_{\text{br}} \)), synchrotron luminosity (\( L_{\text{syn}} \)) emitted from the total surface of the accretion disc, and the radiative luminosity (\( L_{\text{rad}} \)) emitted from the surface of the entire computational region. The monochromatic radiation flux \( F_{\nu} \) at a given radius \( R \) on the disc surface is given by \citep{Manmoto-etal-97}
\begin{eqnarray} \label{flux_eqn}
 F_{\rm \nu }(R) = {{2\pi}\over \sqrt{3}} B_{\rm \nu} \left[1 - {\rm exp} (-2 \sqrt{3} {\tau_{\rm \nu}}^*)\right] 
 \;\; {\rm erg}\;{\rm cm}^{-2}\; {\rm s}^{-1}{\rm Hz}^{-1},
\end{eqnarray}
where $B_{\rm \nu}$ is the Planck function and  ${\tau_{\rm \nu}}^*$ is the optical depth at the accretion disc surface, which is given by 
 
\begin{eqnarray}
 {\tau_{\rm \nu}}^* = (\pi^{1/2}/2) \kappa_{\rm \nu}(0) H,
\end{eqnarray}
 $\kappa_{\rm \nu}(0)$ is the absorption coefficient at frequency ${\rm \nu}$ on the equatorial plane   given by $\kappa_{\rm \nu}(0) = \chi_{\nu}/(4\pi B_{\rm \nu}))$, $\chi_{\nu}$ is the emissivity and $H$ is the disc height.

The bremsstrahlung emissivity $(\chi_{\nu, {\rm br}})$ can be written as 
\begin{align} \label{spec_brem_emisi}
\chi_{\nu, {\rm br}} = q_{\rm br} \overline{G} \exp\left(-\frac{h\nu}{k_{\rm B}T_e}\right),
\end{align}
where $q_{\rm br}$ is the bremsstrahlung cooling rate as depicted in equation 7. Here $\overline{G}$ is the Gaunt factor, which is given by
\[
   \overline{G} = 
\begin{cases}
    \frac{h}{k_{\rm B}T_e} \left(\frac{3}{\pi} \frac{k_{\rm B} T_e}{h \nu}\right)^{1/2}, & \text{if } \frac{k_{\rm B}T_e}{h \nu} < 1\\
    \frac{h}{k_{\rm B}T_e} \frac{\sqrt{3}}{\pi} \ln{\left(\frac{4}{\xi}\frac{k_{\rm B} T_e}{h \nu}\right)}, & \text{if } \frac{k_{\rm B}T_e}{h \nu} > 1.
\end{cases}
\]

 We consider the synchrotron emissivity  $\chi_{\rm \nu,syn}$ by a relativistic Maxwellian distribution of electrons  as follows \citep{Narayan-Yi95}
\begin{eqnarray} \label{spec_sync_emisi}
 \chi_{\nu,syn} = 4.43 \times 10^{-30} { {4\pi n_{\rm e} \nu} \over {K_{2}(1/\theta_{\rm e})}} I(x) 
 \;\; {\rm erg}\;{\rm cm}^{-3}\;{\rm s}^{-1} {\rm Hz}^{-1},
\end{eqnarray}
where $x \equiv {4\pi m_{\rm e}c\nu}/({3eB{\theta_{\rm e}}^2})$,
 \begin{eqnarray}
      I(x) ={4.0505\over x^{1/6}} \left(1+ {0.40 \over x^{1/4}}+{0.5316\over x^{1/2}}\right) {\rm exp} (-1.8899x^{1/3}).     
 \end{eqnarray}

The total radiative luminosity $L_{\rm rad}$ corresponds to the free-free emissions and is evaluated at the outer boundary surface of all computational region, as follows
\begin{align}
%L_{\rm rad} = \int \mathbf{F}_{\rm rad} \cdot \mathrm{d}\mathbf{S}
L_{\rm rad} = \int_S \bm{F}_{\rm rad} \cdot \bm{dS},
\label{rad_lumi}
\end{align}
where $\bm{F}_{\rm rad}$ is the radiation energy flux at the boundary surfaces.

The monochchromatic luminosity $L_{\nu}$ is given by
\begin{align}\label{lumi_nu}
L_{\nu} =\int {F_{\nu}}~ {\rm d}{S}.
\end{align}

The frequency-dependent bremsstrahlung ($L_{\rm br}$) and synchrotron ($L_{\rm syn}$) luminosity can be obtained throughout the disc surface as  
\begin{align}\label{brem_lumi}
L_{\nu},_{\rm br} = \int_S \int_\nu {F_{\nu},_{\rm br}}~d\nu~d{S},
\end{align}
and
\begin{align}\label{sync_lumi}
L_{\nu},_{\rm syn} = \int_S \int_\nu {F_{\nu},_{\rm syn}}~d\nu~ d{S},
\end{align}
where the surface integration is carried out all over the disc surface. ${F_{\nu,\rm syn}}$ and ${F_{\nu,\rm br}}$ can be calculated using synchrotron and bremsstrahlung emissivities using equation (\ref{spec_brem_emisi}) and (\ref{spec_sync_emisi}), respectively.

%%%%%%%%%%%%%%%%%%%%%%%%%%%%%%%%%%%%%%%%%%%%%%%%%%%%%%%%%%%%%%%%%%%%%
%                        Figure 4
%%%%%%%%%%%%%%%%%%%%%%%%%%%%%%%%%%%%%%%%%%%%%%%%%%%%%%%%%%%%%%%%%%%%%
\begin{figure*}
    \begin{tabular}{cc}
     \begin{minipage}{0.5\linewidth}
        \centering
        \includegraphics[width=1.0\textwidth]{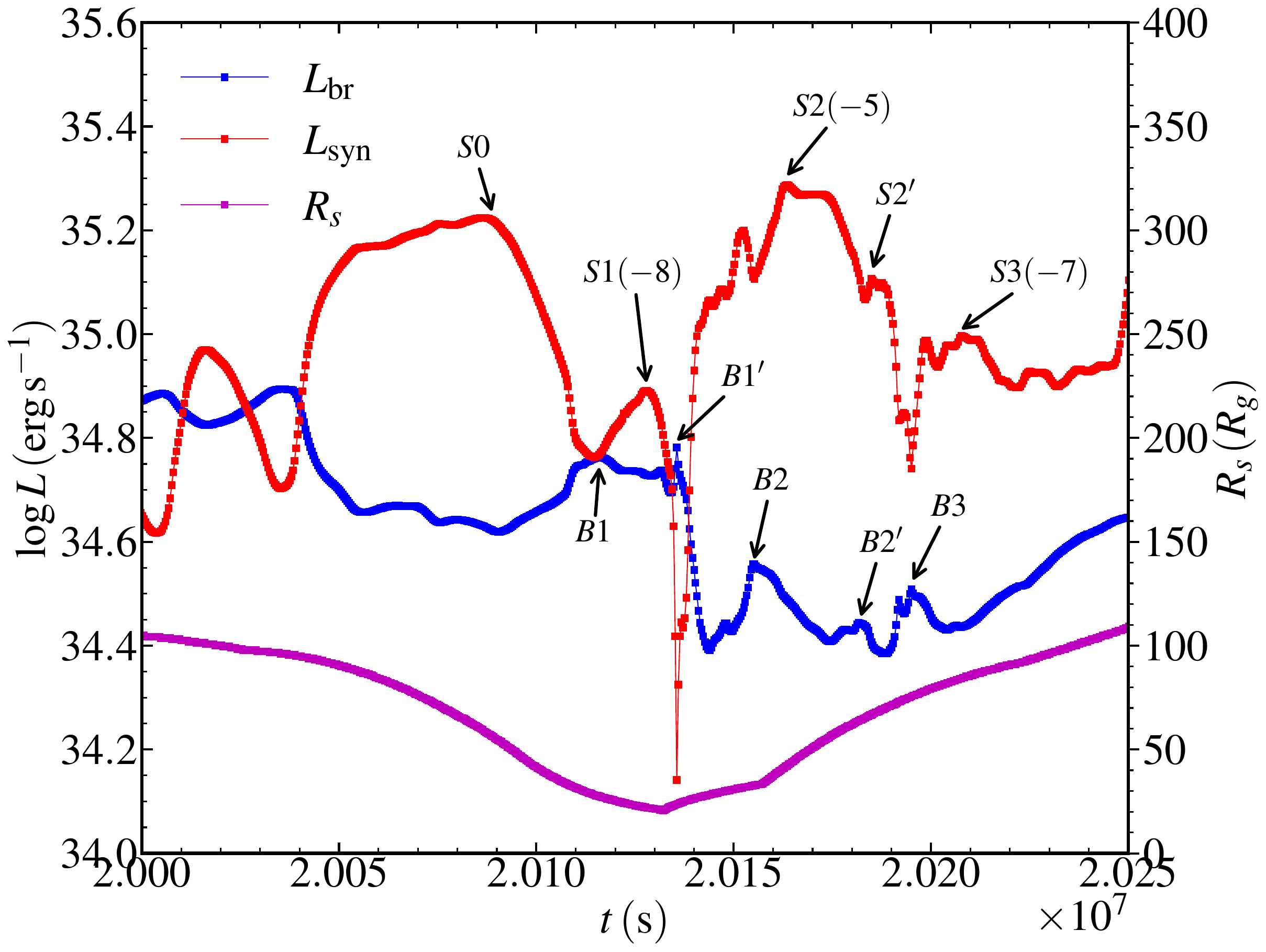}
        \label{fig4a}
      \end{minipage}

      \begin{minipage}{0.5\linewidth}
        \centering
        \includegraphics[width=1.0\textwidth]{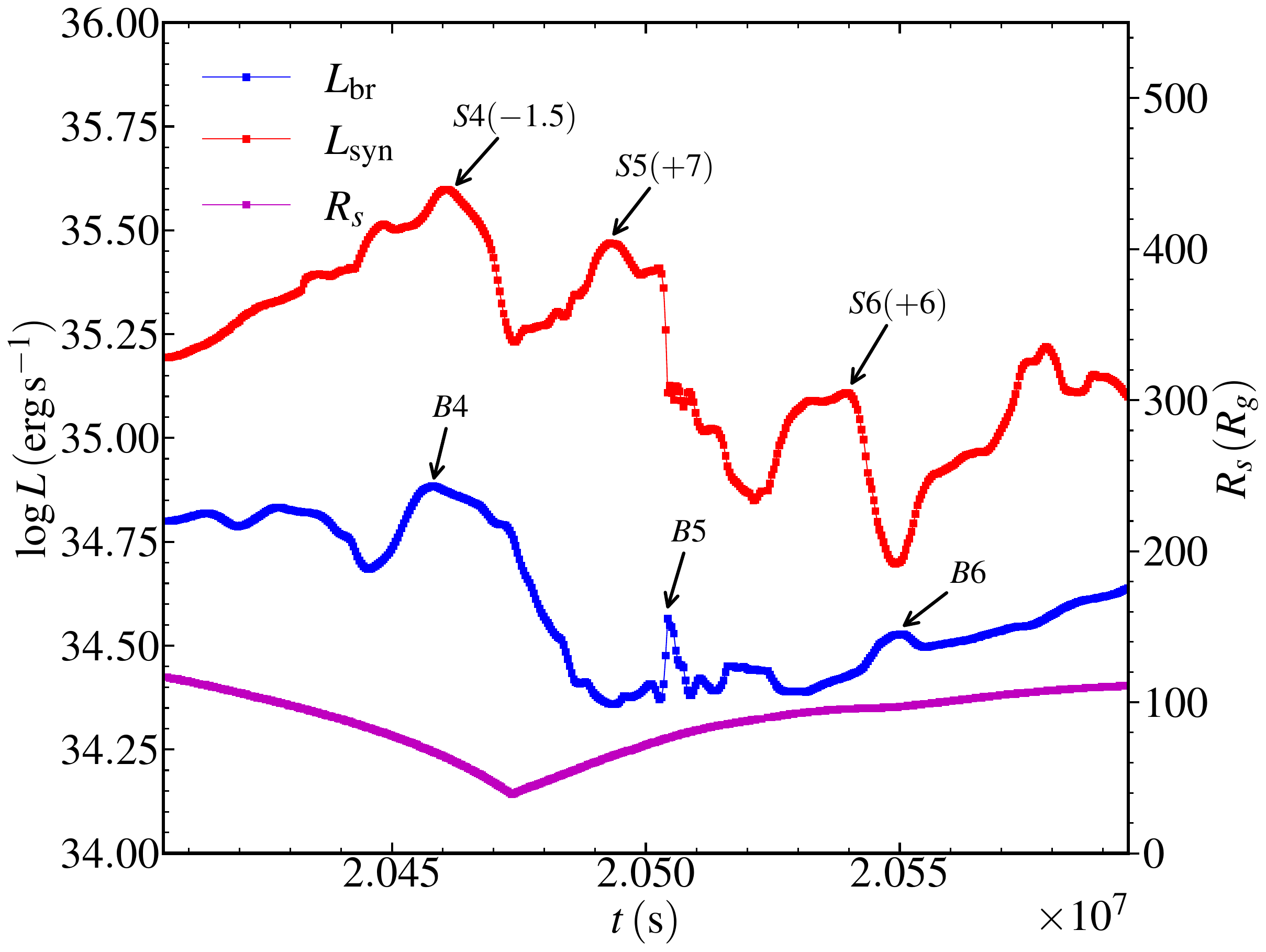}
        \label{fig4b}
      \end{minipage} \\
    \end{tabular}
\caption{Left: The time variations of synchrotron luminosity \(L_{\rm syn}\) (red filled squares) and bremsstrahlung luminosity \(L_{\rm br}\) (blue filled squares) are presented in units of erg s\(^{-1}\). Additionally, the oscillating shock location \(R_{\rm s}\) is shown (magenta line) on the equator during the time interval of \(t = 2.0 - 2.025 \times 10^7\) s. The square points on the curves are plotted at intervals of 26.4 minutes. The arrows indicate several local maxima on the luminosity curve, with the symbols $B1-B6$ representing peaks in bremsstrahlung luminosity and $S1-S6$ symbolizing peaks in synchrotron luminosity. The number in parentheses next to the symbol \(S\) indicates the corresponding delay point number relative to the bremsstrahlung peak, where a negative or positive sign indicates a delay or an inverse delay, respectively. Right: This panel presents the same data as the left-hand panel, but for the time interval of \(t = 2.04 - 2.06 \times 10^7\) s. See the text for details.}
\label{Figure_4}
\end{figure*}
%%%%%%%%%%%%%%%%%%%%%%%%%%%%%%%%%%%%%%%%%%%%%%%%%%%%%%%%%%%%%%%%%%%%%%%  

%%%%%%%%%%%%%%%%%%%%%%%%%%%%%%%%%%%%%%%%%%%%%%%%%%%%%%%%%%%%%%%%%%%%%
%                        Figure 5
%%%%%%%%%%%%%%%%%%%%%%%%%%%%%%%%%%%%%%%%%%%%%%%%%%%%%%%%%%%%%%%%%%%%%
\begin{figure*}
    \begin{tabular}{cc}
           \begin{minipage}{0.5\linewidth}
        \centering
        \includegraphics[width=1.0\textwidth]{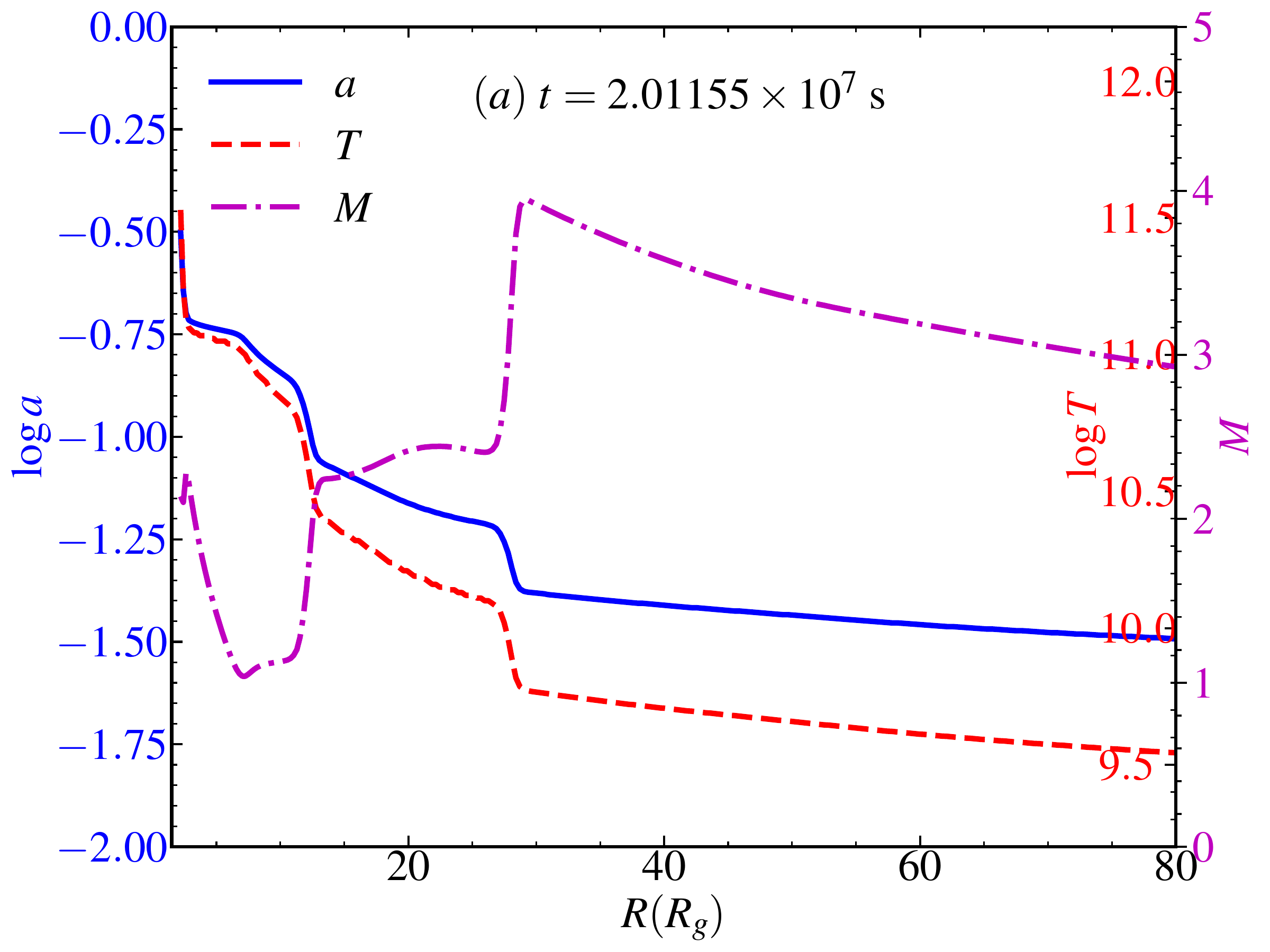}
        \label{fig5-1}
      \end{minipage} 
      
      \begin{minipage}{0.5\linewidth}
        \centering
        \includegraphics[width=1.0\textwidth]{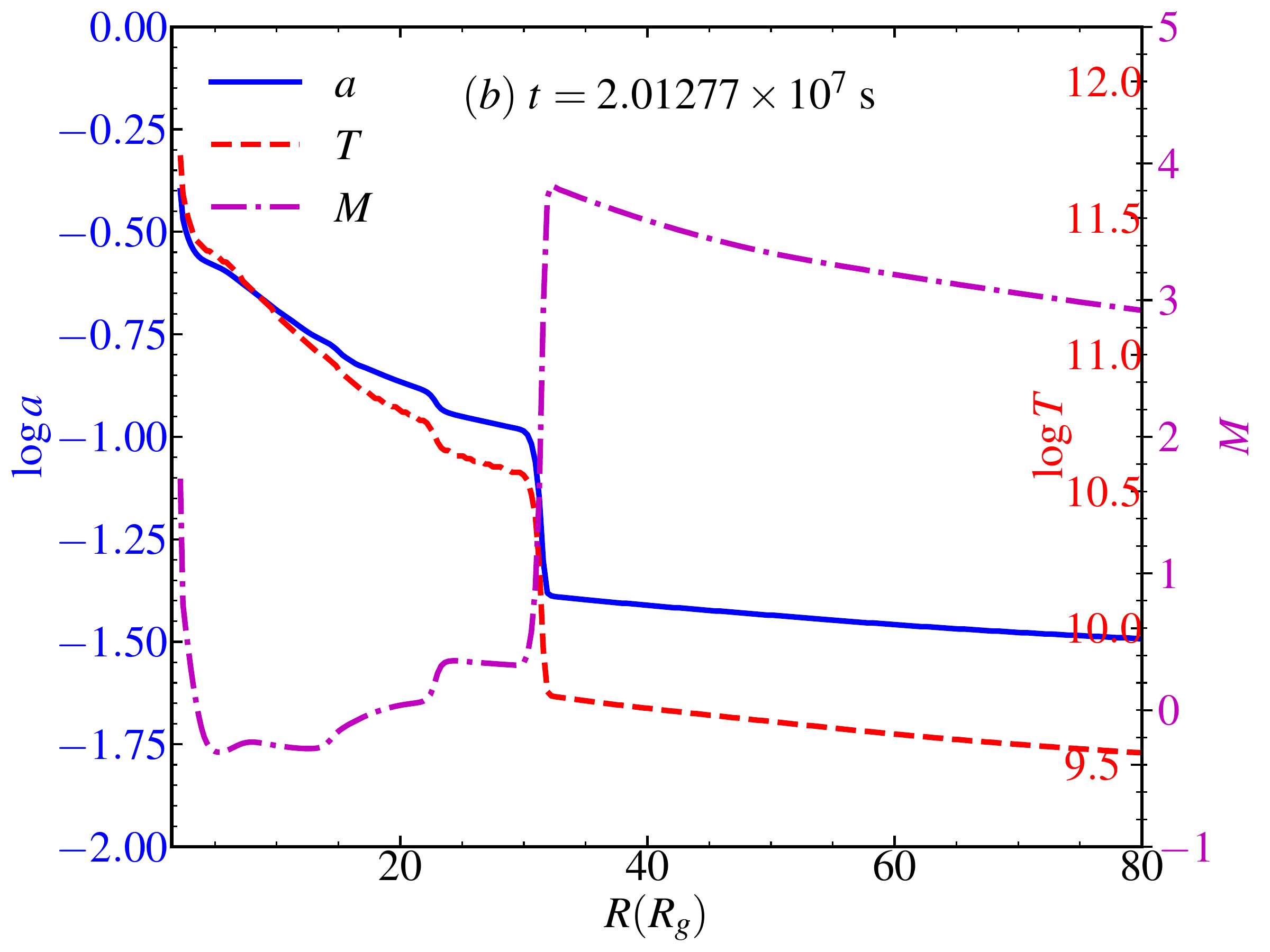}
        \label{fig5-2}
      \end{minipage} \\
    
  \begin{minipage}{0.5\linewidth}
        \centering
        \includegraphics[width=1.0\textwidth]{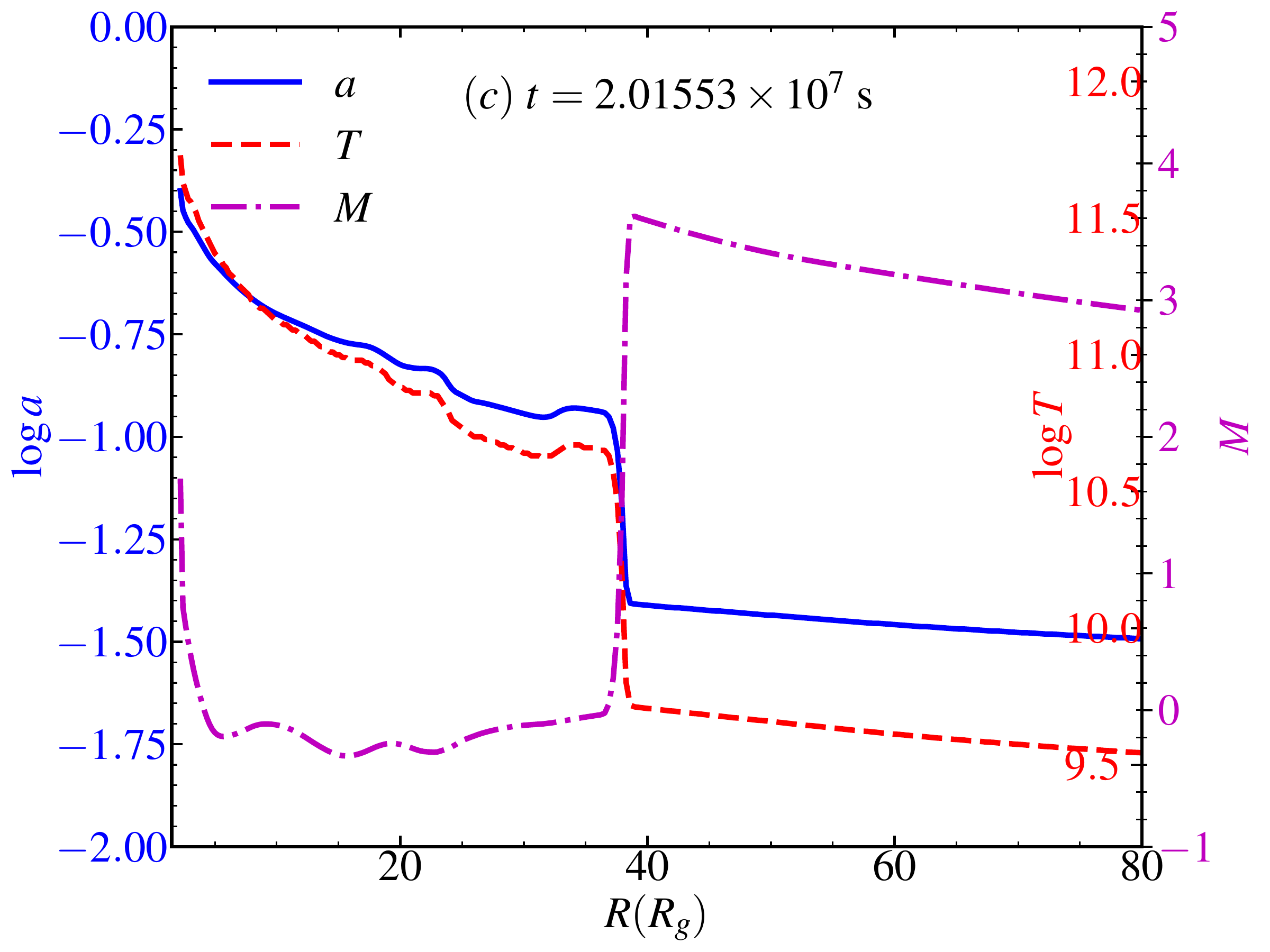}
        \label{fig5-3}
      \end{minipage}

  \begin{minipage}{0.5\linewidth}
        \centering
        \includegraphics[width=1.0\textwidth]{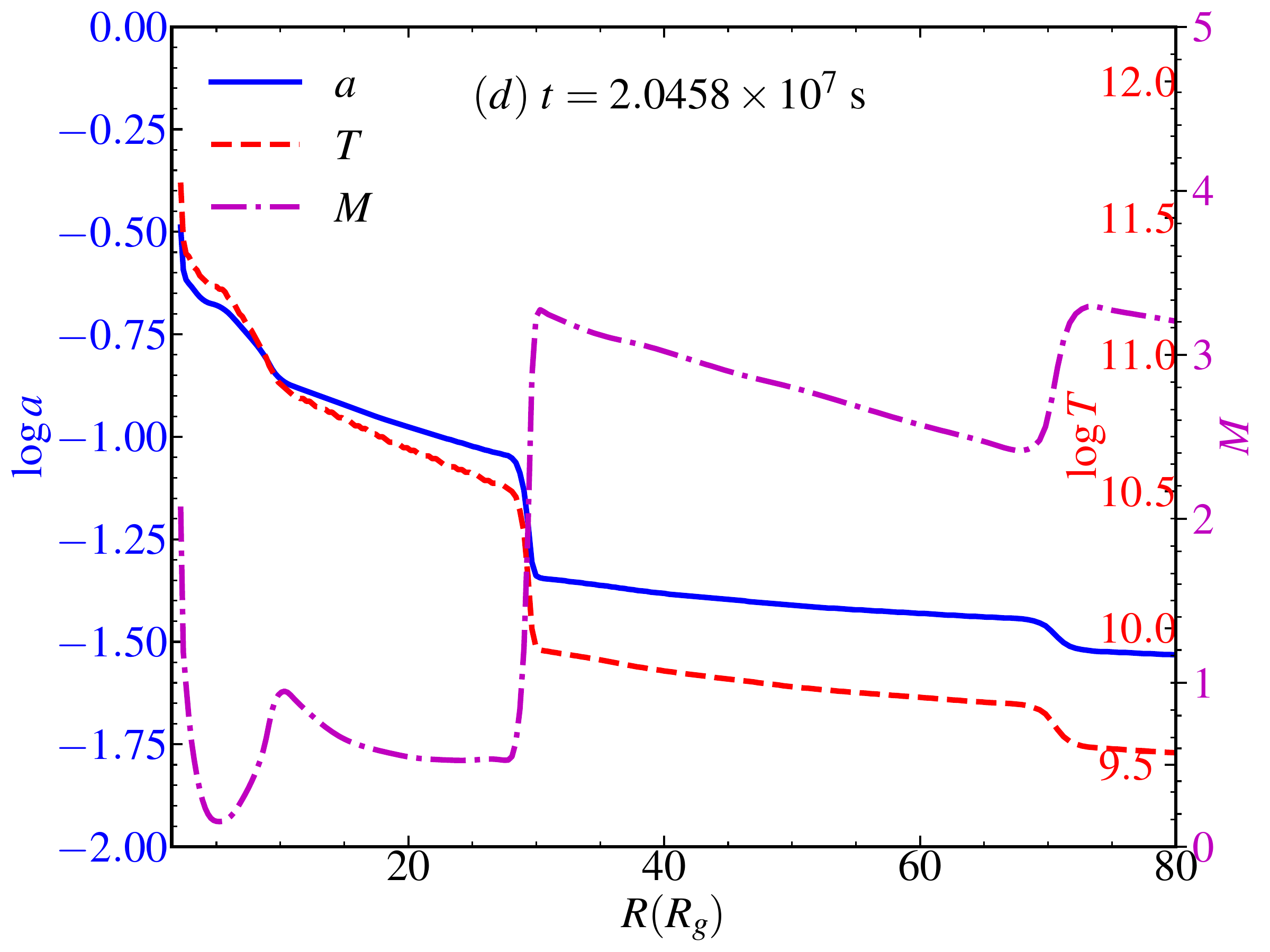}
        \label{fig5-4}
      \end{minipage}
    \end{tabular}
  \caption{The radial distribution of sound speed \( a \) (blue line), Mach number \( M \) (magenta line), and temperature \( T \) (red line) is presented on the equator at times \( t = 2.01155 \) (a), \( 2.01553 \) (b), \( 2.01277 \) (c), and \( 2.01553 \) (d) \(\times 10^7\) s. These correspond to the peak points \( B1 \), \( B2 \), \( B3 \), and \( B4 \) in Figure \ref{Figure_4}, respectively. The oscillating shock wave fronts appear as sharp vertical lines in each curve. See the text for details. }
   \label{Figure_5}
  \end{figure*}
%%%%%%%%%%%%%%%%%%%%%%%%%%%%%%%%%%%%%%%%%%%%%%%%%%%%%%%%%%%%%%%%%%%%%

%%%%%%%%%%%%%%%%%%%%%%%%%%%%%%%%%%%%%%%%%%%%%%%%%%%%%%%%%%%%%%%%%%%%%
\section{Numerical results}
%%%%%%%%%%%%%%%%%%%%%%%%%%%%%%%%%%%%%%%%%%%%%%%%%%%%%%%%%%%%%%%%%%%%%

 We apply the present two-temperature model to the numerical results of 2D relativistic radiation MHD simulations for  Sgr A* flares. We consider the mass of Sgr A* as $M = 4\times 10^6 M_{\odot}$. Here, the unit of distance $R_{\rm g}$ is $2GM/c^2 \sim 1.19\times 10^{12}$ cm. To examine the time delay between the synchrotron and bremsstrahlung emissions from the radiation MHD simulations, we calculate the above simulations, using a smaller time step of 10$R_{\rm g}/c$ ( $\sim$ 6.6 min) in order to get better time resolution for the time delay analysis. Time variations of bremsstrahlung $L_{\rm br}$ (blue line), synchrotron $L_{\rm syn}$ (red line), radiative $L_{\rm rad}$ (cyan line)  luminosities and shock  location (magenta line) are shown in Figure \ref{Figure_1}. Here, during four succcesive shock oscillations, the luminosities vary drastically with a time scale of several hours.

%----------------------------------------------------------------------
\subsection{Compton scattering effect}
%----------------------------------------------------------------------
In the previous papers \citet{Okuda-etal22, Okuda-etal23}, we did not consider the effect of Compton scattering. However, Compton scattering impacts the spectral energy density ${F}_{\rm \nu}$ across all frequencies. To account for this, we utilize the Compton scattering probability function from \citet{Coppi-Blandford-90}, which describes the interaction between an incoming photon and an electron, and apply this to the input energy flux ${F}_{\rm \nu}$. In Figure \ref{Figure_2}, we illustrate the influence of Compton scattering on the spectral energy density (SED) and the delay time of the luminosity curves. We present the spectral energy density of Sgr A* in our simulations at an evolutionary time of $t= 2.0\times 10^7$ s. The red and cyan lines represent the synchrotron and bremsstrahlung SED without Compton scattering, respectively, while the green line shows the total SED that has been affected by Compton scattering. The black solid circles with error bars indicate the observed SED from the Event Horizon Telescope (EHT) observations conducted between April 5 and 11, 2017 \citep{Event-Horizon-etal-2022b}. Our findings reveal a significant enhancement in the energy density spectrum due to the effects of Compton scattering.

%%%%%%%%%%%%%%%%%%%%%%%%%%%%%%%%%%%%%%%%%%%%%%%%%%%%%%%%%%%%%%%%%%%%%%%%%%%%%%%%%
\begin{table}
\centering
\caption{Time delay in unit of hour between  $L_{\rm syn}$ to $L_{\rm brem}$ luminosity peaks in Figure \ref{Figure_4}. The negative and positive signs indicate delay and inverse delay, respectively.
}
\label{Table_2}
\begin{tabular}{@{}ccccccc} 
%&&&&&& \\
\hline \hline
  ${\rm Synchrotron \;peak \;point}$ & ${\rm S1}$ & ${\rm S2}$ & ${\rm S3}$ & ${\rm S4}$ & ${\rm S5}$ & ${\rm S6}$  \\
 \hline
$L_{\rm brem}$     & - & - & - & - & - & - \\ 
 $L_{\rm syn}$    &  -3.5 &  -2.2   & -3.1 & -0.7& +3.1& +2.6  \\ 

\hline
\end{tabular}
\end{table}
%%%%%%%%%%%%%%%%%%%%%%%%%%%%%%%%%%%%%%%%%%%%%%%%%%%%%%%%%%%%%%%%%%%%%%%%%%%%%%%%%% 

To investigate the effect of Compton scattering on bremsstrahlung luminosity, we present the time variations of radiative luminosity (\( L_{\rm rad} \)), Comptonized bremsstrahlung luminosity (\( L_{\rm br}^{\rm CS} \)), and non-Comptonized bremsstrahlung luminosity (\( L_{\rm br} \)) in the left panel of Figure \ref{Figure_3}. The luminosity curves are plotted at intervals of 13.2 minutes. We observe that the peak of the radiative luminosity lags behind the bremsstrahlung peak by approximately 10 points intervals (around 2.2 hours). If there is no absorbing or emitting matter between the disc surface and the outer \( z \)-boundary, this delay should correspond to the light transit time from the outer \( z \)-boundary to the equator, which is roughly \( 200 R_{\rm g}/c \sim 2.2 \) hours. This delay time indeed aligns with the calculated transit duration. The peak of radiative luminosity serves as a useful marker for detecting the corresponding peaks of synchrotron and bremsstrahlung emissions in subsequent time delay analyses, as identifying the luminosity peaks can sometimes be challenging. From the figure, we note that the Comptonized bremsstrahlung luminosity is approximately three times greater than that observed without Compton scattering, yet there is no significant time delay between the luminosity peaks.

In the right panel of Figure \ref{Figure_3}, we display the time variations of monochromatic luminosities \( L_{\rm 22} \) (blue line) at 22 GHz and \( L_{\rm 350} \) (red line) at 350 GHz during the time interval \( t = 2.000-2.021 \times 10^7 \) s. The dotted lines represent the luminosities without Compton scattering. In both \( L_{\rm 22} \) and \( L_{\rm 350} \) curves, the Comptonized luminosities are enhanced by a factor of approximately five compared to those without Compton scattering, and the time delays between the Comptonized and non-Comptonized luminosities at 22 GHz and 350 GHz are minimal, ranging from 7 to 13 minutes.

%----------------------------------------------------------------------
\subsection{Time delay between synchrotron and bremsstrahlung flares}
%----------------------------------------------------------------------

%%%%%%%%%%%%%%%%%%%%%%%%%%%%%%%%%%%%%%%%%%%%%%%%%%%%%%%%%%%%%%%%%%%%%
%                        Figure 6
%%%%%%%%%%%%%%%%%%%%%%%%%%%%%%%%%%%%%%%%%%%%%%%%%%%%%%%%%%%%%%%%%%%%%
 \begin{figure*}
    \begin{tabular}{cc}

      \begin{minipage}{0.5\linewidth}
        \centering
        \includegraphics[width=1.0\textwidth]{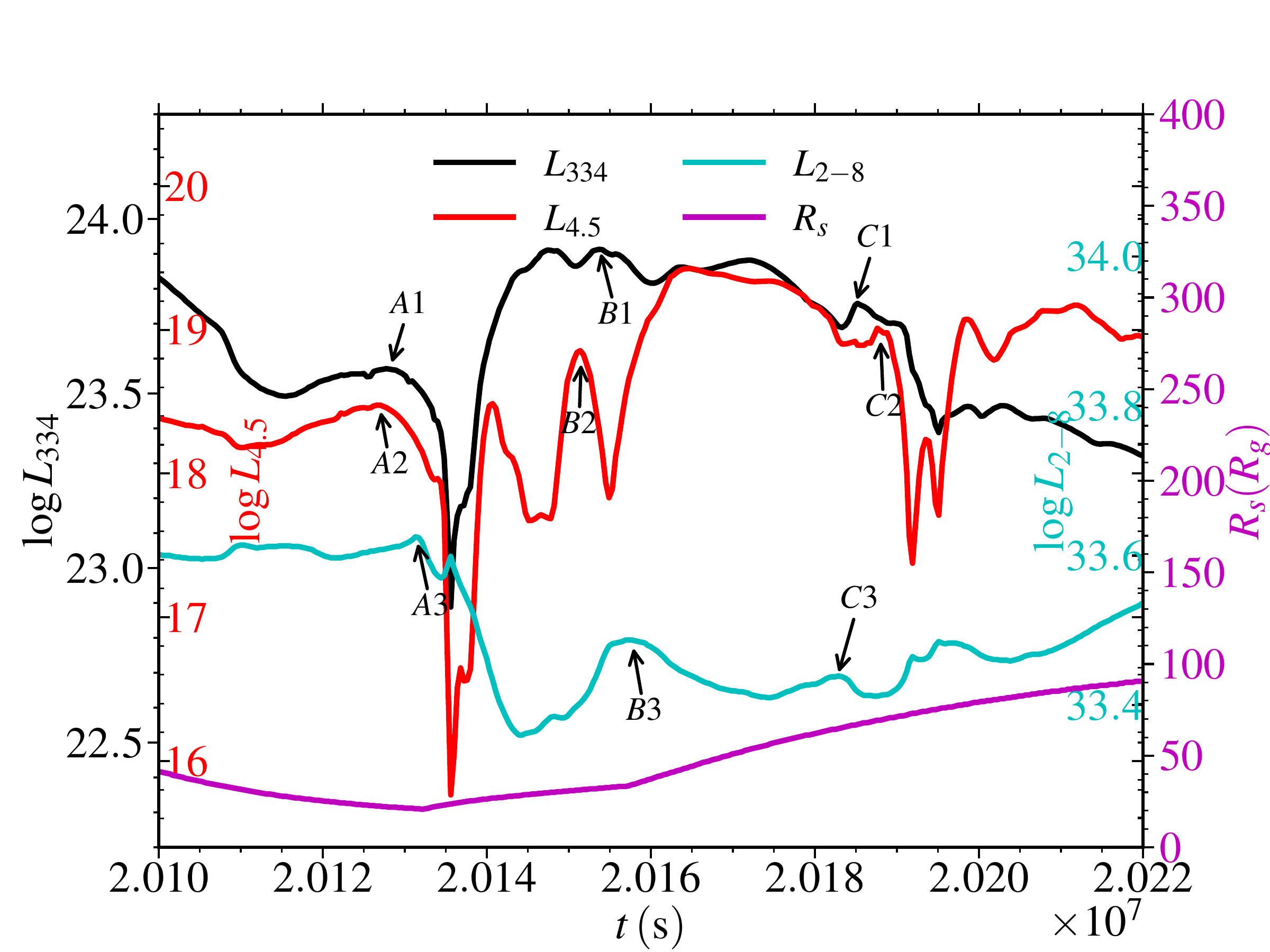}
        \label{fig:6a}
      \end{minipage}

      \begin{minipage}{0.5\linewidth}
        \centering
        \includegraphics[width=1.0\textwidth]{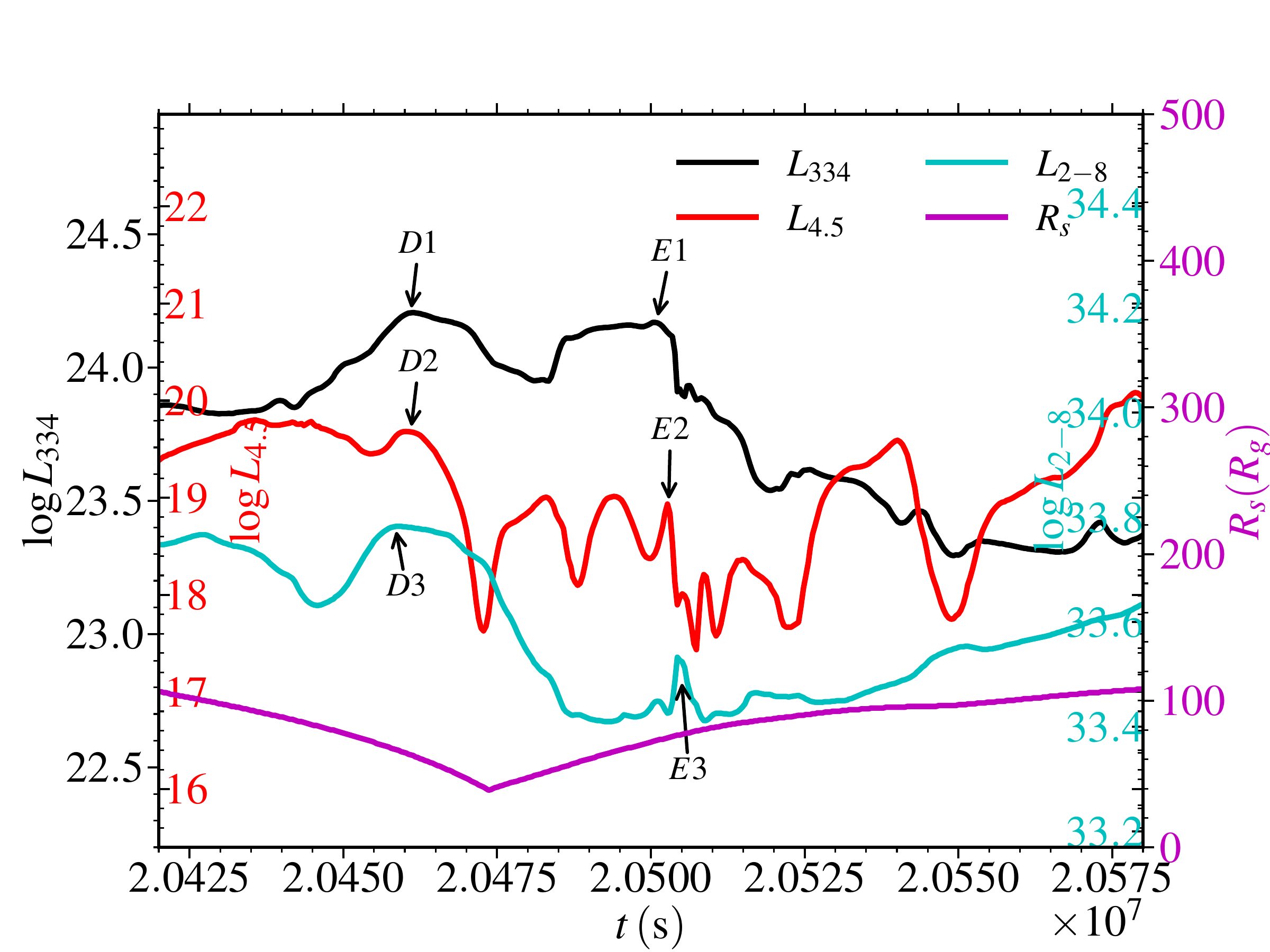}
        \label{fig:6b}
      \end{minipage} \\
    \end{tabular}
\caption {The luminosity curves of  $L_{\rm 334}$ at 334 GHz (black line), $L_{\rm 4.5}$ at 4.5$\mu$m (red line) , $L_{\rm 2-8}$ at 2-8 KeV (cyan line) in units of erg ${\rm s}^{-1} {\rm Hz}^{-1}$ and oscillating shock location $R_{\rm s}$ (magenta line) in units of $R_g$ on the equator during $t= 2.010 - 2.058 \times 10^7$ s. The arrow shows local peaks on the luminosity curves where symbols $A, B, C, D$ and $E$ are the group name  of the selected peaks and the attached number 1, 2 and 3 correspond to those for 334 GHz, 4.5$\mu$m and 2-8 KeV luminosity, respectively. The scales for \( L_{\rm 334} \) and \( L_{\rm 4.5} \) are displayed on the left-hand and right-hand sides of the left \( Y \)-axis, respectively, while \( L_{\rm 2-8} \) is shown on the left-hand side of the right \( Y \)-axis. See the text for details.
}
  \label{Figure_6}
  \end{figure*}
%%%%%%%%%%%%%%%%%%%%%%%%%%%%%%%%%%%%%%%%%%%%%%%%%%%%%%%%%%%%%%%%%%%%%

 Figure \ref{Figure_4} illustrates the variations in synchrotron luminosity \(L_{\rm syn}\) (red filled squares), bremsstrahlung luminosity \(L_{\rm br}\) (blue filled squares), and the oscillating shock location \(R_{\rm s}\) (magenta line) at the equator during two time intervals: \(t = 2.0 - 2.025 \times 10^7\) s in the left panel and \(t = 2.04 - 2.06 \times 10^7\) s in the right panel. The square points are plotted at intervals of 26.4 minutes. The arrows indicate the positions of local peaks on the luminosity curve, while the labels \(B1-B6\) and \(S1-S6\) denote the groups of peak points for bremsstrahlung and synchrotron flares, respectively. The number in parentheses next to the symbol \(S\) indicates the delay point number relative to the corresponding bremsstrahlung peak, where a negative or positive sign signifies a delay or inverse delay, respectively. Table \ref{Table_2} provides the delay times, in hours, between the peaks of \(L_{\rm syn}\) and \(L_{\rm br}\) as derived from Figure \ref{Figure_4}. A negative number indicates a delay, while a positive number indicates an inverse delay.
 
We observe that all synchrotron peaks show a delay of 1.0 to 3.5 hours compared to the bremsstrahlung peaks, as illustrated in the left panel of Figure \ref{Figure_4} and summarized in Table \ref{Table_2}. For example, the synchrotron peak at \(S2(-5)\) exhibits a delay of 2.2 hours relative to the \(B2\) peak, where each point interval corresponds to 26.4 minutes, and the time resolution of the simulation is 6.6 minutes. Conversely, the synchrotron emission peaks \(S5\) and \(S6\) on the shock ascending branch, depicted in the right panel of Figure \ref{Figure_4}, display an inverse delay compared to the corresponding bremsstrahlung emissions. It is essential to note that the observed delays can vary depending on the selection of peak pairs. For instance, if we select the pair \((S2, B3)\) instead of \((B3, S3)\) in the left panel of Figure \ref{Figure_4}, the synchrotron emission peak \(S2\) will show an inverse delay relative to \(B3\). Similarly, if we choose the pair \((B5, S6)\) in the right panel instead of \((S6, B6)\), the peak \(S6\) will demonstrate a delay in relation to peak \(B5\).

We explain the observed delay phenomena using the luminosity peak points \( B1 \), \( S1 \), \( B2 \), and \( S2 \) illustrated in the left panel of Figure \ref{Figure_4}, based on the shock oscillation model. In our previous studies \citep{Okuda-etal22, Okuda-etal23}, we confirmed that synchrotron-dominant emissions consistently arise from the innermost core region (\( R \le 3R_{\rm g} \)). In contrast, the brightest bremsstrahlung luminosity is emitted near the minimum location of the oscillating shock, while the bremsstrahlung emission from the more distant shock is negligible. To provide a clearer interpretation of the results, we present the radial distribution of sound speed ($a$, shown by the blue line), Mach number ($M$, represented by the magenta line), and temperature ($T$, indicated by the red line) at the equator, as illustrated in Figure \ref{Figure_5}. These measurements are taken at the following times: $t$ = 2.01155 ($a$), 2.01277 ($b$), 2.01553 ($c$), and 2.04578 ($d$) $\times 10^7$ s. These times correspond to the peak points $B1$, $B2$, and $B3$ in the left panel of Figure \ref{Figure_4}, and $B4$ in the right panel of Figure \ref{Figure_4}. The oscillating shock fronts appear as sharp vertical lines in each curve. In Figure \ref{Figure_5}$d$, three shock waves are visible, with the outermost shock being the oscillating shock. The inner shocks produce from the interaction between this outermost shock and the burst phenomena originating from the inner core region \citep{Okuda-etal23}. The bremsstrahlung peak, labeled as $B1$, occurs when the shock moves near to the horizon at approximately \(R_{\rm s} \sim 28 R_{\rm g}\). This formation of shock near to the horizon leads to high temperatures behind the shock, resulting in significant bremsstrahlung emission. The acoustic perturbation generated behind the shock travels inward to the inner core region, where the magnetic field is the strongest (around \(3R_{\rm g}\)). This interaction produces a bright synchrotron emission peak, labeled as $S1$ after a transit time of \(25R_{\rm g}/0.08c\) (\(\sim 3.5h\)), with the average acoustic speed taken to be \(0.08c\). Subsequently, the strong acoustic perturbation generated at the $S1$ point propagates outward and collides with the ascending shock at \(R_{\rm s} = 33\). This occurs after a transit time of \(33R_{\rm g}/0.05c\) (\(\sim 7h\)), causing a flare-up that results in another bremsstrahlung peak, \(B2\), because the acoustic wave travels much faster than the ascending shock velocity. The acoustic perturbation generated behind the shock at the $B2$ point travels back towards the event horizon, resulting in a synchrotron flare occurring at the $S2$ point after a transit time of \(30R_{\rm g}/0.15c\) (\( \sim 2.2h\)). In this context, the acoustic velocity at \(R \sim 30\) is estimated to be about \(0.08c\) and \(0.15c\), respectively. In this analysis of time delays, we revise the previous interpretation that suggested the delay time from bremsstrahlung emission to synchrotron emission was equivalent to the transit time of the Alfvén wave from the bremsstrahlung peak to the synchrotron emission peak location. It is essential to note that the Alfvén wave must be influenced by the acoustic wave, as synchrotron emission is significantly dependent on both the magnetic field and the electron temperature. Additionally, the acoustic perturbation propagates much faster than the Alfvén wave.

%%%%%%%%%%%%%%%%%%%%%%%%%%%%%%%%%%%%%%%%%%%%%%%%%%%%%%%%%%%%%%%%%%%%%
%                        Table 3
%%%%%%%%%%%%%%%%%%%%%%%%%%%%%%%%%%%%%%%%%%%%%%%%%%%%%%%%%%%%%%%%%%%%%
\begin{table}
\centering
\caption{Time delay in unit of min of 4.5$\mu$m and 2-8 KeV to 334 GHz luminosity peaks in Figure \ref{Figure_6}. The negative and positive signs indicate delay and inverse delay, respectively.
}
\label{Table_3}
\begin{tabular}{@{}cccccc} 
%&&&&&& \\
\hline \hline
${\rm luminosity \; peak \; group}$ & ${\rm A}$ & ${\rm B}$ & ${\rm C}$ & ${\rm D}$ & ${\rm E}$   \\
 \hline

 $334 \; {\rm GHz}$     & - & - & - & - & - \\ 
 $4.5 \; \mu$m     &  +20 &  +40 & -40 & +20 &  -20 \\ 
 $2-8$ \; ${\rm KeV}$  & -53    &  -53 & +40 &  +40  &  -33  \\ 

\hline
\end{tabular}
\end{table}
%%%%%%%%%%%%%%%%%%%%%%%%%%%%%%%%%%%%%%%%%%%%%%%%%%%%%%%%%%%%%%%%%%%%%

The delay time of the luminosity peaks in the right-hand panel is explained in a manner similar to that in the left-hand panel of Figure \ref{Figure_4}. However, for the peak point B4, we identify the origin of the bremsstrahlung flare as the innermost shock, which occurs at $R_{\rm s} \sim 10$, rather than the outermost oscillating shock. By taking the average sound velocity to be 0.15 for $R \le 10$ (as shown in Figure \ref{Figure_5}$d$), we calculate the transit time from $B4$ to $S5$ to be approximately $10R_{\rm g}/0.15c$ (about 44 min). This is comparable to the simulation delay time of 40 min. This interpretation aligns qualitatively with the delay times observed between the luminosity peak points in Figure \ref{Figure_4}. The shock oscillation model predicts that bright bremsstrahlung emission peaks, such as point $B1$ in the left-hand panel and point $B4$ in the right-hand panel of Figure \ref{Figure_4}, will necessarily occur. The time delay phenomena of synchrotron emission following the bremsstrahlung peaks appear successively as long as the shock location is not far from the event horizon. However, when the shock reaches locations near the maximum at $R_{\rm s} > 90$, both synchrotron and bremsstrahlung emissions become weak, and we do not observe clear and cyclic delay phenomena. This is because the transit time exceeds the time scale of the shock oscillation.

On the other hand, we observe that an additional synchrotron peak emerges independently in the oscillating shock model when a chaotic and significant magnetic burst occurs due to the MRI activity near the event horizon. In this scenario, the strong synchrotron emission peak initially appears near the event horizon, followed by a bremsstrahlung emission peak resulting from the interaction between the expanding burst and the surrounding matter. We note an inverse delay between the synchrotron emission and the bremsstrahlung emission, as indicated by $(S5,B5)$ and $(S6,B6)$ in Figure \ref{Figure_4}. This type of flare event has been demonstrated in previous simulations by \citet{Okuda-etal23} and has been observationally reported by \citet{Ponti-etal-17}.

%%%%%%%%%%%%%%%%%%%%%%%%%%%%%%%%%%%%%%%%%%%%%%%%%%%%%%%%%%%%%%%%%%%%%
\subsection{Time delay between 334 GHz, 4.5 $\mu$m and 2-8 Kev flares}
%%%%%%%%%%%%%%%%%%%%%%%%%%%%%%%%%%%%%%%%%%%%%%%%%%%%%%%%%%%%%%%%%%%%%

Recent observations by \citet{Michail-etal-24} have reported a time delay between flares at 334 GHz, 4.5 µm, and 2-8 keV in Sgr A*. Motivated by these findings, we also investigate this time delay using the luminosity curves of \( L_{\rm 334} \) at 334 GHz (black line), \( L_{\rm 4.5} \) at 4.5 µm (red line), \( L_{\rm 2-8} \) at 2-8 keV (cyan line), and \( R_{\rm s} \) (magenta line). We analyze the data for the time intervals \( t = 2.010 - 2.022 \times 10^7 \)s and \( t = 2.040 - 2.058 \times 10^7 \)s, as shown in Figure \ref{Figure_6}. In this figure, an arrow indicates local peaks in the luminosity curves, with symbols labeled A, B, C, D, and E representing selected luminosity peaks. The attached numbers 1, 2, and 3 correspond to those observed at 334 GHz, 4.5 µm, and 2-8 keV luminosities, respectively. We summarize the time delay data from the peaks of the 4.5 µm and 2-8 keV luminosities relative to the 334 GHz peaks in Table \ref{Table_3}. Here, negative values indicate delays, while positive values indicate inverse delays. The numbers in parentheses represent the delay times in minutes. Our observations reveal that \( L_{\rm 334} \) delays \( L_{\rm 4.5} \) by approximately ±20 minutes and ±40 minutes, which are comparable to previous results obtained without considering the Compton scattering effect \citep{Okuda-etal23}. In contrast, \( L_{\rm 334} \) delays \( L_{\rm 2-8} \) by 40 minutes, but also shows inverse delays of 33 and 53 minutes.

%%%%%%%%%%%%%%%%%%%%%%%%%%%%%%%%%%%%%%%%%%%%%%%%%%%%%%%%%%%%%%%%%%%%%
%                        Figure 7
%%%%%%%%%%%%%%%%%%%%%%%%%%%%%%%%%%%%%%%%%%%%%%%%%%%%%%%%%%%%%%%%%%%%% 
\begin{figure*}
    \begin{tabular}{cc}
      \begin{minipage}{0.5\linewidth}
        \centering
        \includegraphics[width=1.0\textwidth]{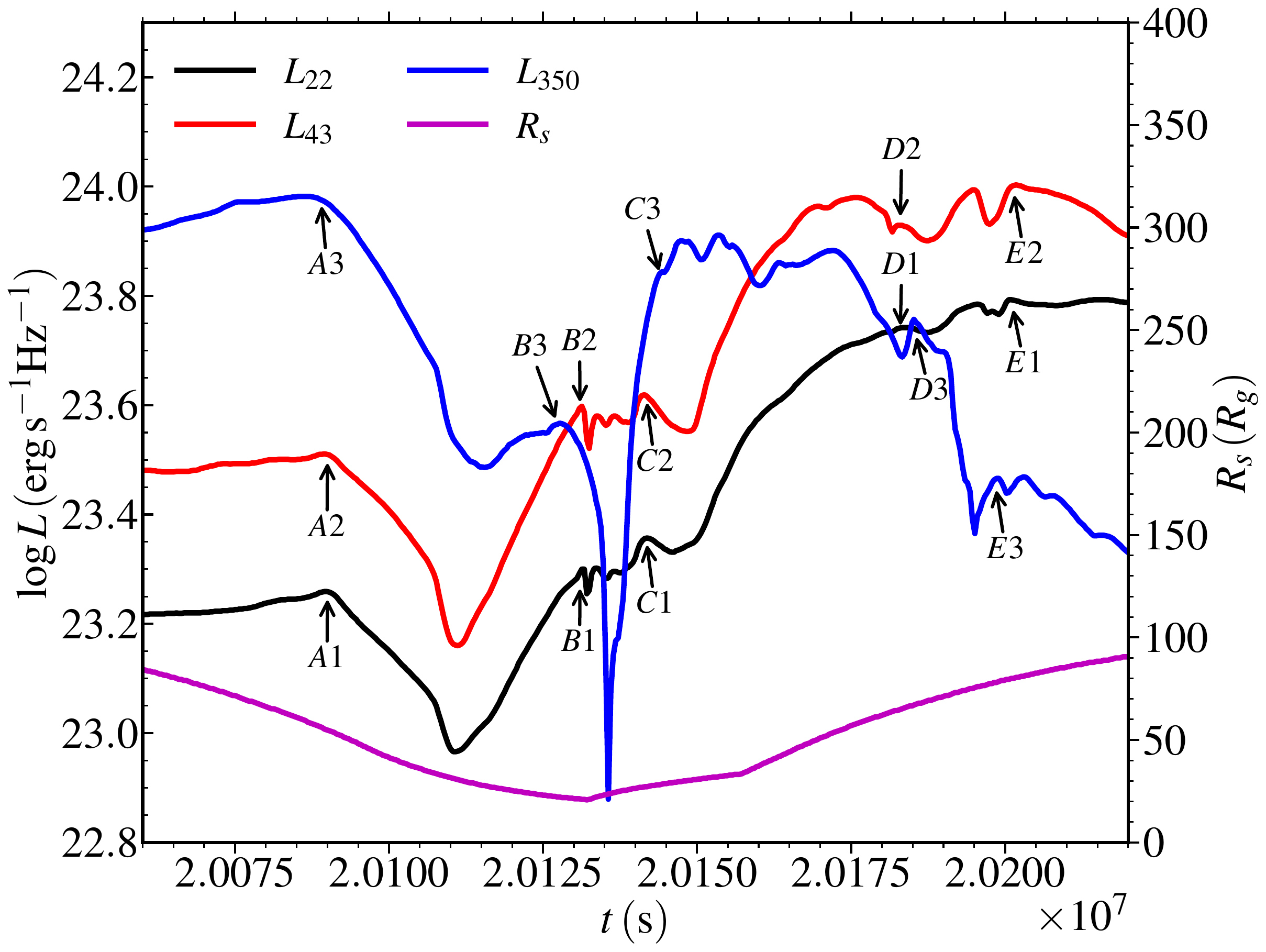}
        \label{fig:7a}
      \end{minipage}

      \begin{minipage}{0.5\linewidth}
        \centering
        \includegraphics[width=1.0\textwidth]{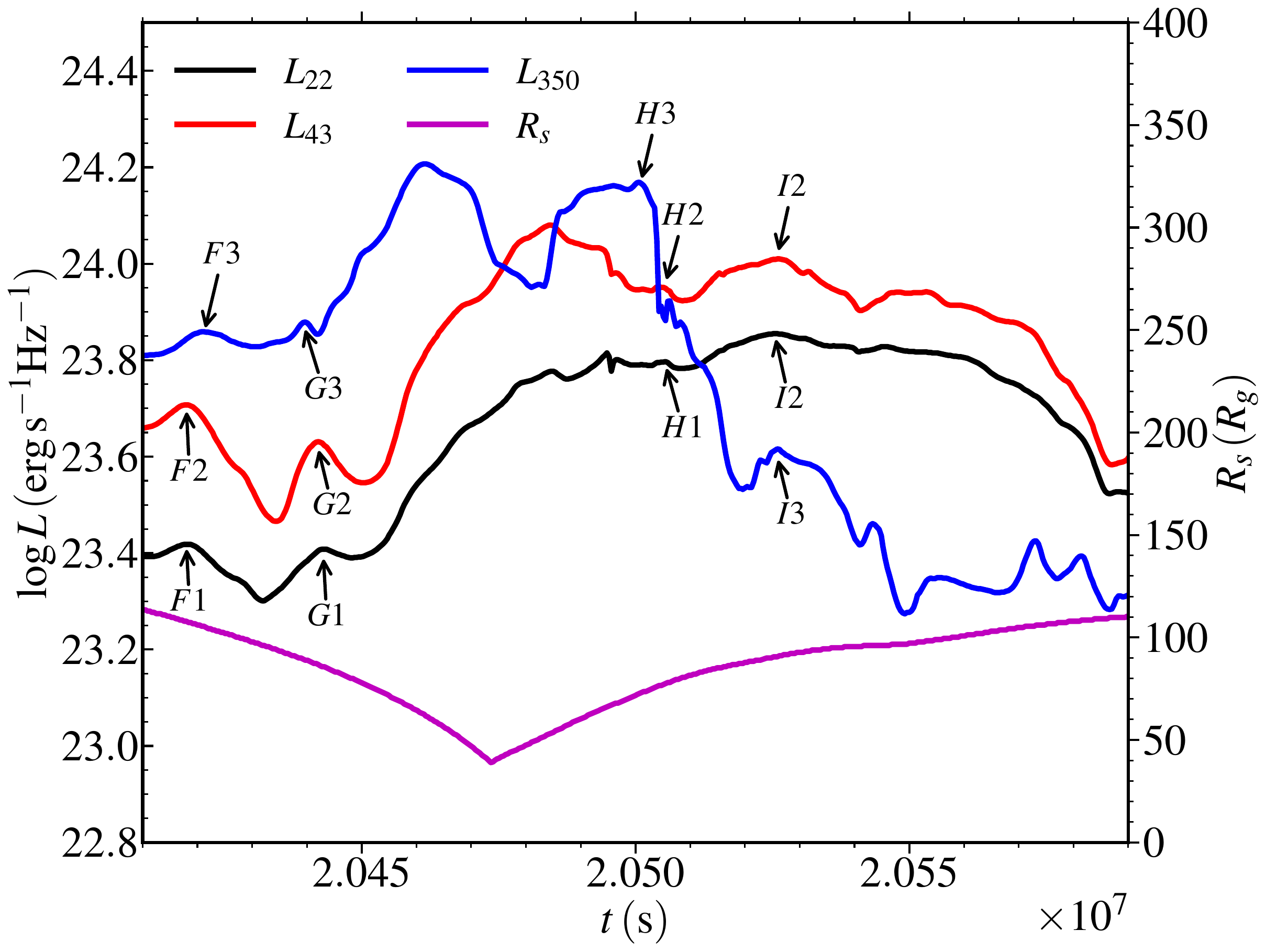}
        \label{fig:7b}
      \end{minipage} \\
   \end{tabular}
\caption { Same as Figure \ref{Figure_6} but monochromatic luminosity curves for $L_{\rm 22}$ at 22 GHz (black line), $L_{\rm 43}$ at 43 GHz (red line), $L_{\rm 350}$ at 350 GHz (blue line) in units of erg ${\rm s}^{-1} {\rm Hz}^{-1}$ and oscillating shock location $R_{\rm s}$ (green line) in units of $R_g$ on the equator during $t= 2.010 - 2.058 \times 10^7$ s. The symbols $A$ to $I$ are the group name  of the selected peaks and the attached number 1, 2 and 3 correspond to those for 22 GHz, 43 GHz  and 350 GHz luminosity, respectively.
  }
  \label{Figure_7}
  \end{figure*}
%%%%%%%%%%%%%%%%%%%%%%%%%%%%%%%%%%%%%%%%%%%%%%%%%%%%%%%%%%%%%%%%%%%%%

%%%%%%%%%%%%%%%%%%%%%%%%%%%%%%%%%%%%%%%%%%%%%%%%%%%%%%%%%%%%%%%%%%%%%
%                        Figure 8
%%%%%%%%%%%%%%%%%%%%%%%%%%%%%%%%%%%%%%%%%%%%%%%%%%%%%%%%%%%%%%%%%%%%% 
\begin{figure*}
    \begin{tabular}{cc}
      \begin{minipage}{0.5\linewidth}
        \centering
        \includegraphics[width=1.0\textwidth]{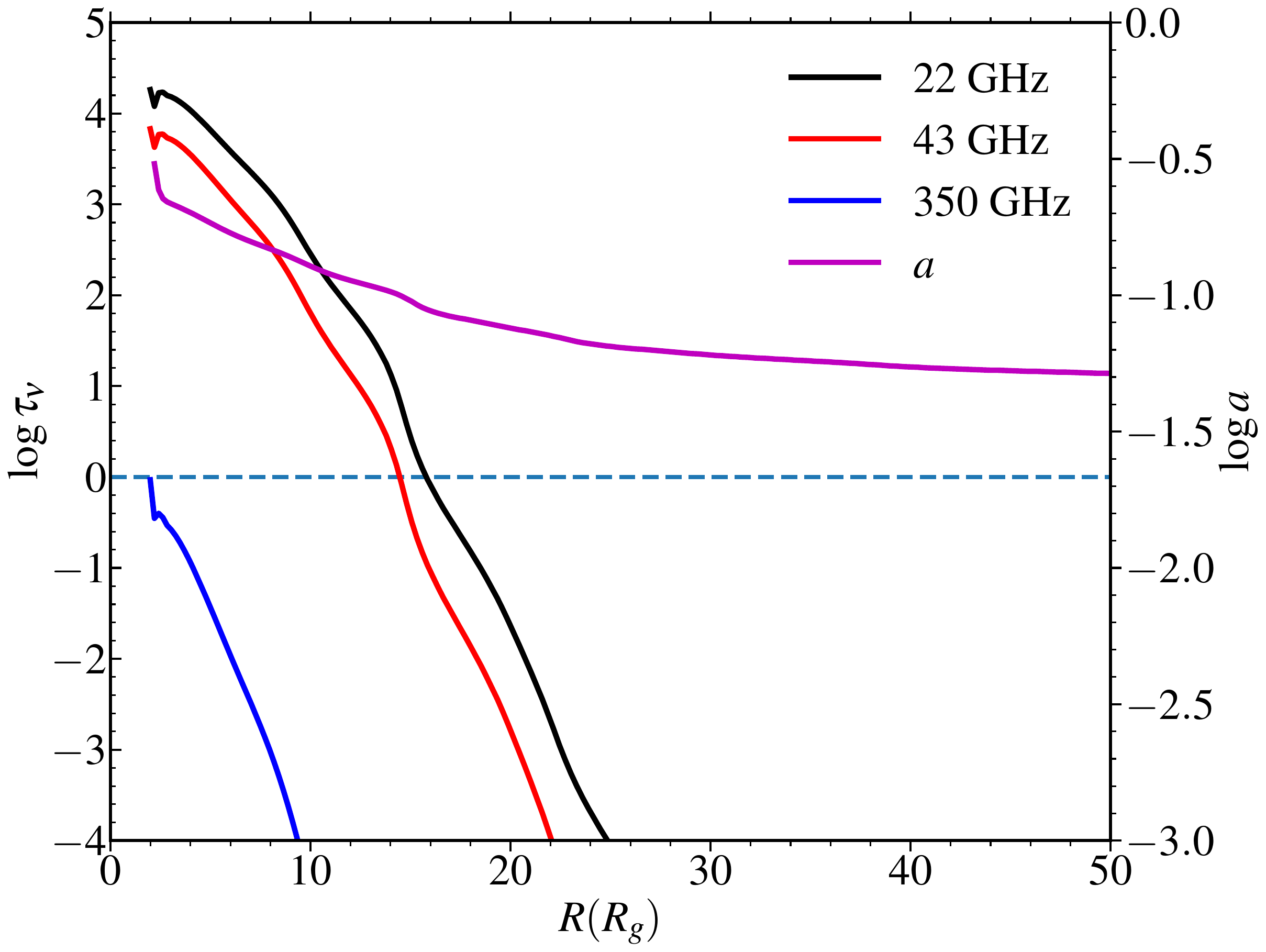}
        \label{fig:8a}
      \end{minipage}

      \begin{minipage}{0.5\linewidth}
        \centering
        \includegraphics[width=1.0\textwidth]{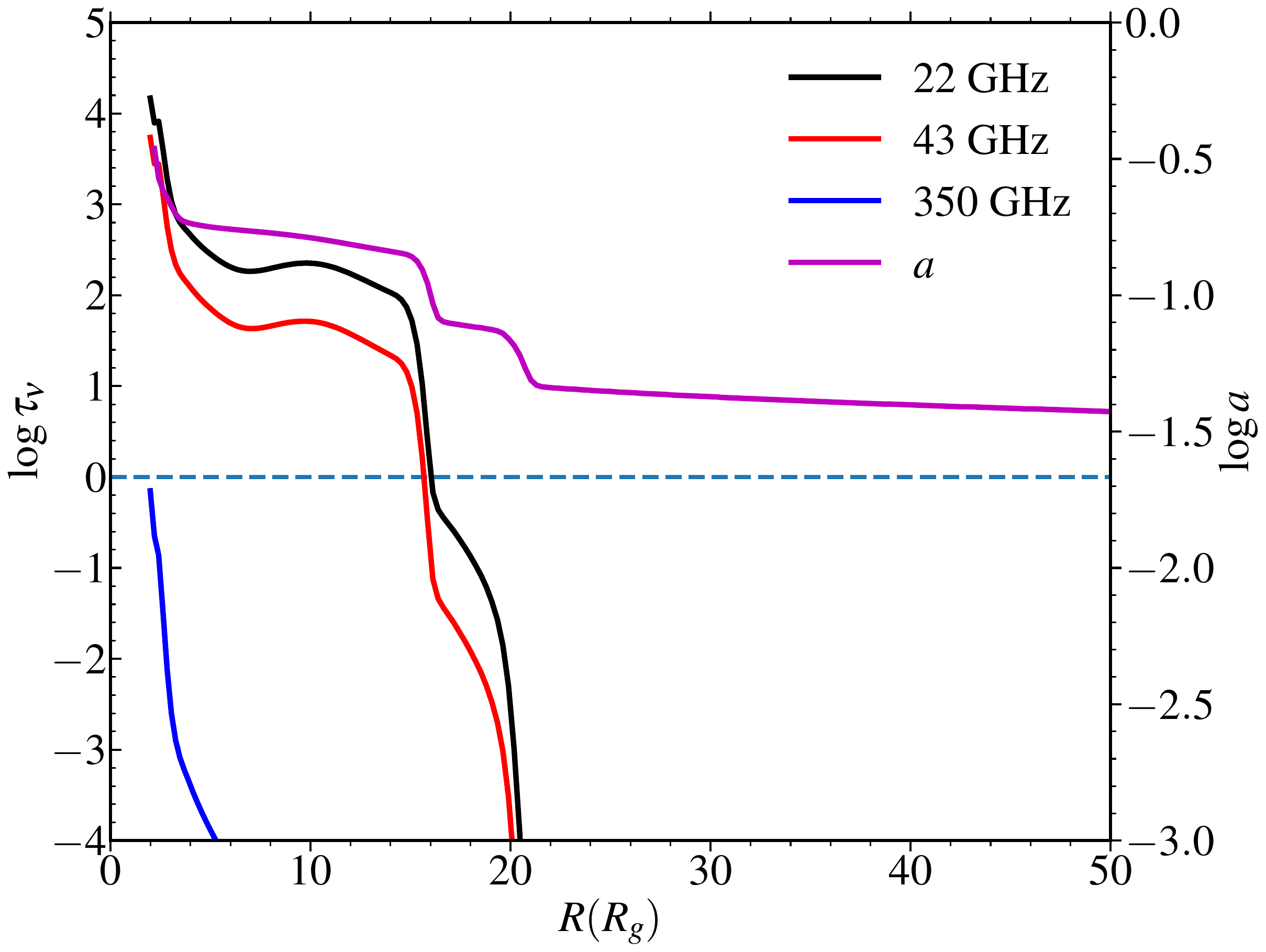}
        \label{fig:8b}
      \end{minipage} \\
   \end{tabular}
\caption { Left: optical depth ${\tau}_{\rm \nu}$ at the disc surface in the frequency 22 GHz (black line), 43 GHz (red line) and 350 GHz (blue line) and  sound speed $a$ (magenta line) at $t=2.00896 \times 10^7$ s corresponding to  group $A$ in Figure \ref{Figure_7}. Right: same as left-hand panel but for $t=2.01312 \times 10^7$ s corresponding to group $B$ in Figure \ref{Figure_7}. See the text for details.}
  \label{Figure_8}
  \end{figure*}
%%%%%%%%%%%%%%%%%%%%%%%%%%%%%%%%%%%%%%%%%%%%%%%%%%%%%%%%%%%%%%%%%%%%% 

%%%%%%%%%%%%%%%%%%%%%%%%%%%%%%%%%%%%%%%%%%%%%%%%%%%%%%%%%%%%%%%%%%%%% 
%                       Table 4
%%%%%%%%%%%%%%%%%%%%%%%%%%%%%%%%%%%%%%%%%%%%%%%%%%%%%%%%%%%%%%%%%%%%% 
\begin{table*}
\centering
\caption {Time delay in unit of min  for  43 GHz and 350 GHz to 22 GHz luminosity peaks
in Fig.~7.
The negative and positive sign mean delay and inverse delay, respectively.  }
\label{Table_4}
\begin{tabular}{@{}cccccccccc} 
&&&&&& \\
\hline \hline
${\rm luminosity \; peak \; group}$ & ${\rm A}$ & ${\rm B}$ & ${\rm C}$ & ${\rm D}$ & ${\rm E}$ & ${\rm F}$ & ${\rm G}$ & ${\rm H}$ & ${\rm I}$   \\
 \hline

 $22 \; {\rm GHz}$     & - & - & - & - & -& - & - & - & -\\ 
 $43 \; {\rm GHz}$     &  +7 & +7   & +7 & +7 &-20 & -7& +20 & +20  & -7 \\ 
 $350\; {\rm GHz}$   & +53 & +59 &-40 &-26&+33 &-46 & +59 & +73 & -7\\ 

\hline
\end{tabular}
\end{table*}
%%%%%%%%%%%%%%%%%%%%%%%%%%%%%%%%%%%%%%%%%%%%%%%%%%%%%%%%%%%%%%%%%%%%% 

%%%%%%%%%%%%%%%%%%%%%%%%%%%%%%%%%%%%%%%%%%%%%%%%%%%%%%%%%%%%%%%%%%%%% 
\subsection{Time delay between 22, 43 and 350 GHz flares}
%%%%%%%%%%%%%%%%%%%%%%%%%%%%%%%%%%%%%%%%%%%%%%%%%%%%%%%%%%%%%%%%%%%%% 

 The time delays between the luminosities at 22 GHz, 43 GHz, and 350 GHz, which are affected by Compton scattering, are examined in this study for comparison with previous results \citep{Okuda-etal23}. Figure \ref{Figure_7} displays the monochromatic luminosity curves: \(L_{\rm 22}\) at 22 GHz (black line), \(L_{\rm 43}\) at 43 GHz (red line), and \(L_{\rm 350}\) at 350 GHz (blue line), measured in units of erg ${\rm s}^{-1} {\rm Hz}^{-1}$. Additionally, the oscillating shock location \(R_{\rm s}\) is depicted (magenta line) over the time span from \(t = 2.006\) to \(2.058 \times 10^7\) seconds. The symbols \(A\) to \(I\) represent groups of selected luminosity peaks, with the accompanying numbers 1, 2, and 3 corresponding to the peaks for 22 GHz, 43 GHz, and 350 GHz luminosities, respectively. In Table \ref{Table_4}, we list the delay times of the 43 GHz and 350 GHz luminosity peaks relative to the 22 GHz peaks, similar to Table \ref{Table_3}. The results show that \(L_{\rm 22}\) delays \(L_{\rm 43}\) by approximately ±7 and ±20 minutes. In contrast, \(L_{\rm 22}\) delays \(L_{\rm 350}\) by 33, 53, 59, and 73 minutes, while it exhibits inverse delays of 26, 40, and 46 minutes. A systematic pattern of delay is observed: \(L_{\rm 22}\) delays \(L_{\rm 43}\) and \(L_{\rm 350}\) on the descending branch of the shock, while inverse delays occur on the ascending branch. These delay times are somewhat shorter than those observed between the luminosities without Compton scattering \citep{Okuda-etal23}.

We qualitatively interpret the time delays in relation to the frequency-dependent effective radii \( R_{\rm eff}(\nu) \), where the optical thickness \( \tau_\nu (R_{\rm eff}) \) at frequency \( \nu \) approaches unity \citep{Okuda-etal23}. We explain the occurrences of both positive and negative delays within the oscillating shock model. In equation \ref{flux_eqn}, we have:
\[
F_{\rm \nu}(R) \propto 
\begin{cases} 
     B_{\rm \nu}(R) \tau_{\rm \nu}(R) & \text{if } \tau_{\rm \nu} \ll 1 \\ 
     B_{\rm \nu}(R) & \text{if } \tau_{\rm \nu} \gg 1.
\end{cases}
\] 
This explains how the flux at frequency (\( \nu \)) varies depending on the optical thickness.

The radiative flux \( F_{\nu}(R) \) is very weak compared to black body radiation when \( \tau_{\nu}(R) \ll 1 \). However, it becomes significantly brighter, resembling black body radiation, when \( \tau_{\nu}(R) \gg 1 \). Generally, we know that the optical depth \( \tau_{\nu}(R) \) for synchrotron opacity in the radio region decreases with increasing radius \( R \) at a fixed frequency, and with increasing frequency at a fixed radius. As a result, the radio emission \( F_{\nu} \) sharply declines beyond the frequency-dependent effective radius. In our model, the synchrotron emission, and thus the radio monochromatic emission, begins near the event horizon and expands outward. The bright monochromatic emission is quenched when it reaches the frequency-dependent effective radii. Consequently, the luminosity peak at lower frequencies occurs later than at higher frequencies; that is, the peak \( L_{22} \) should occur after the peak \( L_{350} \). Conversely, strong acoustic perturbations from the upstream shock-heated region can sometimes be transmitted downstream. In this scenario, the initially weak radio emission flares up when the perturbations reach \( R_{\text{eff}}(\nu) \), resulting in bright radio emission that appears as the frequency increases. By considering the frequency-dependent effective radii, we can roughly estimate the delay time between the monochromatic luminosity peaks as the transit time of the sound wave between the radii.

%This means that the radiaive flux $F_{\rm \nu }(R)$ is very weak compared with the black body radiation when $\tau_{\rm \nu} (\rm R) \ll 1$ while it brightens effectively as the balck body radiation if $\tau_{\rm \nu} (\rm R) \gg 1$. Generally we know that $\tau_{\rm \nu}(\rm R)$ for the synchrotron opacity in radio region considered here decreases with increasing $R$ at a fixed frequency and with increasing frequency at a fixed radius. Therefore the radio emission $F_{\rm \nu}$ decreases sharply beyond the frequency-dependent effective radius. In our model, the synchrotron emission and accordingly radio monochromatic emission peak always starts near the event horizon and expand outward and then the bright monochromatic emission is quenched when it attains the frequency-dependent effective radii. As a result the luminosity peak at smaller frequency delays that at larger frequency, that is, $L_{\rm 22}$ peak should delay $L_{\rm 350}$. Conversely, the strong acoustic perturbation from the upstream shock heated region is sometimems transmitted downstream. In this case, the initially weak radio emission flares up when the perturbations attains ${\rm R}_{\rm eff}(\nu)$ and the bright radio emission appears in turn with increasing frequency. With the frequency-dependent effective radii, we can estimate roughly the delay time between the monochromatic luminosity peaks as the transit time of  sound wave between the radii. 

Figure \ref{Figure_8} illustrates the radial dependence of optical thickness \(\tau_{\nu}(\text{R})\) at frequencies of 22, 43, and 359 GHz for time instances \(t = 2.009 \times 10^7\) s and \(2.013 \times 10^7\) s, which correspond to groups $A$ and $B$ in Figure \ref{Figure_7}, respectively. In both groups $A$ and $B$, the 22 GHz flares lead the 350 GHz flares in terms of peak delays. From the analysis of Figure \ref{Figure_8}, we find the frequency-dependent effective radii \(R_{\text{eff}} (22)\), \(R_{\text{eff}} (43)\), and \(R_{\text{eff}} (350)\) to be approximately 15.9, 13.0, and $\sim 2$ for group $A$, and 16.1, 15.1, and $\sim 2$ for group $B$, respectively. The delay times between the 22 GHz and 43 GHz flares, as well as between the 22 GHz and 350 GHz flares, are estimated using the transit time of the sound speed. This results in delays of \([R_{\text{eff}} (22) - R_{\text{eff}} (43)]/a\) and \([R_{\text{eff}} (22) - R_{\text{eff}} (350)]/a\), which yield estimates of 12 and 56 minutes with \(a = 0.16\) for group $A$, and 4 and 59 minutes with \(a \sim 0.16\) for group $B$, respectively. These delay times are in reasonable agreement with the numerical values reported at $A$ and $B$ in Table \ref{Table_4}. This interpretation is also applicable to groups $C$ and $D$, which show inverse delays in Figure \ref{Figure_7}. The estimated delay times from the 22 GHz peak (\(L_{22}\)) to the 350 GHz peak (\(L_{350}\)) in groups $C$ and $D$ are approximately 60 and 47 minutes with \(a = 0.2\), respectively. The emission peaks corresponding to radio emissions $A$, $B$, $C$, and $D$ are attributed to the synchrotron \(S0\), bremsstrahlung \(B1^{\prime}\), bremsstrahlung \(B2^{\prime}\), and synchrotron \(S2^{\prime}\) peaks on the luminosity curves presented in Figure \ref{Figure_4}.

\section{Comparison of simulation results with observations of Sgr A*}
%%%%%%%%%%%%%%%%%%%%%%%%%%%%%%%%%%%%%%%%%%%%%%%%%%%%%%%%%%%%%%%%%%%%% 

We compare the time lag between synchrotron and bremsstrahlung emissions in the shock oscillation model with the time differences observed between radio and X-ray emissions from Sgr A$^*$. A simultaneous observation campaign involving radio, near-infrared (NIR), and X-ray data collected from HST NICMOS, XMM-Newton, and VLA has revealed that radio flares delay X-ray flares by approximately 2 hours in July 2006 \citep{Yusef-Zadeh-etal-08}, about 5 hours in April 2007 \citep{Yusef-Zadeh-etal-09}, less than 80 minutes in July 2013, roughly 2 hours in September 2013, and between 7.5 hours and 3.9 hours in October 2013. In May 2014, the delay was approximately 30 minutes \citep{Capellupo-etal-17}, followed by a similar delay of about 30 minutes in July 2019 \citep{Michail-etal-21b}, and a 20 to 30-minute delay also in July 2019 \citep{Boyce-etal-22}. The observed delay times are approximately 30 minutes and between 1 to 5 hours, which aligns roughly with the time delays of around 40 minutes and 1 to 4 hours obtained from current numerical simulations, although the delay times can vary depending on the luminosity peak point.

Recently, multiwavelength observations of Sgr A$^*$ conducted on July 21 and 26, 2019 \citep{Michail-etal-24}, revealed that the radio emission at 334 GHz ($L_{\rm 334}$) is delayed relative to the flux at 4.5 µm ($L_{\rm 4.5}$) by 9, 39, and -37 minutes. Additionally, $L_{\rm 334}$ is delayed with respect to the 2-8 keV emissions ($L_{\rm 2-8}$) by 26 and 42 minutes. These delay times align with those found in numerical simulations. \citet{Yusef-Zadeh-etal-08} reported that the 22 GHz flares are delayed by 20 and 30 minutes compared to the 43 GHz flares, and they delay the 335 GHz flares by 65 minutes. These results are comparable to the delay times shown in Table 4. It is important to note that these delay times are not fixed; they vary within the previously mentioned range depending on the evolutionary time of the system. Observations also indicate that the luminosity peaks at 334 GHz and 4.5 µm occur cyclically with a short interval of approximately 40 minutes, while the luminosities in the simulations increase with much longer intervals. Furthermore, the latest observations \citep{Michail-etal-24} clearly demonstrate an inverse delay of the 334 GHz emission relative to the 4.5 $\mu$m emission. We emphasize that such an inverse delay would not be detected using only the expanding hot plasmon model. However, it may occur naturally within the shock oscillation model without requiring any additional assumptions.

% Recently, the multiwavelength observations of Sgr A$^*$ on 2019 July 21 and 26 \citep{Michail-etal-24} shows that the radio emission $L_{\rm 334}$ delays $L_{\rm 4.5}$ at 4.5$\mu$m  by 9, 39 and -37 min and delays $L_{\rm 2-8}$ at 2-8 Kev by 26 and 42 min, respectively. These delay times are also within those in the numerical simulations. \citet{Yusef-Zadeh-etal-08} reports that  22 GHz flare delays 43 GHz flare by 20 and 30 min and delays 335 GHz flare by 65 min which are comparable to  the delay times in Table 4. These delay times are not fixed but are variable between the above range of times depending on the evolutionary time. However, the observation shows that the 335 GHz and 4.5$\mu$m luminosity peaks appear cyclicaly with short interval of $\sim$ 40 min but the luminosities in the simulation flare up with far longer intervals. The latest observation \citet{Michail-etal-24} also shows clearly that the inverse delay of 334 GHz to 4.5$\mu$m occurs. We emphasize that such inverse delay would be never found only by the expanding hot plasmon model. However  in the shock oscillation model it may naturally occurs without any assumption.

%%%%%%%%%%%%%%%%%%%%%%%%%%%%%%%%%%%%%%%%%%%%%%%%%%%%%%%%%%%%%%%%%%%%% 
\section{Summary and Discussion}
%%%%%%%%%%%%%%%%%%%%%%%%%%%%%%%%%%%%%%%%%%%%%%%%%%%%%%%%%%%%%%%%%%%%% 

We investigated the time delay relationships of flares between radio and X-ray emissions, as well as between narrow radio frequency bands in Sgr A*. This was based on our analysis of synchrotron and bremsstrahlung luminosity, along with monochromatic luminosity \(L_{\rm \nu}\) curves derived from time-dependent 2D relativistic radiation MHD simulations based on the shock oscillation model. The following summarizes our time delay analysis, providing a more reasonable interpretation of the time delays observed.

(i) The delay time between synchrotron and bremsstrahlung emissions, corresponding to radio and X-ray emissions respectively, ranges from 1 to 4 hours, depending on the phase of the shock oscillation. This observation is in rough agreement with recent simultaneous radio and X-ray observations. The delay time is not significantly different from our previous simulation results that did not include the effects of Compton scattering. However, it should be noted that the luminosity, when accounting for Compton scattering, is enhanced by a factor of three.
 
(ii) The bremsstrahlung emission peak occurs when the shock wave contracts near its minimum location. At this point, an acoustic perturbation generated behind the shock is transmitted inward and impacts the inner core, which has a strong magnetic field. This interaction causes a bright synchrotron emission peak. Following this, the acoustic perturbation, triggered by the bright synchrotron emission, propagates outward. It then collides with the ascending or descending shock, after the transit time of the sound from the core to the shock location, resulting in another flare-up of the bremsstrahlung emission peak. These transit times represent the delays between the radio and X-ray emissions. This cycle continues to repeat as long as the shock location remains close to the event horizon.

(iii) The 22 GHz emission experiences delays of ±7 to ±20 minutes for the 43 GHz emission and delays of 33 to 73 minutes for the 350 GHz emission. Additionally, it shows an inverse delay of 26, 40, and 46 minutes. A systematic pattern of these delays has been established: the delay from 22 GHz to 43 GHz and 350 GHz emissions occurs during the shock descending branch, while the inverse delay happens during the ascending branch.

(iv) In our model, the synchrotron emission, and consequently the radio monochromatic emission peak, always occurs near the event horizon. The acoustic perturbation excited by the synchrotron emission expands outward. When it reaches the frequency-dependent effective radius, where the optical thickness on the disc surface is approximately 1, the bright radio emission is quenched. Meanwhile, the strong acoustic perturbation from the upstream shock-heated region is transmitted downstream, causing the initially weak radio emission to flare up once the perturbation reaches the frequency-dependent effective radius. The delay time between the flares at different radio wavelengths corresponds to the transit time of sound between the frequency-dependent effective radii. In the first case, the luminosity peak at smaller frequencies occurs after that at larger frequencies, while in the latter case, it precedes the luminosity peak at larger frequencies.

(v) Time delays for emissions at 334 GHz to 4.5 µm and for 2-8 keV flares are estimated to be ±20 minutes and ±40 minutes, and 33 to 53 minutes, respectively, based on numerical simulations. These estimates are comparable to the recent observational findings of 9, 39, and -37 minutes, as well as 26 and 42 minutes \citep{Michail-etal-24}. It's important to note that the time delay between the luminosity peaks at 334 GHz and 4.5 µm is not interpreted in the same way as described above, since the optical thickness at 4.5 µm is consistently much less than one. Conversely, the time delay between the luminosity peaks at 334 GHz and 2-8 keV is interpreted similarly to the relationship between synchrotron and bremsstrahlung emission peaks. Additionally, it should be pointed out that the inverse delay was detected for the first time in this observation \citep{Michail-etal-24}, and it occurs naturally in the shock oscillation model.

It should be noted that in our model, the sound wave travels much faster than the Alfv\'en wave \citep{Okuda-etal23}. As a result, faster magneto-sonic waves are present in the flow, which are a combination of sound waves and Alfv\'en waves. Clearly, the fast magneto-sonic waves are more dominant than both sound waves and Alfv\'en waves in the current analysis. In this paper, we do not compare the flaring events with the fast magneto-sonic waves, but we hope to make this comparison in future works.

The present shock oscillation model effectively explains the phenomenon of delay time between radio and X-ray emissions, as well as between different radio wavelengths, observed during short-term flares of Sgr A$^*$, which span from a few minutes to several hours. This model also accounts for longer time-scale flares lasting 5 to 10 days, as detailed in previous studies \citep{Okuda-etal19, Singh-etal-21, Okuda-etal22, Okuda-etal23}. Regarding the time delay phenomena, \citet{Yusef-Zadeh-etal-08} and \citet{Michail-etal-24} propose a hot expanding plasma model \citep{van-der-Laan-66} or an occultation model involving an adiabatically expanding hot spot. They explain the observed flux values in radio wavelengths by fitting appropriate parameters to their observed light curves. The monochromatic emission peaks at lower frequencies are expected to be delayed compared to those at higher frequencies; however, observations also reveal an opposite delay. \citet{Witzel-etal-21} propose a simple source model based on a three-step process involving electron injection with a non-thermal energy distribution into a spherical central region, compression of magnetic field lines with increasing magnetic flux, and adiabatic expansion. In contrast to their model, the shock oscillation model can account for both the delay and the inverse delay without making assumptions about typical accretion flow—this is applicable even when gas flows toward the event horizon. This model is particularly advantageous for explaining the delay phenomena observed at radio wavelengths. Additionally, observations indicate that radio emissions at 335 GHz and 4.5 $\mu $m occur more frequently, with short intervals of approximately 40 minutes. We speculate that these frequently occurring radio emission peaks are due to a significant enhancement of the magnetic field near the event horizon. We believe that a 3D simulation, instead of a 2D MHD simulation of the shock oscillation model, may better replicate this behavior, especially since we have confirmed that magnetorotational instability (MRI) activity is stronger in 3D simulations than in 2D ones \citep{Aktar-etal-25}. Therefore, a 3D simulation of the shock oscillation model may offer a promising explanation for the timing delay phenomena observed in Sgr A* flares.

In this paper, we focus on a single-temperature model and consider bremsstrahlung opacity as the sole source of radiation opacity in the PLUTO code. It is important to note that our two-temperature model relies on simplified assumptions (see subsection \ref{two-temp}). However, to accurately and consistently estimate variations in luminosity and spectral evolution, a proper two-temperature model is essential. In the future, we plan to develop a comprehensive two-temperature radiation model to further investigate the flow properties and time delay analysis of Sgr A*.

%%%%%%%%%%%%%%%%%%%%%%%%%%%%%%%%%%%%%%%%%%%%%%%%%%%%%%%%
\section*{Acknowledgments}

%This work is supported by the National Science and Technology Council of Taiwan through grant NSTC 112-2811-M-007-038 and 112-2112-M-007-040, and by the Center for Informatics and Computation in Astronomy (CICA) at National Tsing Hua University through a grant from the Ministry of Education of Taiwan. The simulations and data analysis have been carried out on the CICA Cluster at National Tsing Hua University.

We sincerely appreciate the anonymous referee for their insightful suggestions and comments, which have improved the quality of the manuscript. This work is supported by the National Science and Technology Council of Taiwan through grants NSTC 114-2112-M-001-018 and 114-2811-M-001-022. Simulations were also carried out on the {\it Kawas} cluster at ASIAA. The authors acknowledge the access to high-performance computing facilities (Theory cluster and storage) provided by ASIAA. 

%%%%%%%%%%%%%%%%%%%%%%%%%%%%%%%%%%%%%%%%%%%%%%%%%%%%%%%%
\section*{Data Availability}

The data and code underlying this article will be shared on reasonable request to the corresponding author.

%%%%%%%%%%%%%%%%%%%%%%%%%%%%%%%%%%%%%%%%%%%%%%%%%%%%%%%%
\bibliographystyle{mnras}
\bibliography{refs}

@ARTICLE{Chakrabarti-90b,
       author = {{Chakrabarti}, Sandip K.},
        title = "{Standing Rankine-Hugoniot Shocks in the Hybrid Model Flows. II. Nonaxisymmetric Self-similar Solution}",
      journal = {\apj},
         year = 1990,
        month = oct,
       volume = {362},
        pages = {406},
          doi = {10.1086/169278},
       adsurl = {https://ui.adsabs.harvard.edu/abs/1990ApJ...362..406C}
}

@ARTICLE{Shakura-Sunyaev73,
       author = {{Shakura}, N.~I. and {Sunyaev}, R.~A.},
        title = "{Reprint of 1973A\&A....24..337S. Black holes in binary systems. Observational appearance.}",
      journal = {\aap},
         year = 1973,
        month = jun,
       volume = {500},
        pages = {33-51},
       adsurl = {https://ui.adsabs.harvard.edu/abs/1973A&A....24..337S}
}

@ARTICLE{Spruit-87,
       author = {{Spruit}, H.~C.},
        title = "{Stationary shocks in accretion disks}",
      journal = {\aap},
         year = 1987,
        month = oct,
       volume = {184},
       number = {1-2},
        pages = {173-184},
       adsurl = {https://ui.adsabs.harvard.edu/abs/1987A&A...184..173S}
}

@BOOK{Chakrabarti-90a,
       author = {{Chakrabarti}, Sandip K.},
        title = "{Theory of Transonic Astrophysical Flows}",
         year = 1990,
          doi = {10.1142/1091},
       adsurl = {https://ui.adsabs.harvard.edu/abs/1990ttaf.book.....C}
}

@ARTICLE{Das-etal01,
       author = {{Das}, Santabrata and {Chattopadhyay}, Indranil and {Chakrabarti}, Sand
        ip K.},
        title = "{Standing Shocks around Black Holes: An Analytical Study}",
      journal = {\apj},
         year = 2001,
        month = aug,
       volume = {557},
       number = {2},
        pages = {983-989},
          doi = {10.1086/321692},
archivePrefix = {arXiv},
       eprint = {astro-ph/0107046},
 primaryClass = {astro-ph},
       adsurl = {https://ui.adsabs.harvard.edu/abs/2001ApJ...557..983D}
}

@ARTICLE{Balbus-Hawley91,
       author = {{Balbus}, Steven A. and {Hawley}, John F.},
        title = "{A Powerful Local Shear Instability in Weakly Magnetized Disks. I. Linear Analysis}",
      journal = {\apj},
         year = 1991,
        month = jul,
       volume = {376},
        pages = {214},
          doi = {10.1086/170270},
       adsurl = {https://ui.adsabs.harvard.edu/abs/1991ApJ...376..214B}
}

@ARTICLE{Chakrabarti89,
   author = {{Chakrabarti}, S.~K.},
    title = "{Standing Rankine-Hugoniot shocks in the hybrid model flows of the black hole accretion and winds}",
  journal = {\apj},
     year = 1989,
    month = dec,
   volume = 347,
    pages = {365-372},
      doi = {10.1086/168125},
   adsurl = {http://adsabs.harvard.edu/abs/1989ApJ...347..365C}
}

@ARTICLE{Fukue-87,
   author = {{Fukue}, J.},
    title = "{Transonic disk accretion revisited}",
  journal = {\pasj},
     year = 1987,
   volume = 39,
    pages = {309-327},
   adsurl = {http://adsabs.harvard.edu/abs/1987PASJ...39..309F}
}

@ARTICLE{Lu-etal99,
   author = {{Lu}, J.-F. and {Gu}, W.-M. and {Yuan}, F.},
    title = "{Global Dynamics of Advection-dominated Accretion Revisited}",
  journal = {\apj},
   eprint = {astro-ph/9905099},
     year = 1999,
    month = sep,
   volume = 523,
    pages = {340-349},
      doi = {10.1086/307725},
   adsurl = {http://adsabs.harvard.edu/abs/1999ApJ...523..340L}
}

@ARTICLE{Fukumura-Tsuruta04,
   author = {{Fukumura}, K. and {Tsuruta}, S.},
    title = "{Isothermal Shock Formation in Nonequatorial Accretion Flows around Kerr Black Holes}",
  journal = {\apj},
   eprint = {astro-ph/0405269},
     year = 2004,
    month = aug,
   volume = 611,
    pages = {964-976},
      doi = {10.1086/422243},
   adsurl = {http://adsabs.harvard.edu/abs/2004ApJ...611..964F}
}

@ARTICLE{Chakrabarti-Das04,
   author = {{Chakrabarti}, S.~K. and {Das}, S.},
    title = "{Properties of accretion shock waves in viscous flows around black holes}",
  journal = {\mnras},
   eprint = {astro-ph/0402561},
     year = 2004,
    month = apr,
   volume = 349,
    pages = {649-664},
      doi = {10.1111/j.1365-2966.2004.07536.x},
   adsurl = {http://adsabs.harvard.edu/abs/2004MNRAS.349..649C}
}

@ARTICLE{Aktar-etal15,
   author = {{Aktar}, R. and {Das}, S. and {Nandi}, A.},
    title = "{Mass-loss from advective accretion disc around rotating black holes}",
  journal = {\mnras},
archivePrefix = "arXiv",
   eprint = {1508.02571},
 primaryClass = "astro-ph.HE",
     year = 2015,
    month = nov,
   volume = 453,
    pages = {3414-3427},
      doi = {10.1093/mnras/stv1874},
   adsurl = {http://adsabs.harvard.edu/abs/2015MNRAS.453.3414A}
}

@ARTICLE{Aktar-etal17,
   author = {{Aktar}, R. and {Das}, S. and {Nandi}, A. and {Sreehari}, H.
	},
    title = "{Estimation of mass outflow rates from dissipative accretion disc around rotating black holes}",
  journal = {\mnras},
archivePrefix = "arXiv",
   eprint = {1707.07511},
 primaryClass = "astro-ph.HE",
     year = 2017,
    month = nov,
   volume = 471,
    pages = {4806-4819},
      doi = {10.1093/mnras/stx1893},
   adsurl = {http://adsabs.harvard.edu/abs/2017MNRAS.471.4806A}
}

@ARTICLE{Sarkar-Das16,
   author = {{Sarkar}, B. and {Das}, S.},
    title = "{Dynamical structure of magnetized dissipative accretion flow around black holes}",
  journal = {\mnras},
archivePrefix = "arXiv",
   eprint = {1606.00526},
 primaryClass = "astro-ph.HE",
     year = 2016,
    month = sep,
   volume = 461,
    pages = {190-201},
      doi = {10.1093/mnras/stw1327},
   adsurl = {http://adsabs.harvard.edu/abs/2016MNRAS.461..190S}
}

@ARTICLE{Dihingia-etal18,
   author = {{Dihingia}, I.~K. and {Das}, S. and {Mandal}, S.},
    title = "{Properties of two-temperature dissipative accretion flow around black holes}",
  journal = {\mnras},
archivePrefix = "arXiv",
   eprint = {1712.05534},
 primaryClass = "astro-ph.HE",
     year = 2018,
    month = apr,
   volume = 475,
    pages = {2164-2177},
      doi = {10.1093/mnras/stx3269},
   adsurl = {http://adsabs.harvard.edu/abs/2018MNRAS.475.2164D}
}

@ARTICLE{Chakrabarti-Titarchuk95,
   author = {{Chakrabarti}, S. and {Titarchuk}, L.~G.},
    title = "{Spectral Properties of Accretion Disks around Galactic and Extragalactic Black Holes}",
  journal = {\apj},
   eprint = {astro-ph/9510005},
     year = 1995,
    month = dec,
   volume = 455,
    pages = {623},
      doi = {10.1086/176610},
   adsurl = {http://adsabs.harvard.edu/abs/1995ApJ...455..623C}
}

@ARTICLE{Das-etal14,
   author = {{Das}, S. and {Chattopadhyay}, I. and {Nandi}, A. and {Molteni}, D.
	},
    title = "{Periodic mass loss from viscous accretion flows around black holes}",
  journal = {\mnras},
archivePrefix = "arXiv",
   eprint = {1405.4415},
 primaryClass = "astro-ph.HE",
     year = 2014,
    month = jul,
   volume = 442,
    pages = {251-258},
      doi = {10.1093/mnras/stu864},
   adsurl = {http://adsabs.harvard.edu/abs/2014MNRAS.442..251D}
}

@ARTICLE{Okuda-Das15,
   author = {{Okuda}, T. and {Das}, S.},
    title = "{Unstable mass-outflows in geometrically thick accretion flows around black holes}",
  journal = {\mnras},
archivePrefix = "arXiv",
   eprint = {1507.04326},
 primaryClass = "astro-ph.HE",
     year = 2015,
    month = oct,
   volume = 453,
    pages = {147-156},
      doi = {10.1093/mnras/stv1626},
   adsurl = {http://adsabs.harvard.edu/abs/2015MNRAS.453..147O}
}

@ARTICLE{Aktar-etal18,
   author = {{Aktar}, R. and {Das}, S. and {Nandi}, A. and {Sreehari}, H.},
    title = "{Advective accretion flow properties around rotating black holes - application to GRO J1655-40}",
  journal = {Journal of Astrophysics and Astronomy},
archivePrefix = "arXiv",
   eprint = {1801.04116},
 primaryClass = "astro-ph.HE",
     year = 2018,
    month = feb,
   volume = 39,
      eid = {17},
    pages = {17},
      doi = {10.1007/s12036-017-9507-0},
   adsurl = {http://adsabs.harvard.edu/abs/2018JApA...39...17A}
}

@ARTICLE{Aktar-etal19,
       author = {{Aktar}, Ramiz and {Nandi}, Anuj and {Das}, Santabrata},
        title = "{Accretion-ejection in rotating black holes: a model for `outliers' track of radio-X-ray correlation in X-ray binaries}",
      journal = {\apss},
         year = 2019,
        month = feb,
       volume = {364},
       number = {2},
          eid = {22},
        pages = {22},
          doi = {10.1007/s10509-019-3509-0},
archivePrefix = {arXiv},
       eprint = {1901.10091},
 primaryClass = {astro-ph.HE},
       adsurl = {https://ui.adsabs.harvard.edu/abs/2019Ap&SS.364...22A}
}

@ARTICLE{Okuda-etal04,
   author = {{Okuda}, T. and {Teresi}, V. and {Toscano}, E. and {Molteni}, D.
	},
    title = "{Radiative Shocks in Rotating Accretion Flows around Black Holes}",
  journal = {\pasj},
   eprint = {astro-ph/0403115},
     year = 2004,
    month = jun,
   volume = 56,
    pages = {547-552},
      doi = {10.1093/pasj/56.3.547},
   adsurl = {http://adsabs.harvard.edu/abs/2004PASJ...56..547O}
}

@ARTICLE{Okuda-etal07,
   author = {{Okuda}, T. and {Teresi}, V. and {Molteni}, D.},
    title = "{Shock oscillation model for quasi-periodic oscillations in stellar mass and supermassive black holes}",
  journal = {\mnras},
   eprint = {astro-ph/0703465},
     year = 2007,
    month = jun,
   volume = 377,
    pages = {1431-1438},
      doi = {10.1111/j.1365-2966.2007.11736.x},
   adsurl = {http://adsabs.harvard.edu/abs/2007MNRAS.377.1431O}
}

@article{Aktar-etal21,
	doi = {10.3847/1538-4357/ac24fd},
	url = {https://doi.org/10.3847/1538-4357/ac24fd},
	year = 2021,
	month = {nov},
	publisher = {American Astronomical Society},
	volume = {922},
	number = {2},
	pages = {120},
	author = {Ramiz Aktar and Li Xue and Tong Liu},
	title = {Point-wise Self-similar Solution for Spiral Shocks in an Accretion Disk with Mass Outflow in a Binary},
	journal = {The Astrophysical Journal}
}

@ARTICLE{Okuda-etal19,
       author = {{Okuda}, Toru and {Singh}, Chandra B. and {Das}, Santabrata and {Aktar}, Ramiz and {Nandi}, Anuj and {Dal Pino}, Elisabete M. de Gouveia},
        title = "{A possible model for the long-term flares of Sgr A*}",
      journal = {\pasj},
         year = 2019,
        month = jun,
       volume = {71},
       number = {3},
          eid = {49},
        pages = {49},
          doi = {10.1093/pasj/psz021},
archivePrefix = {arXiv},
       eprint = {1902.02933},
 primaryClass = {astro-ph.HE},
       adsurl = {https://ui.adsabs.harvard.edu/abs/2019PASJ...71...49O}
}

@ARTICLE{Okuda-etal22,
       author = {{Okuda}, Toru and {Singh}, Chandra B. and {Aktar}, Ramiz},
        title = "{Radiative shock oscillation model for the long-term flares of Sgr A*}",
      journal = {\mnras},
         year = 2022,
        month = aug,
       volume = {514},
       number = {4},
        pages = {5074-5084},
          doi = {10.1093/mnras/stac1630},
archivePrefix = {arXiv},
       eprint = {2206.04919},
 primaryClass = {astro-ph.HE},
       adsurl = {https://ui.adsabs.harvard.edu/abs/2022MNRAS.514.5074O}
}

@ARTICLE{Migone-etal07,
       author = {{Mignone}, A. and {Bodo}, G. and {Massaglia}, S. and {Matsakos}, T. and {Tesileanu}, O. and {Zanni}, C. and {Ferrari}, A.},
        title = "{PLUTO: A Numerical Code for Computational Astrophysics}",
      journal = {\apjs},
         year = 2007,
        month = may,
       volume = {170},
       number = {1},
        pages = {228-242},
          doi = {10.1086/513316},
archivePrefix = {arXiv},
       eprint = {astro-ph/0701854},
 primaryClass = {astro-ph},
       adsurl = {https://ui.adsabs.harvard.edu/abs/2007ApJS..170..228M}
}

@ARTICLE{Machida-etal01,
       author = {{Machida}, Mami and {Matsumoto}, Ryoji and {Mineshige}, Shin},
        title = "{Convection-Dominated, Magnetized Accretion Flows into Black Holes}",
      journal = {\pasj},
         year = 2001,
        month = feb,
       volume = {53},
       number = {1},
        pages = {L1-L4},
          doi = {10.1093/pasj/53.1.L1},
       adsurl = {https://ui.adsabs.harvard.edu/abs/2001PASJ...53L...1M}
}

@ARTICLE{Proga-Begelman-03,
       author = {{Proga}, Daniel and {Begelman}, Mitchell C.},
        title = "{Accretion of Low Angular Momentum Material onto Black Holes: Two-dimensional Hydrodynamical Inviscid Case}",
      journal = {\apj},
         year = 2003,
        month = jan,
       volume = {582},
       number = {1},
        pages = {69-81},
          doi = {10.1086/344537},
archivePrefix = {arXiv},
       eprint = {astro-ph/0208517},
 primaryClass = {astro-ph},
       adsurl = {https://ui.adsabs.harvard.edu/abs/2003ApJ...582...69P}
}

@ARTICLE{Okuda-etal23,
       author = {{Okuda}, Toru and {Singh}, Chandra B. and {Aktar}, Ramiz},
        title = "{Time delays between radio and X-ray and between narrow radio bands of Sgr A* flares in the shock oscillation model}",
      journal = {\mnras},
         year = 2023,
        month = jun,
       volume = {522},
       number = {2},
        pages = {1814-1825},
          doi = {10.1093/mnras/stad1096},
archivePrefix = {arXiv},
       eprint = {2304.03925},
 primaryClass = {astro-ph.HE},
       adsurl = {https://ui.adsabs.harvard.edu/abs/2023MNRAS.522.1814O}
}

@ARTICLE{Yuan-Narayan14,
       author = {{Yuan}, Feng and {Narayan}, Ramesh},
        title = "{Hot Accretion Flows Around Black Holes}",
      journal = {\araa},
         year = 2014,
        month = aug,
       volume = {52},
        pages = {529-588},
          doi = {10.1146/annurev-astro-082812-141003},
archivePrefix = {arXiv},
       eprint = {1401.0586},
 primaryClass = {astro-ph.HE},
       adsurl = {https://ui.adsabs.harvard.edu/abs/2014ARA&A..52..529Y}
}

@ARTICLE{Machida-etal00,
       author = {{Machida}, M. and {Hayashi}, Mitsuru R. and {Matsumoto}, R.},
        title = "{Global Simulations of Differentially Rotating Magnetized Disks: Formation of Low-{\ensuremath{\beta}} Filaments and Structured Coronae}",
      journal = {\apjl},
         year = 2000,
        month = mar,
       volume = {532},
       number = {1},
        pages = {L67-L70},
          doi = {10.1086/312553},
archivePrefix = {arXiv},
       eprint = {astro-ph/9911291},
 primaryClass = {astro-ph},
       adsurl = {https://ui.adsabs.harvard.edu/abs/2000ApJ...532L..67M}
}

@ARTICLE{Stone-Pringle01,
       author = {{Stone}, James M. and {Pringle}, James E.},
        title = "{Magnetohydrodynamical non-radiative accretion flows in two dimensions}",
      journal = {\mnras},
         year = 2001,
        month = apr,
       volume = {322},
       number = {3},
        pages = {461-472},
          doi = {10.1046/j.1365-8711.2001.04138.x},
archivePrefix = {arXiv},
       eprint = {astro-ph/0009233},
 primaryClass = {astro-ph},
       adsurl = {https://ui.adsabs.harvard.edu/abs/2001MNRAS.322..461S}
}

@ARTICLE{Narayan-etal12,
       author = {{Narayan}, Ramesh and {S{\"A} dowski}, Aleksander and {Penna}, Robert F. and {Kulkarni}, Akshay K.},
        title = "{GRMHD simulations of magnetized advection-dominated accretion on a non-spinning black hole: role of outflows}",
      journal = {\mnras},
         year = 2012,
        month = nov,
       volume = {426},
       number = {4},
        pages = {3241-3259},
          doi = {10.1111/j.1365-2966.2012.22002.x},
archivePrefix = {arXiv},
       eprint = {1206.1213},
 primaryClass = {astro-ph.HE},
       adsurl = {https://ui.adsabs.harvard.edu/abs/2012MNRAS.426.3241N}
}

@ARTICLE{Yuan-etal-12,
       author = {{Yuan}, Feng and {Bu}, Defu and {Wu}, Maochun},
        title = "{Numerical Simulation of Hot Accretion Flows. II. Nature, Origin, and Properties of Outflows and their Possible Observational Applications}",
      journal = {\apj},
         year = 2012,
        month = dec,
       volume = {761},
       number = {2},
          eid = {130},
        pages = {130},
          doi = {10.1088/0004-637X/761/2/130},
archivePrefix = {arXiv},
       eprint = {1206.4173},
 primaryClass = {astro-ph.HE},
       adsurl = {https://ui.adsabs.harvard.edu/abs/2012ApJ...761..130Y}
}

@ARTICLE{Igumenshchev-etal-03,
       author = {{Igumenshchev}, Igor V. and {Narayan}, Ramesh and {Abramowicz}, Marek A.},
        title = "{Three-dimensional Magnetohydrodynamic Simulations of Radiatively Inefficient Accretion Flows}",
      journal = {\apj},
         year = 2003,
        month = aug,
       volume = {592},
       number = {2},
        pages = {1042-1059},
          doi = {10.1086/375769},
archivePrefix = {arXiv},
       eprint = {astro-ph/0301402},
 primaryClass = {astro-ph},
       adsurl = {https://ui.adsabs.harvard.edu/abs/2003ApJ...592.1042I}
}

@ARTICLE{Kim-etal-19,
       author = {{Kim}, Jinho and {Garain}, Sudip K. and {Chakrabarti}, Sandip K. and {Balsara}, Dinshaw S.},
        title = "{General relativistic numerical simulation of sub-Keplerian transonic accretion flows on to rotating black holes: Kerr space-time}",
      journal = {\mnras},
         year = 2019,
        month = jan,
       volume = {482},
       number = {3},
        pages = {3636-3645},
          doi = {10.1093/mnras/sty2953},
archivePrefix = {arXiv},
       eprint = {1810.12469},
 primaryClass = {astro-ph.HE},
       adsurl = {https://ui.adsabs.harvard.edu/abs/2019MNRAS.482.3636K}
}

@ARTICLE{Narayan-etal-03,
       author = {{Narayan}, Ramesh and {Igumenshchev}, Igor V. and {Abramowicz}, Marek A.},
        title = "{Magnetically Arrested Disk: an Energetically Efficient Accretion Flow}",
      journal = {\pasj},
         year = 2003,
        month = dec,
       volume = {55},
        pages = {L69-L72},
          doi = {10.1093/pasj/55.6.L69},
archivePrefix = {arXiv},
       eprint = {astro-ph/0305029},
 primaryClass = {astro-ph},
       adsurl = {https://ui.adsabs.harvard.edu/abs/2003PASJ...55L..69N}
}

@ARTICLE{Event-Horizon-etal-2022a,
       author = {{Event Horizon Telescope Collaboration} and {Akiyama}, Kazunori and {Alberdi}, Antxon and {Alef}, Walter and {Algaba}, Juan Carlos and {Anantua}, Richard and {Asada}, Keiichi and {Azulay}, Rebecca and {Bach}, Uwe and {Baczko}, Anne-Kathrin and {Ball}, David and {Balokovi{\'c}}, Mislav and {Barrett}, John and {Baub{\"o}ck}, Michi and {Benson}, Bradford A. and {Bintley}, Dan and {Blackburn}, Lindy and {Blundell}, Raymond and {Bouman}, Katherine L. and {Bower}, Geoffrey C. and {Boyce}, Hope and {Bremer}, Michael and {Brinkerink}, Christiaan D. and {Brissenden}, Roger and {Britzen}, Silke and {Broderick}, Avery E. and {Broguiere}, Dominique and {Bronzwaer}, Thomas and {Bustamante}, Sandra and {Byun}, Do-Young and {Carlstrom}, John E. and {Ceccobello}, Chiara and {Chael}, Andrew and {Chan}, Chi-kwan and {Chatterjee}, Koushik and {Chatterjee}, Shami and {Chen}, Ming-Tang and {Chen}, Yongjun and {Cheng}, Xiaopeng and {Cho}, Ilje and {Christian}, Pierre and {Conroy}, Nicholas S. and {Conway}, John E. and {Cordes}, James M. and {Crawford}, Thomas M. and {Crew}, Geoffrey B. and {Cruz-Osorio}, Alejandro and {Cui}, Yuzhu and {Davelaar}, Jordy and {De Laurentis}, Mariafelicia and {Deane}, Roger and {Dempsey}, Jessica and {Desvignes}, Gregory and {Dexter}, Jason and {Dhruv}, Vedant and {Doeleman}, Sheperd S. and {Dougal}, Sean and {Dzib}, Sergio A. and {Eatough}, Ralph P. and {Emami}, Razieh and {Falcke}, Heino and {Farah}, Joseph and {Fish}, Vincent L. and {Fomalont}, Ed and {Ford}, H. Alyson and {Fraga-Encinas}, Raquel and {Freeman}, William T. and {Friberg}, Per and {Fromm}, Christian M. and {Fuentes}, Antonio and {Galison}, Peter and {Gammie}, Charles F. and {Garc{\'\i}a}, Roberto and {Gentaz}, Olivier and {Georgiev}, Boris and {Goddi}, Ciriaco and {Gold}, Roman and {G{\'o}mez-Ruiz}, Arturo I. and {G{\'o}mez}, Jos{\'e} L. and {Gu}, Minfeng and {Gurwell}, Mark and {Hada}, Kazuhiro and {Haggard}, Daryl and {Haworth}, Kari and {Hecht}, Michael H. and {Hesper}, Ronald and {Heumann}, Dirk and {Ho}, Luis C. and {Ho}, Paul and {Honma}, Mareki and {Huang}, Chih-Wei L. and {Huang}, Lei and {Hughes}, David H. and {Ikeda}, Shiro and {Impellizzeri}, C.~M. Violette and {Inoue}, Makoto and {Issaoun}, Sara and {James}, David J. and {Jannuzi}, Buell T. and {Janssen}, Michael and {Jeter}, Britton and {Jiang}, Wu and {Jim{\'e}nez-Rosales}, Alejandra and {Johnson}, Michael D. and {Jorstad}, Svetlana and {Joshi}, Abhishek V. and {Jung}, Taehyun and {Karami}, Mansour and {Karuppusamy}, Ramesh and {Kawashima}, Tomohisa and {Keating}, Garrett K. and {Kettenis}, Mark and {Kim}, Dong-Jin and {Kim}, Jae-Young and {Kim}, Jongsoo and {Kim}, Junhan and {Kino}, Motoki and {Koay}, Jun Yi and {Kocherlakota}, Prashant and {Kofuji}, Yutaro and {Koch}, Patrick M. and {Koyama}, Shoko and {Kramer}, Carsten and {Kramer}, Michael and {Krichbaum}, Thomas P. and {Kuo}, Cheng-Yu and {La Bella}, Noemi and {Lauer}, Tod R. and {Lee}, Daeyoung and {Lee}, Sang-Sung and {Leung}, Po Kin and {Levis}, Aviad and {Li}, Zhiyuan and {Lico}, Rocco and {Lindahl}, Greg and {Lindqvist}, Michael and {Lisakov}, Mikhail and {Liu}, Jun and {Liu}, Kuo and {Liuzzo}, Elisabetta and {Lo}, Wen-Ping and {Lobanov}, Andrei P. and {Loinard}, Laurent and {Lonsdale}, Colin J. and {Lu}, Ru-Sen and {Mao}, Jirong and {Marchili}, Nicola and {Markoff}, Sera and {Marrone}, Daniel P. and {Marscher}, Alan P. and {Mart{\'\i}-Vidal}, Iv{\'a}n and {Matsushita}, Satoki and {Matthews}, Lynn D. and {Medeiros}, Lia and {Menten}, Karl M. and {Michalik}, Daniel and {Mizuno}, Izumi and {Mizuno}, Yosuke and {Moran}, James M. and {Moriyama}, Kotaro and {Moscibrodzka}, Monika and {M{\"u}ller}, Cornelia and {Mus}, Alejandro and {Musoke}, Gibwa and {Myserlis}, Ioannis and {Nadolski}, Andrew and {Nagai}, Hiroshi and {Nagar}, Neil M. and {Nakamura}, Masanori and {Narayan}, Ramesh and {Narayanan}, Gopal and {Natarajan}, Iniyan and {Nathanail}, Antonios and {Fuentes}, Santiago Navarro and {Neilsen}, Joey and {Neri}, Roberto and {Ni}, Chunchong and {Noutsos}, Aristeidis and {Nowak}, Michael A. and {Oh}, Junghwan and {Okino}, Hiroki and {Olivares}, H{\'e}ctor and {Ortiz-Le{\'o}n}, Gisela N. and {Oyama}, Tomoaki and {{\"O}zel}, Feryal and {Palumbo}, Daniel C.~M. and {Paraschos}, Georgios Filippos and {Park}, Jongho and {Parsons}, Harriet and {Patel}, Nimesh and {Pen}, Ue-Li and {Pesce}, Dominic W. and {Pi{\'e}tu}, Vincent and {Plambeck}, Richard and {PopStefanija}, Aleksandar and {Porth}, Oliver and {P{\"o}tzl}, Felix M. and {Prather}, Ben and {Preciado-L{\'o}pez}, Jorge A. and {Psaltis}, Dimitrios and {Pu}, Hung-Yi and {Ramakrishnan}, Venkatessh and {Rao}, Ramprasad and {Rawlings}, Mark G. and {Raymond}, Alexander W. and {Rezzolla}, Luciano and {Ricarte}, Angelo and {Ripperda}, Bart and {Roelofs}, Freek and {Rogers}, Alan and {Ros}, Eduardo and {Romero-Ca{\~n}izales}, Cristina and {Roshanineshat}, Arash and {Rottmann}, Helge and {Roy}, Alan L. and {Ruiz}, Ignacio and {Ruszczyk}, Chet and {Rygl}, Kazi L.~J. and {S{\'a}nchez}, Salvador and {S{\'a}nchez-Arg{\"u}elles}, David and {S{\'a}nchez-Portal}, Miguel and {Sasada}, Mahito and {Satapathy}, Kaushik and {Savolainen}, Tuomas and {Schloerb}, F. Peter and {Schonfeld}, Jonathan and {Schuster}, Karl-Friedrich and {Shao}, Lijing and {Shen}, Zhiqiang and {Small}, Des and {Sohn}, Bong Won and {SooHoo}, Jason and {Souccar}, Kamal and {Sun}, He and {Tazaki}, Fumie and {Tetarenko}, Alexandra J. and {Tiede}, Paul and {Tilanus}, Remo P.~J. and {Titus}, Michael and {Torne}, Pablo and {Traianou}, Efthalia and {Trent}, Tyler and {Trippe}, Sascha and {Turk}, Matthew and {van Bemmel}, Ilse and {van Langevelde}, Huib Jan and {van Rossum}, Daniel R. and {Vos}, Jesse and {Wagner}, Jan and {Ward-Thompson}, Derek and {Wardle}, John and {Weintroub}, Jonathan and {Wex}, Norbert and {Wharton}, Robert and {Wielgus}, Maciek and {Wiik}, Kaj and {Witzel}, Gunther and {Wondrak}, Michael F. and {Wong}, George N. and {Wu}, Qingwen and {Yamaguchi}, Paul and {Yoon}, Doosoo and {Young}, Andr{\'e} and {Young}, Ken and {Younsi}, Ziri and {Yuan}, Feng and {Yuan}, Ye-Fei and {Zensus}, J. Anton and {Zhang}, Shuo and {Zhao}, Guang-Yao and {Zhao}, Shan-Shan and {Agurto}, Claudio and {Allardi}, Alexander and {Amestica}, Rodrigo and {Araneda}, Juan Pablo and {Arriagada}, Oriel and {Berghuis}, Jennie L. and {Bertarini}, Alessandra and {Berthold}, Ryan and {Blanchard}, Jay and {Brown}, Ken and {C{\'a}rdenas}, Mauricio and {Cantzler}, Michael and {Caro}, Patricio and {Castillo-Dom{\'\i}nguez}, Edgar and {Chan}, Tin Lok and {Chang}, Chih-Cheng and {Chang}, Dominic O. and {Chang}, Shu-Hao and {Chang}, Song-Chu and {Chen}, Chung-Chen and {Chilson}, Ryan and {Chuter}, Tim C. and {Ciechanowicz}, Miroslaw and {Colin-Beltran}, Edgar and {Coulson}, Iain M. and {Crowley}, Joseph and {Degenaar}, Nathalie and {Dornbusch}, Sven and {Dur{\'a}n}, Carlos A. and {Everett}, Wendeline B. and {Faber}, Aaron and {Forster}, Karl and {Fuchs}, Miriam M. and {Gale}, David M. and {Geertsema}, Gertie and {Gonz{\'a}lez}, Edouard and {Graham}, Dave and {Gueth}, Fr{\'e}d{\'e}ric and {Halverson}, Nils W. and {Han}, Chih-Chiang and {Han}, Kuo-Chang and {Hasegawa}, Yutaka and {Hern{\'a}ndez-Rebollar}, Jos{\'e} Luis and {Herrera}, Cristian and {Herrero-Illana}, Ruben and {Heyminck}, Stefan and {Hirota}, Akihiko and {Hoge}, James and {Hostler Schimpf}, Shelbi R. and {Howie}, Ryan E. and {Huang}, Yau-De and {Jiang}, Homin and {Jinchi}, Hao and {John}, David and {Kimura}, Kimihiro and {Klein}, Thomas and {Kubo}, Derek and {Kuroda}, John and {Kwon}, Caleb and {Lacasse}, Richard and {Laing}, Robert and {Leitch}, Erik M. and {Li}, Chao-Te and {Liu}, Ching-Tang and {Liu}, Kuan-Yu and {Lin}, Lupin C. -C. and {Lu}, Li-Ming and {Mac-Auliffe}, Felipe and {Martin-Cocher}, Pierre and {Matulonis}, Callie and {Maute}, John K. and {Messias}, Hugo and {Meyer-Zhao}, Zheng and {Monta{\~n}a}, Alfredo and {Montenegro-Montes}, Francisco and {Montgomerie}, William and {Moreno Nolasco}, Marcos Emir and {Muders}, Dirk and {Nishioka}, Hiroaki and {Norton}, Timothy J. and {Nystrom}, George and {Ogawa}, Hideo and {Olivares}, Rodrigo and {Oshiro}, Peter and {P{\'e}rez-Beaupuits}, Juan Pablo and {Parra}, Rodrigo and {Phillips}, Neil M. and {Poirier}, Michael and {Pradel}, Nicolas and {Qiu}, Richard and {Raffin}, Philippe A. and {Rahlin}, Alexandra S. and {Ram{\'\i}rez}, Jorge and {Ressler}, Sean and {Reynolds}, Mark and {Rodr{\'\i}guez-Montoya}, Iv{\'a}n and {Saez-Madain}, Alejandro F. and {Santana}, Jorge and {Shaw}, Paul and {Shirkey}, Leslie E. and {Silva}, Kevin M. and {Snow}, William and {Sousa}, Don and {Sridharan}, T.~K. and {Stahm}, William and {Stark}, Anthony A. and {Test}, John and {Torstensson}, Karl and {Venegas}, Paulina and {Walther}, Craig and {Wei}, Ta-Shun and {White}, Chris and {Wieching}, Gundolf and {Wijnands}, Rudy and {Wouterloot}, Jan G.~A. and {Yu}, Chen-Yu and {Yu (于威)}, Wei and {Zeballos}, Milagros},
        title = "{First Sagittarius A* Event Horizon Telescope Results. I. The Shadow of the Supermassive Black Hole in the Center of the Milky Way}",
      journal = {\apjl},
         year = 2022,
        month = may,
       volume = {930},
       number = {2},
          eid = {L12},
        pages = {L12},
          doi = {10.3847/2041-8213/ac6674},
       adsurl = {https://ui.adsabs.harvard.edu/abs/2022ApJ...930L..12E}
}

@ARTICLE{Event-Horizon-etal-2022b,
       author = {{Event Horizon Telescope Collaboration} and {Akiyama}, Kazunori and {Alberdi}, Antxon and {Alef}, Walter and {Algaba}, Juan Carlos and {Anantua}, Richard and {Asada}, Keiichi and {Azulay}, Rebecca and {Bach}, Uwe and {Baczko}, Anne-Kathrin and {Ball}, David and {Balokovi{\'c}}, Mislav and {Barrett}, John and {Baub{\"o}ck}, Michi and {Benson}, Bradford A. and {Bintley}, Dan and {Blackburn}, Lindy and {Blundell}, Raymond and {Bouman}, Katherine L. and {Bower}, Geoffrey C. and {Boyce}, Hope and {Bremer}, Michael and {Brinkerink}, Christiaan D. and {Brissenden}, Roger and {Britzen}, Silke and {Broderick}, Avery E. and {Broguiere}, Dominique and {Bronzwaer}, Thomas and {Bustamante}, Sandra and {Byun}, Do-Young and {Carlstrom}, John E. and {Ceccobello}, Chiara and {Chael}, Andrew and {Chan}, Chi-kwan and {Chatterjee}, Koushik and {Chatterjee}, Shami and {Chen}, Ming-Tang and {Chen}, Yongjun and {Cheng}, Xiaopeng and {Cho}, Ilje and {Christian}, Pierre and {Conroy}, Nicholas S. and {Conway}, John E. and {Cordes}, James M. and {Crawford}, Thomas M. and {Crew}, Geoffrey B. and {Cruz-Osorio}, Alejandro and {Cui}, Yuzhu and {Davelaar}, Jordy and {De Laurentis}, Mariafelicia and {Deane}, Roger and {Dempsey}, Jessica and {Desvignes}, Gregory and {Dexter}, Jason and {Dhruv}, Vedant and {Doeleman}, Sheperd S. and {Dougal}, Sean and {Dzib}, Sergio A. and {Eatough}, Ralph P. and {Emami}, Razieh and {Falcke}, Heino and {Farah}, Joseph and {Fish}, Vincent L. and {Fomalont}, Ed and {Ford}, H. Alyson and {Fraga-Encinas}, Raquel and {Freeman}, William T. and {Friberg}, Per and {Fromm}, Christian M. and {Fuentes}, Antonio and {Galison}, Peter and {Gammie}, Charles F. and {Garc{\'\i}a}, Roberto and {Gentaz}, Olivier and {Georgiev}, Boris and {Goddi}, Ciriaco and {Gold}, Roman and {G{\'o}mez-Ruiz}, Arturo I. and {G{\'o}mez}, Jos{\'e} L. and {Gu}, Minfeng and {Gurwell}, Mark and {Hada}, Kazuhiro and {Haggard}, Daryl and {Haworth}, Kari and {Hecht}, Michael H. and {Hesper}, Ronald and {Heumann}, Dirk and {Ho}, Luis C. and {Ho}, Paul and {Honma}, Mareki and {Huang}, Chih-Wei L. and {Huang}, Lei and {Hughes}, David H. and {Ikeda}, Shiro and {Impellizzeri}, C.~M. Violette and {Inoue}, Makoto and {Issaoun}, Sara and {James}, David J. and {Jannuzi}, Buell T. and {Janssen}, Michael and {Jeter}, Britton and {Jiang}, Wu and {Jim{\'e}nez-Rosales}, Alejandra and {Johnson}, Michael D. and {Jorstad}, Svetlana and {Joshi}, Abhishek V. and {Jung}, Taehyun and {Karami}, Mansour and {Karuppusamy}, Ramesh and {Kawashima}, Tomohisa and {Keating}, Garrett K. and {Kettenis}, Mark and {Kim}, Dong-Jin and {Kim}, Jae-Young and {Kim}, Jongsoo and {Kim}, Junhan and {Kino}, Motoki and {Koay}, Jun Yi and {Kocherlakota}, Prashant and {Kofuji}, Yutaro and {Koch}, Patrick M. and {Koyama}, Shoko and {Kramer}, Carsten and {Kramer}, Michael and {Krichbaum}, Thomas P. and {Kuo}, Cheng-Yu and {La Bella}, Noemi and {Lauer}, Tod R. and {Lee}, Daeyoung and {Lee}, Sang-Sung and {Leung}, Po Kin and {Levis}, Aviad and {Li}, Zhiyuan and {Lico}, Rocco and {Lindahl}, Greg and {Lindqvist}, Michael and {Lisakov}, Mikhail and {Liu}, Jun and {Liu}, Kuo and {Liuzzo}, Elisabetta and {Lo}, Wen-Ping and {Lobanov}, Andrei P. and {Loinard}, Laurent and {Lonsdale}, Colin J. and {Lu}, Ru-Sen and {Mao}, Jirong and {Marchili}, Nicola and {Markoff}, Sera and {Marrone}, Daniel P. and {Marscher}, Alan P. and {Mart{\'\i}-Vidal}, Iv{\'a}n and {Matsushita}, Satoki and {Matthews}, Lynn D. and {Medeiros}, Lia and {Menten}, Karl M. and {Michalik}, Daniel and {Mizuno}, Izumi and {Mizuno}, Yosuke and {Moran}, James M. and {Moriyama}, Kotaro and {Moscibrodzka}, Monika and {M{\"u}ller}, Cornelia and {Mus}, Alejandro and {Musoke}, Gibwa and {Myserlis}, Ioannis and {Nadolski}, Andrew and {Nagai}, Hiroshi and {Nagar}, Neil M. and {Nakamura}, Masanori and {Narayan}, Ramesh and {Narayanan}, Gopal and {Natarajan}, Iniyan and {Nathanail}, Antonios and {Fuentes}, Santiago Navarro and {Neilsen}, Joey and {Neri}, Roberto and {Ni}, Chunchong and {Noutsos}, Aristeidis and {Nowak}, Michael A. and {Oh}, Junghwan and {Okino}, Hiroki and {Olivares}, H{\'e}ctor and {Ortiz-Le{\'o}n}, Gisela N. and {Oyama}, Tomoaki and {{\"O}zel}, Feryal and {Palumbo}, Daniel C.~M. and {Paraschos}, Georgios Filippos and {Park}, Jongho and {Parsons}, Harriet and {Patel}, Nimesh and {Pen}, Ue-Li and {Pesce}, Dominic W. and {Pi{\'e}tu}, Vincent and {Plambeck}, Richard and {PopStefanija}, Aleksandar and {Porth}, Oliver and {P{\"o}tzl}, Felix M. and {Prather}, Ben and {Preciado-L{\'o}pez}, Jorge A. and {Psaltis}, Dimitrios and {Pu}, Hung-Yi and {Ramakrishnan}, Venkatessh and {Rao}, Ramprasad and {Rawlings}, Mark G. and {Raymond}, Alexander W. and {Rezzolla}, Luciano and {Ricarte}, Angelo and {Ripperda}, Bart and {Roelofs}, Freek and {Rogers}, Alan and {Ros}, Eduardo and {Romero-Ca{\~n}izales}, Cristina and {Roshanineshat}, Arash and {Rottmann}, Helge and {Roy}, Alan L. and {Ruiz}, Ignacio and {Ruszczyk}, Chet and {Rygl}, Kazi L.~J. and {S{\'a}nchez}, Salvador and {S{\'a}nchez-Arg{\"u}elles}, David and {S{\'a}nchez-Portal}, Miguel and {Sasada}, Mahito and {Satapathy}, Kaushik and {Savolainen}, Tuomas and {Schloerb}, F. Peter and {Schonfeld}, Jonathan and {Schuster}, Karl-Friedrich and {Shao}, Lijing and {Shen}, Zhiqiang and {Small}, Des and {Sohn}, Bong Won and {SooHoo}, Jason and {Souccar}, Kamal and {Sun}, He and {Tazaki}, Fumie and {Tetarenko}, Alexandra J. and {Tiede}, Paul and {Tilanus}, Remo P.~J. and {Titus}, Michael and {Torne}, Pablo and {Traianou}, Efthalia and {Trent}, Tyler and {Trippe}, Sascha and {Turk}, Matthew and {van Bemmel}, Ilse and {van Langevelde}, Huib Jan and {van Rossum}, Daniel R. and {Vos}, Jesse and {Wagner}, Jan and {Ward-Thompson}, Derek and {Wardle}, John and {Weintroub}, Jonathan and {Wex}, Norbert and {Wharton}, Robert and {Wielgus}, Maciek and {Wiik}, Kaj and {Witzel}, Gunther and {Wondrak}, Michael F. and {Wong}, George N. and {Wu}, Qingwen and {Yamaguchi}, Paul and {Yoon}, Doosoo and {Young}, Andr{\'e} and {Young}, Ken and {Younsi}, Ziri and {Yuan}, Feng and {Yuan}, Ye-Fei and {Zensus}, J. Anton and {Zhang}, Shuo and {Zhao}, Guang-Yao and {Zhao}, Shan-Shan and {Agurto}, Claudio and {Araneda}, Juan Pablo and {Arriagada}, Oriel and {Bertarini}, Alessandra and {Berthold}, Ryan and {Blanchard}, Jay and {Brown}, Ken and {C{\'a}rdenas}, Mauricio and {Cantzler}, Michael and {Caro}, Patricio and {Chuter}, Tim C. and {Ciechanowicz}, Miroslaw and {Coulson}, Iain M. and {Crowley}, Joseph and {Degenaar}, Nathalie and {Dornbusch}, Sven and {Dur{\'a}n}, Carlos A. and {Forster}, Karl and {Geertsema}, Gertie and {Gonz{\'a}lez}, Edouard and {Graham}, Dave and {Gueth}, Fr{\'e}d{\'e}ric and {Han}, Chih-Chiang and {Herrera}, Cristian and {Herrero-Illana}, Ruben and {Heyminck}, Stefan and {Hoge}, James and {Huang}, Yau-De and {Jiang}, Homin and {John}, David and {Klein}, Thomas and {Kubo}, Derek and {Kuroda}, John and {Kwon}, Caleb and {Laing}, Robert and {Liu}, Ching-Tang and {Liu}, Kuan-Yu and {Mac-Auliffe}, Felipe and {Martin-Cocher}, Pierre and {Matulonis}, Callie and {Messias}, Hugo and {Meyer-Zhao}, Zheng and {Montenegro-Montes}, Francisco and {Montgomerie}, William and {Muders}, Dirk and {Nishioka}, Hiroaki and {Norton}, Timothy J. and {Olivares}, Rodrigo and {P{\'e}rez-Beaupuits}, Juan Pablo and {Parra}, Rodrigo and {Poirier}, Michael and {Pradel}, Nicolas and {Raffin}, Philippe A. and {Ram{\'\i}rez}, Jorge and {Reynolds}, Mark and {Saez-Madain}, Alejandro F. and {Santana}, Jorge and {Silva}, Kevin M. and {Sousa}, Don and {Stahm}, William and {Torstensson}, Karl and {Venegas}, Paulina and {Walther}, Craig and {Wieching}, Gundolf and {Wijnands}, Rudy and {Wouterloot}, Jan G.~A.},
        title = "{First Sagittarius A* Event Horizon Telescope Results. II. EHT and Multiwavelength Observations, Data Processing, and Calibration}",
      journal = {\apjl},
         year = 2022,
        month = may,
       volume = {930},
       number = {2},
          eid = {L13},
        pages = {L13},
          doi = {10.3847/2041-8213/ac6675},
       adsurl = {https://ui.adsabs.harvard.edu/abs/2022ApJ...930L..13E}
}

@ARTICLE{Fuksman-Mignone-19,
       author = {{Melon Fuksman}, Julio David and {Mignone}, Andrea},
        title = "{A Radiative Transfer Module for Relativistic Magnetohydrodynamics in the PLUTO Code}",
      journal = {\apjs},
         year = 2019,
        month = jun,
       volume = {242},
       number = {2},
          eid = {20},
        pages = {20},
          doi = {10.3847/1538-4365/ab18ff},
archivePrefix = {arXiv},
       eprint = {1903.10456},
 primaryClass = {astro-ph.IM},
       adsurl = {https://ui.adsabs.harvard.edu/abs/2019ApJS..242...20M}
}

@ARTICLE{Aktar-etal24a,
       author = {{Aktar}, Ramiz and {Pan}, Kuo-Chuan and {Okuda}, Toru},
        title = "{Evolution of MHD torus and mass outflow around spinning AGNs}",
      journal = {\mnras},
         year = 2024,
        month = jan,
       volume = {527},
       number = {2},
        pages = {1745-1759},
          doi = {10.1093/mnras/stad3287},
archivePrefix = {arXiv},
       eprint = {2310.15501},
 primaryClass = {astro-ph.HE},
       adsurl = {https://ui.adsabs.harvard.edu/abs/2024MNRAS.527.1745A}
}

@ARTICLE{Stepney-Guilbert-83,
       author = {{Stepney}, S. and {Guilbert}, P.~W.},
        title = "{Numerical fits to important rates in high temperature astrophysical plasmas.}",
      journal = {\mnras},
         year = 1983,
        month = sep,
       volume = {204},
        pages = {1269-1277},
          doi = {10.1093/mnras/204.4.1269},
       adsurl = {https://ui.adsabs.harvard.edu/abs/1983MNRAS.204.1269S}
}

@ARTICLE{Esin-etal96,
       author = {{Esin}, Ann A. and {Narayan}, Ramesh and {Ostriker}, Eve and {Yi}, Insu},
        title = "{Hot One-Temperature Accretion Flows around Black Holes}",
      journal = {\apj},
         year = 1996,
        month = jul,
       volume = {465},
        pages = {312},
          doi = {10.1086/177421},
archivePrefix = {arXiv},
       eprint = {astro-ph/9601074},
 primaryClass = {astro-ph},
       adsurl = {https://ui.adsabs.harvard.edu/abs/1996ApJ...465..312E}
}

@ARTICLE{Narayan-Yi95,
       author = {{Narayan}, Ramesh and {Yi}, Insu},
        title = "{Advection-dominated Accretion: Underfed Black Holes and Neutron Stars}",
      journal = {\apj},
         year = 1995,
        month = oct,
       volume = {452},
        pages = {710},
          doi = {10.1086/176343},
archivePrefix = {arXiv},
       eprint = {astro-ph/9411059},
 primaryClass = {astro-ph},
       adsurl = {https://ui.adsabs.harvard.edu/abs/1995ApJ...452..710N}
}

@ARTICLE{Manmoto-etal-97,
       author = {{Manmoto}, T. and {Mineshige}, S. and {Kusunose}, M.},
        title = "{Spectrum of Optically Thin Advection-dominated Accretion Flow around a Black Hole: Application to Sagittarius A*}",
      journal = {\apj},
         year = 1997,
        month = nov,
       volume = {489},
       number = {2},
        pages = {791-803},
          doi = {10.1086/304817},
archivePrefix = {arXiv},
       eprint = {astro-ph/9708234},
 primaryClass = {astro-ph},
       adsurl = {https://ui.adsabs.harvard.edu/abs/1997ApJ...489..791M}
}

@ARTICLE{Garain-Kim-23,
       author = {{Garain}, Sudip K. and {Kim}, Jinho},
        title = "{Three-dimensional simulations of advective, sub-Keplerian accretion flow on to non-rotating black holes}",
      journal = {\mnras},
         year = 2023,
        month = mar,
       volume = {519},
       number = {3},
        pages = {4550-4563},
          doi = {10.1093/mnras/stac3736},
archivePrefix = {arXiv},
       eprint = {2212.08310},
 primaryClass = {astro-ph.HE},
       adsurl = {https://ui.adsabs.harvard.edu/abs/2023MNRAS.519.4550G}
}

@ARTICLE{Aktar-etal24b,
       author = {{Aktar}, Ramiz and {Pan}, Kuo-Chuan and {Okuda}, Toru},
        title = "{Radiation RMHD Accretion Flows around Spinning AGNs: A Comparative Study of MAD and SANE State}",
      journal = {\apj},
         year = 2024,
        month = sep,
       volume = {972},
       number = {1},
          eid = {18},
        pages = {18},
          doi = {10.3847/1538-4357/ad5a8a},
archivePrefix = {arXiv},
       eprint = {2406.10496},
 primaryClass = {astro-ph.HE},
       adsurl = {https://ui.adsabs.harvard.edu/abs/2024ApJ...972...18A}
}

@ARTICLE{Genzel-etal-10,
       author = {{Genzel}, Reinhard and {Eisenhauer}, Frank and {Gillessen}, Stefan},
        title = "{The Galactic Center massive black hole and nuclear star cluster}",
      journal = {Reviews of Modern Physics},
         year = 2010,
        month = oct,
       volume = {82},
       number = {4},
        pages = {3121-3195},
          doi = {10.1103/RevModPhys.82.3121},
archivePrefix = {arXiv},
       eprint = {1006.0064},
 primaryClass = {astro-ph.GA},
       adsurl = {https://ui.adsabs.harvard.edu/abs/2010RvMP...82.3121G}
}

@ARTICLE{Genzel-etal-03,
       author = {{Genzel}, R. and {Sch{\"o}del}, R. and {Ott}, T. and {Eckart}, A. and {Alexander}, T. and {Lacombe}, F. and {Rouan}, D. and {Aschenbach}, B.},
        title = "{Near-infrared flares from accreting gas around the supermassive black hole at the Galactic Centre}",
      journal = {\nat},
         year = 2003,
        month = oct,
       volume = {425},
       number = {6961},
        pages = {934-937},
          doi = {10.1038/nature02065},
archivePrefix = {arXiv},
       eprint = {astro-ph/0310821},
 primaryClass = {astro-ph},
       adsurl = {https://ui.adsabs.harvard.edu/abs/2003Natur.425..934G}
}

@ARTICLE{Narayan-McClintock-08,
       author = {{Narayan}, Ramesh and {McClintock}, Jeffrey E.},
        title = "{Advection-dominated accretion and the black hole event horizon}",
      journal = {\nar},
         year = 2008,
        month = may,
       volume = {51},
       number = {10-12},
        pages = {733-751},
          doi = {10.1016/j.newar.2008.03.002},
archivePrefix = {arXiv},
       eprint = {0803.0322},
 primaryClass = {astro-ph},
       adsurl = {https://ui.adsabs.harvard.edu/abs/2008NewAR..51..733N}
}

@ARTICLE{Yuan-etal-15,
       author = {{Yuan}, Feng and {Gan}, Zhaoming and {Narayan}, Ramesh and {Sadowski}, Aleksander and {Bu}, Defu and {Bai}, Xue-Ning},
        title = "{Numerical Simulation of Hot Accretion Flows. III. Revisiting Wind Properties Using the Trajectory Approach}",
      journal = {\apj},
         year = 2015,
        month = may,
       volume = {804},
       number = {2},
          eid = {101},
        pages = {101},
          doi = {10.1088/0004-637X/804/2/101},
archivePrefix = {arXiv},
       eprint = {1501.01197},
 primaryClass = {astro-ph.HE},
       adsurl = {https://ui.adsabs.harvard.edu/abs/2015ApJ...804..101Y}
}

@ARTICLE{Chan-etal-09,
       author = {{Chan}, Chi-kwan and {Liu}, Siming and {Fryer}, Christopher L. and {Psaltis}, Dimitrios and {{\"O}zel}, Feryal and {Rockefeller}, Gabriel and {Melia}, Fulvio},
        title = "{MHD Simulations of Accretion onto Sgr A*: Quiescent Fluctuations, Outbursts, and Quasiperiodicity}",
      journal = {\apj},
         year = 2009,
        month = aug,
       volume = {701},
       number = {1},
        pages = {521-534},
          doi = {10.1088/0004-637X/701/1/521},
archivePrefix = {arXiv},
       eprint = {astro-ph/0611269},
 primaryClass = {astro-ph},
       adsurl = {https://ui.adsabs.harvard.edu/abs/2009ApJ...701..521C}
}

@ARTICLE{Dexter-etal-09,
       author = {{Dexter}, Jason and {Agol}, Eric and {Fragile}, P. Chris},
        title = "{Millimeter Flares and VLBI Visibilities from Relativistic Simulations of Magnetized Accretion Onto the Galactic Center Black Hole}",
      journal = {\apjl},
         year = 2009,
        month = oct,
       volume = {703},
       number = {2},
        pages = {L142-L146},
          doi = {10.1088/0004-637X/703/2/L142},
archivePrefix = {arXiv},
       eprint = {0909.0267},
 primaryClass = {astro-ph.HE},
       adsurl = {https://ui.adsabs.harvard.edu/abs/2009ApJ...703L.142D}
}

@ARTICLE{Yuan-etal-09,
       author = {{Yuan}, Feng and {Lin}, Jun and {Wu}, Kinwah and {Ho}, Luis C.},
        title = "{A magnetohydrodynamical model for the formation of episodic jets}",
      journal = {\mnras},
         year = 2009,
        month = jun,
       volume = {395},
       number = {4},
        pages = {2183-2188},
          doi = {10.1111/j.1365-2966.2009.14673.x},
archivePrefix = {arXiv},
       eprint = {0811.2893},
 primaryClass = {astro-ph},
       adsurl = {https://ui.adsabs.harvard.edu/abs/2009MNRAS.395.2183Y}
}

@ARTICLE{Dodds-Eden-etal-10,
       author = {{Dodds-Eden}, Katie and {Sharma}, Prateek and {Quataert}, Eliot and {Genzel}, Reinhard and {Gillessen}, Stefan and {Eisenhauer}, Frank and {Porquet}, Delphine},
        title = "{Time-Dependent Models of Flares from Sagittarius A*}",
      journal = {\apj},
         year = 2010,
        month = dec,
       volume = {725},
       number = {1},
        pages = {450-465},
          doi = {10.1088/0004-637X/725/1/450},
archivePrefix = {arXiv},
       eprint = {1005.0389},
 primaryClass = {astro-ph.GA},
       adsurl = {https://ui.adsabs.harvard.edu/abs/2010ApJ...725..450D}
}

@ARTICLE{Ball-etal-16,
       author = {{Ball}, David and {{\"O}zel}, Feryal and {Psaltis}, Dimitrios and {Chan}, Chi-kwan},
        title = "{Particle Acceleration and the Origin of X-Ray Flares in GRMHD Simulations of SGR A}",
      journal = {\apj},
         year = 2016,
        month = jul,
       volume = {826},
       number = {1},
          eid = {77},
        pages = {77},
          doi = {10.3847/0004-637X/826/1/77},
archivePrefix = {arXiv},
       eprint = {1602.05968},
 primaryClass = {astro-ph.HE},
       adsurl = {https://ui.adsabs.harvard.edu/abs/2016ApJ...826...77B}
}

@ARTICLE{Ressler-etal-17,
       author = {{Ressler}, S.~M. and {Tchekhovskoy}, A. and {Quataert}, E. and {Gammie}, C.~F.},
        title = "{The disc-jet symbiosis emerges: modelling the emission of Sagittarius A* with electron thermodynamics}",
      journal = {\mnras},
         year = 2017,
        month = may,
       volume = {467},
       number = {3},
        pages = {3604-3619},
          doi = {10.1093/mnras/stx364},
archivePrefix = {arXiv},
       eprint = {1611.09365},
 primaryClass = {astro-ph.HE},
       adsurl = {https://ui.adsabs.harvard.edu/abs/2017MNRAS.467.3604R}
}

@ARTICLE{Roberts-etal-17,
       author = {{Roberts}, Shawn R. and {Jiang}, Yan-Fei and {Wang}, Q. Daniel and {Ostriker}, Jeremiah P.},
        title = "{Towards self-consistent modelling of the Sgr A* accretion flow: linking theory and observation}",
      journal = {\mnras},
         year = 2017,
        month = apr,
       volume = {466},
       number = {2},
        pages = {1477-1490},
          doi = {10.1093/mnras/stw2995},
archivePrefix = {arXiv},
       eprint = {1611.00118},
 primaryClass = {astro-ph.HE},
       adsurl = {https://ui.adsabs.harvard.edu/abs/2017MNRAS.466.1477R}
}

@ARTICLE{Li-etal-17,
       author = {{Li}, Ya-Ping and {Yuan}, Feng and {Wang}, Q. Daniel},
        title = "{A magnetohydrodynamic model for multiwavelength flares from Sagittarius A$^{{\ensuremath{\star}}}$ (I): model and the near-infrared and X-ray flares}",
      journal = {\mnras},
         year = 2017,
        month = jul,
       volume = {468},
       number = {3},
        pages = {2552-2568},
          doi = {10.1093/mnras/stx655},
archivePrefix = {arXiv},
       eprint = {1611.02904},
 primaryClass = {astro-ph.HE},
       adsurl = {https://ui.adsabs.harvard.edu/abs/2017MNRAS.468.2552L}
}

@ARTICLE{Singh-etal-21,
       author = {{Singh}, Chandra B. and {Okuda}, Toru and {Aktar}, Ramiz},
        title = "{Effects of resistivity on standing shocks in low angular momentum flows around black holes}",
      journal = {Research in Astronomy and Astrophysics},
         year = 2021,
        month = aug,
       volume = {21},
       number = {6},
          eid = {134},
        pages = {134},
          doi = {10.1088/1674-4527/21/6/134},
archivePrefix = {arXiv},
       eprint = {2101.05474},
 primaryClass = {astro-ph.HE},
       adsurl = {https://ui.adsabs.harvard.edu/abs/2021RAA....21..134S}
}

@ARTICLE{Yusef-Zadeh-etal-06a,
       author = {{Yusef-Zadeh}, F. and {Bushouse}, H. and {Dowell}, C.~D. and {Wardle}, M. and {Roberts}, D. and {Heinke}, C. and {Bower}, G.~C. and {Vila-Vilar{\'o}}, B. and {Shapiro}, S. and {Goldwurm}, A. and {B{\'e}langer}, G.},
        title = "{A Multiwavelength Study of Sgr A*: The Role of Near-IR Flares in Production of X-Ray, Soft {\ensuremath{\gamma}}-Ray, and Submillimeter Emission}",
      journal = {\apj},
         year = 2006,
        month = jun,
       volume = {644},
       number = {1},
        pages = {198-213},
          doi = {10.1086/503287},
archivePrefix = {arXiv},
       eprint = {astro-ph/0510787},
 primaryClass = {astro-ph},
       adsurl = {https://ui.adsabs.harvard.edu/abs/2006ApJ...644..198Y}
}

@ARTICLE{Yusef-Zadeh-etal-09,
       author = {{Yusef-Zadeh}, F. and {Bushouse}, H. and {Wardle}, M. and {Heinke}, C. and {Roberts}, D.~A. and {Dowell}, C.~D. and {Brunthaler}, A. and {Reid}, M.~J. and {Martin}, C.~L. and {Marrone}, D.~P. and {Porquet}, D. and {Grosso}, N. and {Dodds-Eden}, K. and {Bower}, G.~C. and {Wiesemeyer}, H. and {Miyazaki}, A. and {Pal}, S. and {Gillessen}, S. and {Goldwurm}, A. and {Trap}, G. and {Maness}, H.},
        title = "{Simultaneous Multi-Wavelength Observations of Sgr A* During 2007 April 1-11}",
      journal = {\apj},
         year = 2009,
        month = nov,
       volume = {706},
       number = {1},
        pages = {348-375},
          doi = {10.1088/0004-637X/706/1/348},
archivePrefix = {arXiv},
       eprint = {0907.3786},
 primaryClass = {astro-ph.GA},
       adsurl = {https://ui.adsabs.harvard.edu/abs/2009ApJ...706..348Y}
}

@ARTICLE{Degenaar-etal-13,
       author = {{Degenaar}, N. and {Miller}, J.~M. and {Kennea}, J. and {Gehrels}, N. and {Reynolds}, M.~T. and {Wijnands}, R.},
        title = "{The X-Ray Flaring Properties of Sgr A* during Six Years of Monitoring with Swift}",
      journal = {\apj},
         year = 2013,
        month = jun,
       volume = {769},
       number = {2},
          eid = {155},
        pages = {155},
          doi = {10.1088/0004-637X/769/2/155},
archivePrefix = {arXiv},
       eprint = {1210.7237},
 primaryClass = {astro-ph.HE},
       adsurl = {https://ui.adsabs.harvard.edu/abs/2013ApJ...769..155D}
}

@ARTICLE{Neilsen-etal-13,
       author = {{Neilsen}, J. and {Nowak}, M.~A. and {Gammie}, C. and {Dexter}, J. and {Markoff}, S. and {Haggard}, D. and {Nayakshin}, S. and {Wang}, Q.~D. and {Grosso}, N. and {Porquet}, D. and {Tomsick}, J.~A. and {Degenaar}, N. and {Fragile}, P.~C. and {Houck}, J.~C. and {Wijnands}, R. and {Miller}, J.~M. and {Baganoff}, F.~K.},
        title = "{A Chandra/HETGS Census of X-Ray Variability from Sgr A* during 2012}",
      journal = {\apj},
         year = 2013,
        month = sep,
       volume = {774},
       number = {1},
          eid = {42},
        pages = {42},
          doi = {10.1088/0004-637X/774/1/42},
archivePrefix = {arXiv},
       eprint = {1307.5843},
 primaryClass = {astro-ph.HE},
       adsurl = {https://ui.adsabs.harvard.edu/abs/2013ApJ...774...42N}
}

@ARTICLE{Neilsen-etal-15,
       author = {{Neilsen}, J. and {Markoff}, S. and {Nowak}, M.~A. and {Dexter}, J. and {Witzel}, G. and {Barri{\`e}re}, N. and {Li}, Y. and {Baganoff}, F.~K. and {Degenaar}, N. and {Fragile}, P.~C. and {Gammie}, C. and {Goldwurm}, A. and {Grosso}, N. and {Haggard}, D.},
        title = "{The X-Ray Flux Distribution of Sagittarius A* as Seen by Chandra}",
      journal = {\apj},
         year = 2015,
        month = feb,
       volume = {799},
       number = {2},
          eid = {199},
        pages = {199},
          doi = {10.1088/0004-637X/799/2/199},
archivePrefix = {arXiv},
       eprint = {1412.3106},
 primaryClass = {astro-ph.HE},
       adsurl = {https://ui.adsabs.harvard.edu/abs/2015ApJ...799..199N}
}

@ARTICLE{Ponti-etal-15,
       author = {{Ponti}, G. and {De Marco}, B. and {Morris}, M.~R. and {Merloni}, A. and {Mu{\~n}oz-Darias}, T. and {Clavel}, M. and {Haggard}, D. and {Zhang}, S. and {Nandra}, K. and {Gillessen}, S. and {Mori}, K. and {Neilsen}, J. and {Rea}, N. and {Degenaar}, N. and {Terrier}, R. and {Goldwurm}, A.},
        title = "{Fifteen years of XMM-Newton and Chandra monitoring of Sgr A$^{★}$: evidence for a recent increase in the bright flaring rate}",
      journal = {\mnras},
         year = 2015,
        month = dec,
       volume = {454},
       number = {2},
        pages = {1525-1544},
          doi = {10.1093/mnras/stv1537},
archivePrefix = {arXiv},
       eprint = {1507.02690},
 primaryClass = {astro-ph.HE},
       adsurl = {https://ui.adsabs.harvard.edu/abs/2015MNRAS.454.1525P}
}

@ARTICLE{Ponti-etal-17,
       author = {{Ponti}, G. and {George}, E. and {Scaringi}, S. and {Zhang}, S. and {Jin}, C. and {Dexter}, J. and {Terrier}, R. and {Clavel}, M. and {Degenaar}, N. and {Eisenhauer}, F. and {Genzel}, R. and {Gillessen}, S. and {Goldwurm}, A. and {Habibi}, M. and {Haggard}, D. and {Hailey}, C. and {Harrison}, F. and {Merloni}, A. and {Mori}, K. and {Nandra}, K. and {Ott}, T. and {Pfuhl}, O. and {Plewa}, P.~M. and {Waisberg}, I.},
        title = "{A powerful flare from Sgr A* confirms the synchrotron nature of the X-ray emission}",
      journal = {\mnras},
         year = 2017,
        month = jun,
       volume = {468},
       number = {2},
        pages = {2447-2468},
          doi = {10.1093/mnras/stx596},
archivePrefix = {arXiv},
       eprint = {1703.03410},
 primaryClass = {astro-ph.HE},
       adsurl = {https://ui.adsabs.harvard.edu/abs/2017MNRAS.468.2447P}
}

@ARTICLE{Capellupo-etal-17,
       author = {{Capellupo}, Daniel M. and {Haggard}, Daryl and {Choux}, Nicolas and {Baganoff}, Fred and {Bower}, Geoffrey C. and {Cotton}, Bill and {Degenaar}, Nathalie and {Dexter}, Jason and {Falcke}, Heino and {Fragile}, P. Chris and {Heinke}, Craig O. and {Law}, Casey J. and {Markoff}, Sera and {Neilsen}, Joey and {Ponti}, Gabriele and {Rea}, Nanda and {Yusef-Zadeh}, Farhad},
        title = "{Simultaneous Monitoring of X-Ray and Radio Variability in Sagittarius A*}",
      journal = {\apj},
         year = 2017,
        month = aug,
       volume = {845},
       number = {1},
          eid = {35},
        pages = {35},
          doi = {10.3847/1538-4357/aa7da6},
archivePrefix = {arXiv},
       eprint = {1707.01937},
 primaryClass = {astro-ph.HE},
       adsurl = {https://ui.adsabs.harvard.edu/abs/2017ApJ...845...35C}
}

@ARTICLE{Witzel-etal-21,
       author = {{Witzel}, G. and {Martinez}, G. and {Willner}, S.~P. and {Becklin}, E.~E. and {Boyce}, H. and {Do}, T. and {Eckart}, A. and {Fazio}, G.~G. and {Ghez}, A. and {Gurwell}, M.~A. and {Haggard}, D. and {Herrero-Illana}, R. and {Hora}, J.~L. and {Li}, Z. and {Liu}, J. and {Marchili}, N. and {Morris}, Mark R. and {Smith}, Howard A. and {Subroweit}, M. and {Zensus}, J.~A.},
        title = "{Rapid Variability of Sgr A* across the Electromagnetic Spectrum}",
      journal = {\apj},
         year = 2021,
        month = aug,
       volume = {917},
       number = {2},
          eid = {73},
        pages = {73},
          doi = {10.3847/1538-4357/ac0891},
archivePrefix = {arXiv},
       eprint = {2011.09582},
 primaryClass = {astro-ph.HE},
       adsurl = {https://ui.adsabs.harvard.edu/abs/2021ApJ...917...73W}
}

@ARTICLE{Michail-etal-21a,
       author = {{Michail}, Joseph M. and {Yusef-Zadeh}, Farhad and {Wardle}, Mark},
        title = "{Detection of a 20-min time lag observed from Sgr A* between 8 and 10 GHz with the VLA}",
      journal = {\mnras},
         year = 2021,
        month = aug,
       volume = {505},
       number = {3},
        pages = {3616-3623},
          doi = {10.1093/mnras/stab1529},
archivePrefix = {arXiv},
       eprint = {2105.11473},
 primaryClass = {astro-ph.HE},
       adsurl = {https://ui.adsabs.harvard.edu/abs/2021MNRAS.505.3616M}
}

@ARTICLE{Michail-etal-21b,
       author = {{Michail}, Joseph M. and {Wardle}, Mark and {Yusef-Zadeh}, Farhad and {Kunneriath}, Devaky},
        title = "{Multiwavelength Observations of Sgr A*. I. 2019 July 18}",
      journal = {\apj},
         year = 2021,
        month = dec,
       volume = {923},
       number = {1},
          eid = {54},
        pages = {54},
          doi = {10.3847/1538-4357/ac2d2c},
archivePrefix = {arXiv},
       eprint = {2107.09681},
 primaryClass = {astro-ph.HE},
       adsurl = {https://ui.adsabs.harvard.edu/abs/2021ApJ...923...54M}
}

@ARTICLE{Boyce-etal-22,
       author = {{Boyce}, H. and {Haggard}, D. and {Witzel}, G. and {von Fellenberg}, S. and {Willner}, S.~P. and {Becklin}, E.~E. and {Do}, T. and {Eckart}, A. and {Fazio}, G.~G. and {Gurwell}, M.~A. and {Hora}, J.~L. and {Markoff}, S. and {Morris}, M.~R. and {Neilsen}, J. and {Nowak}, M. and {Smith}, H.~A. and {Zhang}, S.},
        title = "{Multiwavelength Variability of Sagittarius A* in 2019 July}",
      journal = {\apj},
         year = 2022,
        month = may,
       volume = {931},
       number = {1},
          eid = {7},
        pages = {7},
          doi = {10.3847/1538-4357/ac6104},
archivePrefix = {arXiv},
       eprint = {2203.13311},
 primaryClass = {astro-ph.HE},
       adsurl = {https://ui.adsabs.harvard.edu/abs/2022ApJ...931....7B}
}

@ARTICLE{Ryu-etal-97,
       author = {{Ryu}, Dongsu and {Chakrabarti}, Sandip K. and {Molteni}, Diego},
        title = "{Zero-Energy Rotating Accretion Flows near a Black Hole}",
      journal = {\apj},
         year = 1997,
        month = jan,
       volume = {474},
       number = {1},
        pages = {378-388},
          doi = {10.1086/303461},
archivePrefix = {arXiv},
       eprint = {astro-ph/9607051},
 primaryClass = {astro-ph},
       adsurl = {https://ui.adsabs.harvard.edu/abs/1997ApJ...474..378R}
}

@ARTICLE{Molteni-etal-94,
       author = {{Molteni}, Diego and {Lanzafame}, Giuseppe and {Chakrabarti}, Sandip K.},
        title = "{Simulation of Thick Accretion Disks with Standing Shocks by Smoothed Particle Hydrodynamics}",
      journal = {\apj},
         year = 1994,
        month = apr,
       volume = {425},
        pages = {161},
          doi = {10.1086/173972},
archivePrefix = {arXiv},
       eprint = {astro-ph/9310047},
 primaryClass = {astro-ph},
       adsurl = {https://ui.adsabs.harvard.edu/abs/1994ApJ...425..161M}
}

@ARTICLE{Molteni-etal-96a,
       author = {{Molteni}, Diego and {Sponholz}, Hanno and {Chakrabarti}, Sandip K.},
        title = "{Resonance Oscillation of Radiative Shock Waves in Accretion Disks around Compact Objects}",
      journal = {\apj},
         year = 1996,
        month = feb,
       volume = {457},
        pages = {805},
          doi = {10.1086/176775},
archivePrefix = {arXiv},
       eprint = {astro-ph/9508022},
 primaryClass = {astro-ph},
       adsurl = {https://ui.adsabs.harvard.edu/abs/1996ApJ...457..805M}
}

@ARTICLE{Molteni-etal-96b,
       author = {{Molteni}, Diego and {Ryu}, Dongsu and {Chakrabarti}, Sandip K.},
        title = "{Numerical Simulations of Standing Shocks in Accretion Flows around Black Holes: A Comparative Study}",
      journal = {\apj},
         year = 1996,
        month = oct,
       volume = {470},
        pages = {460},
          doi = {10.1086/177877},
archivePrefix = {arXiv},
       eprint = {astro-ph/9605116},
 primaryClass = {astro-ph},
       adsurl = {https://ui.adsabs.harvard.edu/abs/1996ApJ...470..460M}
}

@ARTICLE{Aktar-etal-25,
       author = {{Aktar}, Ramiz and {Pan}, Kuo-Chuan and {Okuda}, Toru},
        title = "{Global Resistive Magnetohydrodynamic Accretion Flows around Spinning Active Galactic Nuclei: Impact of Resistivity on the Magnetically Arrested Disk State}",
      journal = {\apj},
         year = 2025,
        month = jul,
       volume = {987},
       number = {2},
          eid = {145},
        pages = {145},
          doi = {10.3847/1538-4357/ade05a},
archivePrefix = {arXiv},
       eprint = {2505.23051},
 primaryClass = {astro-ph.HE},
       adsurl = {https://ui.adsabs.harvard.edu/abs/2025ApJ...987..145A}
}

@ARTICLE{Aktar-etal-26,
       author = {{Aktar}, Ramiz and {Pan}, Kuo-Chuan and {Okuda}, Toru},
        title = "{3D Global Relativistic Radiation Magnetohydrodynamics of Magnetically Arrested Disk Accretion Flows in Active Galactic Nuclei}",
      journal = {\apj},
         year = 2026,
        month = apr,
       volume = {1001},
       number = {1},
          eid = {88},
        pages = {88},
          doi = {10.3847/1538-4357/ae5184},
archivePrefix = {arXiv},
       eprint = {2603.11452},
 primaryClass = {astro-ph.HE},
       adsurl = {https://ui.adsabs.harvard.edu/abs/2026ApJ..1001...88A}
}

@ARTICLE{Coppi-Blandford-90,
       author = {{Coppi}, P.~S. and {Blandford}, R.~D.},
        title = "{Reaction rates and energy distributions for elementary processes in relativistic pair plasmas}",
      journal = {\mnras},
         year = 1990,
        month = aug,
       volume = {245},
       number = {3},
        pages = {453-453},
          doi = {10.1093/mnras/245.3.453},
       adsurl = {https://ui.adsabs.harvard.edu/abs/1990MNRAS.245..453C}
}

@ARTICLE{Michail-etal-24,
       author = {{Michail}, Joseph M. and {Yusef-Zadeh}, Farhad and {Wardle}, Mark and {Kunneriath}, Devaky and {Hora}, Joseph L. and {Bushouse}, Howard and {Fazio}, Giovanni G. and {Markoff}, Sera and {Smith}, Howard A.},
        title = "{Multiwavelength Observations of Sgr A*. II. 2019 July 21 and 26}",
      journal = {\apj},
         year = 2024,
        month = aug,
       volume = {971},
       number = {1},
          eid = {52},
        pages = {52},
          doi = {10.3847/1538-4357/ad5332},
archivePrefix = {arXiv},
       eprint = {2406.01671},
 primaryClass = {astro-ph.HE},
       adsurl = {https://ui.adsabs.harvard.edu/abs/2024ApJ...971...52M}
}

@ARTICLE{Yusef-Zadeh-etal-08,
       author = {{Yusef-Zadeh}, F. and {Wardle}, M. and {Heinke}, C. and {Dowell}, C.~D. and {Roberts}, D. and {Baganoff}, F.~K. and {Cotton}, W.},
        title = "{Simultaneous Chandra, CSO, and VLA Observations of Sgr A*: The Nature of Flaring Activity}",
      journal = {\apj},
         year = 2008,
        month = jul,
       volume = {682},
       number = {1},
        pages = {361-372},
          doi = {10.1086/588803},
archivePrefix = {arXiv},
       eprint = {0712.2882},
 primaryClass = {astro-ph},
       adsurl = {https://ui.adsabs.harvard.edu/abs/2008ApJ...682..361Y}
}

@ARTICLE{van-der-Laan-66,
       author = {{van der Laan}, H.},
        title = "{A Model for Variable Extragalactic Radio Sources}",
      journal = {\nat},
         year = 1966,
        month = sep,
       volume = {211},
       number = {5054},
        pages = {1131-1133},
          doi = {10.1038/2111131a0},
       adsurl = {https://ui.adsabs.harvard.edu/abs/1966Natur.211.1131V}
}

@ARTICLE{Debnath-etal-24,
       author = {{Debnath}, Sanjit and {Chattopadhyay}, Indranil and {Joshi}, Raj Kishor},
        title = "{Oscillating shocks in the transonic viscous, variable {\ensuremath{\Gamma}} accretion flows around black holes}",
      journal = {\mnras},
         year = 2024,
        month = mar,
       volume = {528},
       number = {3},
        pages = {3964-3980},
          doi = {10.1093/mnras/stae181},
archivePrefix = {arXiv},
       eprint = {2401.07786},
 primaryClass = {astro-ph.HE},
       adsurl = {https://ui.adsabs.harvard.edu/abs/2024MNRAS.528.3964D}
}

@ARTICLE{Mao-etal-25,
       author = {{Mao}, Jirong and {Dihingia}, Indu K. and {Mizuno}, Yosuke and {Nagataki}, Shigehiro},
        title = "{Low Angular Momentum Black Hole Accretion: First GRMHD Evidence of Standing Shocks}",
      journal = {arXiv e-prints},
         year = 2025,
        month = jul,
          eid = {arXiv:2507.22506},
        pages = {arXiv:2507.22506},
          doi = {10.48550/arXiv.2507.22506},
archivePrefix = {arXiv},
       eprint = {2507.22506},
 primaryClass = {astro-ph.HE},
       adsurl = {https://ui.adsabs.harvard.edu/abs/2025arXiv250722506M}
}

@ARTICLE{Mitra-etal-22,
       author = {{Mitra}, Samik and {Maity}, Debaprasad and {Dihingia}, Indu Kalpa and {Das}, Santabrata},
        title = "{Study of general relativistic magnetohydrodynamic accretion flow around black holes}",
      journal = {\mnras},
         year = 2022,
        month = nov,
       volume = {516},
       number = {4},
        pages = {5092-5109},
          doi = {10.1093/mnras/stac2431},
archivePrefix = {arXiv},
       eprint = {2204.01412},
 primaryClass = {astro-ph.HE},
       adsurl = {https://ui.adsabs.harvard.edu/abs/2022MNRAS.516.5092M}
}

@ARTICLE{Sarkar-etal-23,
       author = {{Sarkar}, Shilpa and {Singh}, Kuldeep and {Chattopadhyay}, Indranil and {Laurent}, Philippe},
        title = "{Two-temperature accretion flows around strongly magnetized stars and their spectral analysis}",
      journal = {\mnras},
         year = 2023,
        month = jul,
       volume = {522},
       number = {3},
        pages = {3735-3752},
          doi = {10.1093/mnras/stad1064},
archivePrefix = {arXiv},
       eprint = {2304.03329},
 primaryClass = {astro-ph.HE},
       adsurl = {https://ui.adsabs.harvard.edu/abs/2023MNRAS.522.3735S}
}

@ARTICLE{Sarkar-25,
       author = {{Sarkar}, Shilpa},
        title = "{Novel methodology to obtain transonic solutions for dissipative flows around compact objects}",
      journal = {\prd},
         year = 2025,
        month = jul,
       volume = {112},
       number = {2},
          eid = {023018},
        pages = {023018},
          doi = {10.1103/s2js-j6kp},
archivePrefix = {arXiv},
       eprint = {2505.24839},
 primaryClass = {astro-ph.HE},
       adsurl = {https://ui.adsabs.harvard.edu/abs/2025PhRvD.112b3018S}
}

@ARTICLE{Dihingia-etal-25,
       author = {{Dihingia}, Indu K. and {Uniyal}, Akhil and {Mizuno}, Yosuke},
        title = "{Imprints of Different Types of Low-angular-momentum Accretion Flow Solutions in General Relativistic Hydrodynamic Simulations}",
      journal = {\apj},
         year = 2025,
        month = sep,
       volume = {990},
       number = {1},
          eid = {35},
        pages = {35},
          doi = {10.3847/1538-4357/adf221},
archivePrefix = {arXiv},
       eprint = {2506.05750},
 primaryClass = {astro-ph.HE},
       adsurl = {https://ui.adsabs.harvard.edu/abs/2025ApJ...990...35D}
}

@ARTICLE{Aktar-etal-23,
       author = {{Aktar}, Ramiz and {Xue}, Li and {Zhang}, Li-Xin and {Luo}, Jing-Yi},
        title = "{Spiral shocks induced in a galactic gaseous disk: Hydrodynamic understanding of observational properties of spiral galaxies}",
      journal = {\aap},
         year = 2023,
        month = nov,
       volume = {679},
          eid = {A154},
        pages = {A154},
          doi = {10.1051/0004-6361/202346624},
archivePrefix = {arXiv},
       eprint = {2309.17271},
 primaryClass = {astro-ph.GA},
       adsurl = {https://ui.adsabs.harvard.edu/abs/2023A&A...679A.154A}
}

@ARTICLE{Chakrabarti-1990,
       author = {{Chakrabarti}, Sandip K.},
        title = "{Standing Shocks in the Rotating Winds and Accretion in Kerr Spacetime}",
      journal = {\apj},
         year = 1990,
        month = feb,
       volume = {350},
        pages = {275},
          doi = {10.1086/168379},
       adsurl = {https://ui.adsabs.harvard.edu/abs/1990ApJ...350..275C}
}

@ARTICLE{Chakrabarti-Das-01,
       author = {{Chakrabarti}, Sandip K. and {Das}, Santabrata},
        title = "{Model dependence of transonic properties of accretion flows around black holes}",
      journal = {\mnras},
         year = 2001,
        month = nov,
       volume = {327},
       number = {3},
        pages = {808-812},
          doi = {10.1046/j.1365-8711.2001.04758.x},
archivePrefix = {arXiv},
       eprint = {astro-ph/0402553},
 primaryClass = {astro-ph},
       adsurl = {https://ui.adsabs.harvard.edu/abs/2001MNRAS.327..808C}
}

@ARTICLE{Chakrabarti-Molteni-93,
       author = {{Chakrabarti}, Sandip K. and {Molteni}, Diego},
        title = "{Smoothed Particle Hydrodynamics Confronts Theory: Formation of Standing Shocks in Accretion Disks and Winds around Black Holes}",
      journal = {\apj},
         year = 1993,
        month = nov,
       volume = {417},
        pages = {671},
          doi = {10.1086/173345},
archivePrefix = {arXiv},
       eprint = {astro-ph/9310042},
 primaryClass = {astro-ph},
       adsurl = {https://ui.adsabs.harvard.edu/abs/1993ApJ...417..671C}
}

@ARTICLE{Chakrabarti-Molteni-95,
       author = {{Chakrabarti}, Sandip K. and {Molteni}, Diego},
        title = "{Viscosity prescriptions in accretion discs with shock waves}",
      journal = {\mnras},
         year = 1995,
        month = jan,
       volume = {272},
       number = {1},
        pages = {80-88},
          doi = {10.1093/mnras/272.1.80},
       adsurl = {https://ui.adsabs.harvard.edu/abs/1995MNRAS.272...80C}
}

@ARTICLE{Das-Chattopadhyay-Nandi-01,
       author = {{Das}, S. and {Chattopadhyay}, I. and {Nandi}, A. and {Chakrabarti}, S.~K.},
        title = "{Computation of outflow rates from accretion disks around black holes}",
      journal = {\aap},
         year = 2001,
        month = nov,
       volume = {379},
        pages = {683-689},
          doi = {10.1051/0004-6361:20011307},
archivePrefix = {arXiv},
       eprint = {astro-ph/0402555},
 primaryClass = {astro-ph},
       adsurl = {https://ui.adsabs.harvard.edu/abs/2001A&A...379..683D}
}
%%%%%%%%%%%%%%%%%%%%%%%%%%%%%%%%%%%%%%%%%%%%%%%%%%%%%%%%

%%%%%%%%%%%%%%%%%%%%%%%%%%%%%%%%%%%%%%%%%%%%%%%%%%

% Don't change these lines
%\bsp	% typesetting comment
\label{lastpage}
\end{document}